# Energy Saving Techniques for Energy Constrained CMOS Circuits and Systems

A thesis submitted in partial fulfillment of

the requirements for the degree of

Doctor of Philosophy

by

**Sivaneswaran Sankar**

**(Roll No. 14T070003)**

Under the guidance of

**Prof. Maryam Shojaei Baghini**

**Prof. V. Ramgopal Rao**

**Prof. Po-Hung Chen** (External co-supervisor)

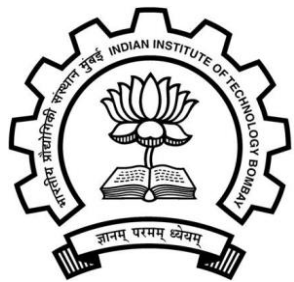

DEPARTMENT OF ELECTRICAL ENGINEERING

INDIAN INSTITUTE OF TECHNOLOGY BOMBAY

2022

# Thesis Approval

This thesis entitled **Energy Saving Techniques for Energy Constrained CMOS Circuits and Systems** by **Sivaneswaran Sankar** (Roll No: 14T070003) is approved for the degree of Doctor of Philosophy.

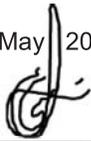
23 May 2022
__________________________
Prof. Maryam Shojaei Baghini
Supervisor

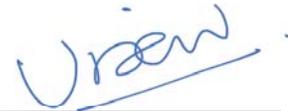
__________________________
Prof. Virendra Singh
Internal Examiner

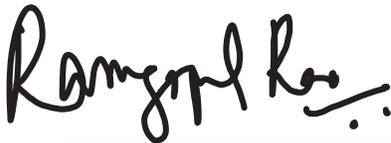
__________________________
Prof. V. Ramgopal Rao
Co-Supervisor

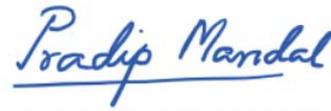
__________________________
Prof. Pradip Mandal
External Examiner

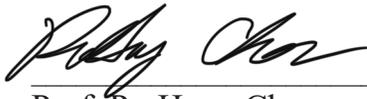
__________________________
Prof. Po-Hung Chen
External Co-Supervisor (NYCU, Taiwan)

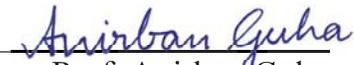
__________________________
Prof. Anirban Guha
Chairman

Date: 23rd May 2022
Place: Mumbai, India

i

# Declaration

I declare that this written submission represents my ideas in my own words and where others ideas or words have been included, I have adequately cited and referenced the original sources. I also declare that I have adhered to all principles of academic honesty and integrity and have not misrepresented or fabricated or falsified any idea/data/fact/source in my submission. I understand that any violation of the above will be cause for disciplinary action by the Institute and can also evoke penal action from the sources which have thus not been properly cited or from whom proper permission has not been taken when needed.

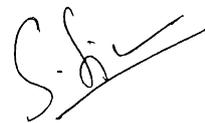

\_\_\_\_\_\_\_\_\_\_\_\_\_\_\_\_\_\_

Sivaneswaran Sankar
Roll No: 14T070003

Date: \_23rd May 2022\_\_

# Abstract

Portable devices like smartphones, tablets, wearable electronic devices, medical implants, wireless sensor nodes, and Internet-of-Things (IoT) devices have tremendous constraints on their energy consumption. Adding more functionalities onto the portable devices increases its energy consumption significantly. However, the energy capacity of the battery does not increase proportionally. Hence, to overcome the constraints on energy consumption, two main approaches are being undertaken by the designers to integrate more functionalities onto the energy-constrained systems. One approach involves reducing the inherent energy consumption of circuits, and the other involves harvesting energy from the ambient sources and utilizing it to power the circuits. In this thesis, both the approaches mentioned above are followed in developing energy-saving techniques for energy-constrained CMOS circuits and systems.

In applications like sensor nodes, medical implants, wearable devices etc., the standby (or) idle time is significantly higher than the active (or) on-time. Hence, the overall energy consumption is dominated by the leakage currents during idle times. In this thesis, two methods of reducing the leakage currents of the large digital blocks are explored. The first method is through the proposed switched-capacitor-assisted power gating (SwCap PG) technique, and the second method is by utilizing nano-electromechanical switches (NEMS) for power gating.

In the proposed switched-capacitor assisted power gating (SwCap PG), the PG switch is biased in the super turn-off and the super turn-on mode during the off-state and the on-state, respectively. The proposed SwCap PG is experimentally validated in the 180 nm CMOS technology. Measurement results of CMOS SwCap PG show that leakage current and $R_{ON}$ reduce by 186-226× and 18%, respectively, compared to the conventional PG technique. An alternate solution for the SwCap network using MEMS devices as the switching elements is implemented for additional benefits. Measurement results of MEMS SwCap PG show that leakage current and $R_{ON}$ reduce by 172× and 26%, respectively, compared to the conventional PG technique. Finally, the applicability of the SwCap PG in the nano-scale CMOS technologies is addressed.

In this thesis, leakage current reduction using nano-electromechanical switches (NEMS)

is also explored. NEMS switches in principal offer close to infinite resistance in the off-state. The power gating technique is analyzed with a NEMS switch using detailed circuit-level simulations to obtain the conditions under which one can obtain net energy savings compared with FinFET-based power gating. Finally, applicability in the energy reduction on a system-on-chip for a mobile platform made using a 14 nm gate length FinFET device is evaluated.

Apart from the leakage currents in the digital circuits, the other primary source of energy consumption is the quiescent currents of the analog front-end circuits. This thesis proposes a novel technique of realizing discrete-time (D-T) signal amplification using nano-electromechanical switches (NEMS). The amplifier uses mechanical switches instead of traditional solid-state devices and consumes only dynamic power. As a proof of concept, the proposed NEMS D-T amplifier is demonstrated in circuit simulations using the calibrated Verilog-A models of the designed NEMS device. The simulated amplifier achieves a gain of 5, handles the maximum differential input signal of 0.65 V, and consumes only 0.6 µW of dynamic power for a sampling frequency of 100 kHz. The non-idealities present in the proposed amplifier are highlighted, and possible ways to overcome them are discussed. Finally, the design considerations required for the NEMS D-T amplifier are also described.

Apart from the three energy reduction techniques mentioned above, harvesting energy from mechanical vibrations is also explored in this thesis. An efficient technique of extracting power from the piezoelectric transducer is proposed. The proposed method allows for extracting the highest possible output power for a given CMOS technology. The operational steps are self-timed, and the negative voltage swing across the transducer is avoided. The proposed inductive rectifier is realized using the buck-boost power stage of the DC energy harvesting system for enabling efficient multi-source harvesting. Test chip is fabricated in the 180 nm CMOS technology, having a $V_{MAX}$ of 3.3 V. The proposed rectifier extracts power in a single stage, even at a lower rectified output voltage $\leq 1$ V. For a piezo open-circuit voltage of 1 V, the proposed rectifier extracts 3.68× more power than the maximum output power of a full-bridge rectifier with ideal diodes. This is despite the voltage loss occurring during accumulation in the power stage implementation.

# Contents

# List of Tables

# List of Figures

# Chapter 1

# Introduction

## 1.1 Motivation

Tremendous integration of functionalities on a chip increases its energy consumption significantly. However, the energy capacity of the battery does not increase proportionally. Due to this, portable electronic devices like smartphones, tablets, wearable devices, medical implants, and sensor nodes have tremendous constraints on their energy consumption. Two main approaches are being undertaken by the designers to integrate more functionalities onto the energy-constrained systems. One approach involves reducing the inherent energy consumption of circuits, and the other involves harvesting energy from the ambient sources and utilizing it to power the circuits [1], [2]. In this thesis, both the approaches mentioned above are followed in developing energy-saving techniques for energy-constrained CMOS circuits and systems.

Consider the generic representation of the signal chain in a portable electronic device shown in Fig. 1.1 [3]. The sensors or actuators along with the analog front-end, interface to the digital processor through the mixed-signal blocks (ADC/DAC). Every portable electronic device interacts with other devices through wireless communication links. All the blocks derive their supply voltages and currents from the power management unit (PMU). The PMU plays a vital role in efficiently delivering the stored energy from the battery to the various blocks. More functionalities are integrated into the portable devices over time. However, the battery's energy capacity is not scaling up proportionally. Hence, the circuits in the portable devices are energy-constrained.

In applications like sensor nodes, medical implants, wearable devices etc., the standby (or) idle time is significantly higher than the active (or) on-time. This could be observed from the typical power dissipation profile [4] for a portable generic sensor node shown in Fig. 1.2. Hence, the overall energy consumption is dominated by the leakage currents during idle times. Thus, reducing the leakage currents during idle times improves the battery life of portable electronic devices.

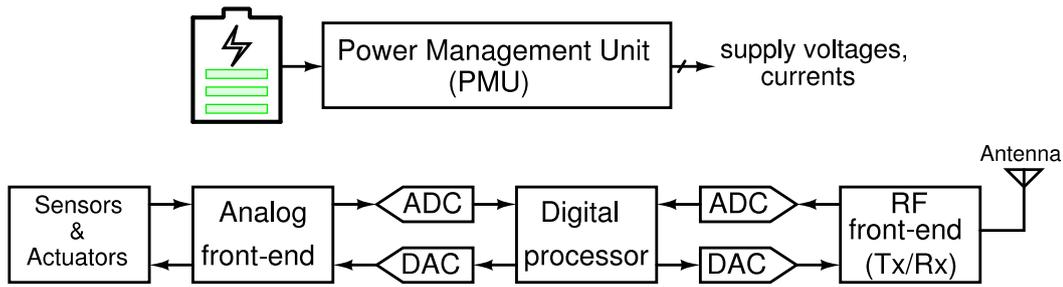

Fig. 1.1. Generic representation of the signal chain in the portable electronic device.

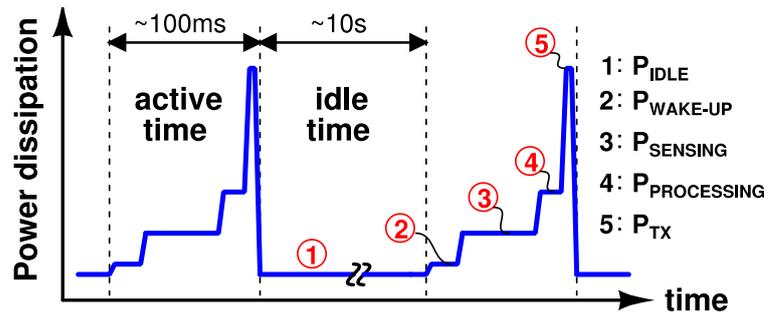

Fig. 1.2. Typical power dissipation profile for a portable generic sensor node.

The other primary source of energy consumption is the quiescent currents drawn from the power supply in the analog front-end block [5] shown in Fig. 1.1. Since the sensor's analog signal amplitude is low enough, it must be amplified before being digitized by the ADC. The function of amplification and sampling in the ADC is combined together in a discrete-time (D-T) signal amplifier [6]. However, the implementation of the D-T amplifier requires an operational amplifier (op-amp). Limitations of the op-amp (finite open-loop DC gain and bandwidth, voltage swing limit, non-linearity, and noise) significantly affect the performance of the D-T amplifier, and it also consumes significant quiescent power [7]. Hence, performing the D-T signal amplification without the necessity of an op-amp greatly aids in saving the energy consumption of the analog front-end circuits.

Energy harvesting from the ambient energy sources in conjunction with the battery's chemical energy can power the circuits [4]. In this way, the need for replacement or frequent charging of the battery can be avoided. Fig. 1.3 highlights the various sources of energy available in the environment. Due to its abundant nature, significant research has been carried out in the development of an efficient energy harvesting interface for harvesting solar and thermal energy [8]. The advent of mobile communication devices and smartphones has resulted in significant penetration of RF signals in our surrounding environment. However, the available RF power levels are quite low. Moreover, harvesting a specific band of RF signals needs a dedicated impedance matching network followed by voltage rectification [9]. Hence, harvesting energy from RF signals is quite different from harvesting DC energy sources (solar and thermal).

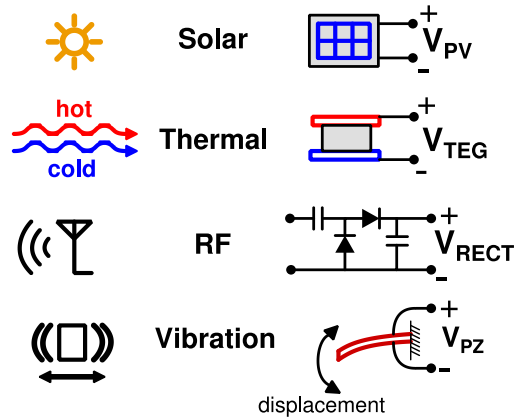

Fig. 1.3. Various energy sources available in the environment for energy harvesting.

Vibrations present in the environment are also a source of a mechanical form of energy. They could be harnessed to power the sensor nodes used for environmental monitoring, structural health monitoring, machine or motor health monitoring, power electronics systems monitoring, etc. Piezoelectric transducers convert mechanical energy into electrical energy. Due to the sinusoidal nature of vibrations, the resulting output current from the transducer is also sinusoidal. Recent advancements in rectifier architectures using the inductor and the switches have enabled efficient piezo energy harvesting (PEH) [10]. Hence, the inductive power stage developed for harvesting the DC energy sources (solar and thermal) can be re-utilized for the piezo energy harvesting. Moreover, the amount of harvested energy depends on the ambient conditions and is not always reliable. Hence, an efficient multi-source energy harvesting system along with the provision to store the excess energy and reuse it later is required.

As motivated by the above problem, the primary focus of this thesis will be on the development of energy-saving techniques through leakage reduction in digital circuits, quiescent current reduction in analog front-end circuits, and harnessing of mechanical energy.

## 1.2 Research Contributions

### 1.2.1 Energy saving by leakage current reduction in digital circuits

Large digital blocks in portable electronic devices remain as one of the top contributors to the leakage currents during idle times. Power gating (PG) is one of the key techniques in reducing the leakage currents during idle time [11]. However, it introduces a trade-off in the amount of leakage power saving during the off-state and the additional delay penalty overhead during the on-state. In this thesis, two possible ways to overcome the trade-off in power gating are presented. One is through the proposed switched-capacitor-assisted power gating (SwCap PG) technique [12], and the other is by utilizing nano-electromechanical switches (NEMS) for power gating [13].

- **Switched-Capacitor Assisted Power Gating (SwCap PG):** Switched-Capacitor assisted Power Gating (SwCap PG) is proposed for reducing the leakage currents of large digital circuits [12]. For the first time, the PG switch is biased in the super turn-off and the super turn-on mode during the off-state and the on-state, respectively. A simple switched-capacitor network reconfigures and biases the PG switch in four possible states with low area and power overhead. During the super turn-off, voltage stress is avoided in the PG switch when the circuit load uses supply voltage equal to the nominal $V_{DD}$ in a given technology, and maximum possible leakage current reduction is achieved by the optimal biasing of the gate voltage. The proposed SwCap PG is experimentally validated in the 180nm CMOS technology. Measurement results of CMOS SwCap PG show that leakage current and $R_{ON}$ reduce by 186-226× and 18%, respectively, as compared to the conventional PG. An alternate solution for the SwCap network using MEMS devices as the switching elements is implemented for additional benefits. Measurement results of MEMS SwCap PG show that leakage current and $R_{ON}$ reduce by 172× and 26%, respectively, compared to the conventional PG. Finally, the applicability of the SwCap PG in the nano-scale CMOS technologies is addressed.

- **NEMS Power Gating:** Nano-electromechanical switches (NEMS) in principal offer close to infinite resistance in the off-state. So by using NEMS switches for power gating, the on-state performance can be improved by adding multiple parallel switches without increasing the leakage current during the off-state. This eventually breaks the trade-off mentioned earlier. NEMS switches have been recently proposed for the power gating application. However, a detailed analysis of the conditions at which the NEMS devices will have an impact is missing. In this thesis, the technique of power gating is analyzed with a NEMS switch using detailed circuit-level simulations to obtain the conditions under which one can obtain net energy savings as compared with FinFET-based power gating [13]. Finally, applicability in the energy reduction on a system-on-chip for a mobile platform made using a 14 nm gate length FinFET device is evaluated.

### 1.2.2 Energy saving by quiescent current reduction in analog front-end

As discussed in Section 1.1, quiescent currents in the analog front-end blocks also remain as one of the significant sources of energy consumption in the generic portable electronic device described in Fig. 1.1. An alternative method of discrete-time (D-T) signal amplification without the need for a power-hungry operational amplifier (op-amp) is proposed in this work.

- **NEMS D-T Signal Amplification:** In this thesis, a novel technique of realizing discrete-time (D-T) signal amplification using nano-electromechanical switches (NEMS) is proposed [14]. The amplifier uses mechanical switches instead of traditional solid-state devices and acts as an inherent sample and hold amplifier. The proposed NEMS D-T amplifier operates on a wide dynamic range of signals without consuming any DC power. Moreover, the proposed amplifier does not suffer from the leakage current and the non-linearity associated with the sampling ohmic switch. As a proof of concept, the proposed NEMS D-T amplifier is demonstrated in circuit simulations using the calibrated Verilog-A models of the NEMS device. The simulated amplifier achieves a gain of ∼5, handles the maximum differential input signal of 0.65 V, and consumes only 0.6 µW of dynamic power for a sampling frequency of 100 kHz. The non-idealities present in the proposed amplifier are highlighted, and possible ways to overcome them are discussed. Finally, the design considerations required for the NEMS D-T amplifier are described.

### 1.2.3 Energy saving through harnessing of mechanical energy

Harvesting energy from the ambient sources and utilizing it to power the circuits can enable energy-autonomous portable electronic devices, as discussed in Section 1.1. In this thesis, an efficient technique of harvesting energy from mechanical vibrations is presented.

- **Efficient Inductive Rectifier for Piezo Energy Harvesting**: In this thesis, a new efficient method of extracting power from the piezoelectric transducer using the inductive rectifier is proposed [15]. In the proposed method, the internal capacitance of the transducer is initially pre-charged, and the generated charges in response to the mechanical vibration are accumulated on the capacitor. When the accumulated voltage reaches a maximum allowed value ($V_{MAX}$), the total energy stored in the capacitor is transferred to the output, and the process is repeated. The proposed method ensures that the voltage swing across the transducer is always maximum for extracting the highest possible output power for given CMOS technology. The operational steps are self-timed, and the negative voltage swing across the transducer is avoided. The proposed inductive rectifier is realized using the buck-boost power stage of the DC energy harvesting system for enabling efficient multi-source harvesting. Test chip is fabricated in the 180 nm CMOS technology, having a $V_{MAX}$ of 3.3 V. The proposed rectifier extracts power in a single stage, even at a lower rectified output voltage $\leq$ 1 V. For a piezo open-circuit voltage of 1 V, the proposed rectifier extracts 3.68× more power than the maximum output power of a full-bridge rectifier with ideal diodes. This is despite the voltage loss occurring during accumulation in the power stage implementation.

### 1.2.4 Summary of the research contributions

Fig. 1.4 highlights the various research contributions made in this thesis. Both the directions involving the reduction of inherent energy consumption and harvesting of energy from the ambient source are followed to achieve energy saving in the energy-constrained CMOS circuits and systems.

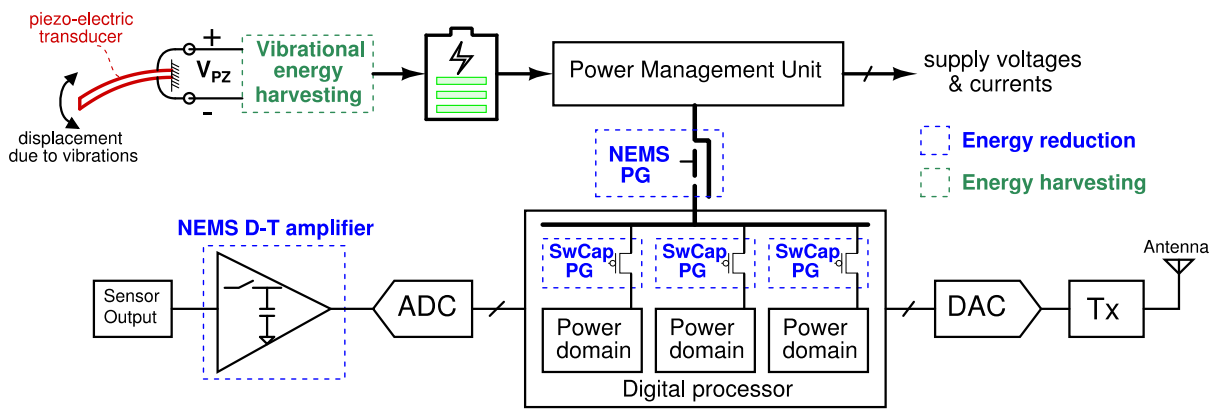

Fig. 1.4. Summary of the research contributions by this thesis (highlighted in blue and green color).

## 1.3 Thesis Organization

This thesis is organized as follows:

Chapter 2 presents the work on switched-capacitor assisted power gating (SwCap PG). A review of the existing super cut-off PG methods is initially discussed. Later, the concept of the proposed SwCap PG is presented. The circuit realization of the CMOS SwCap PG is provided. Benefits and the realization of MEMS SwCap PG are discussed. Chip measurement results of the proposed technique is presented in this chapter. Applicability of the SwCap PG in the nano-scale CMOS technologies is also addressed in this chapter.

Chapter 3 presents the work on NEMS power gating. The description of the considered NEMS switch and its simplified circuit model is initially provided. The expression for net energy gain ($E_G$) that is obtained using NEMS power gating is presented. The details of the benchmark circuit used in the simulations are provided. Later, the NEMS power gating is analyzed with respect to the FinFET power gating. Finally, the applicability of NEMS power gating to SoCs is addressed.

Chapter 4 presents the work on NEMS D-T signal amplifier. The NEMS devices and their characteristics are presented. The principle of operation of the proposed NEMS D-T amplifier is described. The details of the NEMS D-T amplifier along with the simulation results are presented. Non-idealities present in the NEMS D-T amplifier are discussed. Finally, the design guidelines required for the NEMS D-T amplifier are provided.

Chapter 5 presents the work on the vibrational energy harvesting using the proposed inductive rectifier. Prior approaches to piezo energy harvesting are summarized. The concept of the proposed inductive rectifier is then presented. The circuit implementation details of the proposed inductive rectifier is discussed. Chip measurement results of the proposed inductive rectifier are presented. Finally, our main conclusions presented in this thesis is summarized.

Chapter 6 provides the conclusion of our work by highlighting the key contributions made in this thesis. It also suggests the future scope of research that could be pursued.

# 1.4 References

[1] L. Lin, S. Jain and M. Alioto, "Integrated Power Management for Battery-Indifferent Systems With Ultra-Wide Adaptation Down to nW," in *IEEE Journal of Solid-State Circuits*, vol. 55, no. 4, pp. 967-976, April 2020.

[2] L. Lin, S. Jain and M. Alioto, "Sub-nW Microcontroller With Dual-Mode Logic and Self-Startup for Battery-Indifferent Sensor Nodes," in *IEEE Journal of Solid-State Circuits*, vol. 56, no. 5, pp. 1618-1629, May 2021.

[3] "Integrated Connections: The Signal Chain Guide", *White paper*, Avnet, 2017.

[4] S. S. Amin and P. P. Mercier, "MISIMO: A Multi-Input Single-Inductor Multi-Output Energy Harvesting Platform in 28-nm FDSOI for Powering Net-Zero-Energy Systems," in *IEEE Journal of Solid-State Circuits*, vol. 53, no. 12, pp. 3407-3419, Dec. 2018.

[5] M. Maruyama, S. Taguchi, M. Yamanoue and K. Iizuka, "An Analog Front-End for a Multifunction Sensor Employing a Weak-Inversion Biasing Technique With 26 nVrms, 25 aCrms, and 19 fArms Input-Referred Noise," in *IEEE Journal of Solid-State Circuits*, vol. 51, no. 10, pp. 2252-2261, Oct. 2016.

[6] Y. Huang et al., "A Self-Powered CMOS Reconfigurable Multi-Sensor SoC for Biomedical Applications," in *IEEE Journal of Solid-State Circuits*, vol. 49, no. 4, pp. 851-866, April 2014.

[7] B. Razavi, "Design of Analog CMOS Integrated circuits", McGraw-Hill, 2002.

[8] P.-H. Chen, H.-C. Cheng, and C.-L. Lo, "A Single-Inductor Triple-Source Quad-Mode Energy-Harvesting Interface With Automatic Source Selection and Reversely Polarized Energy Recycling," in *IEEE J. Solid-State Circuits*, vol. 54, no. 10, pp. 2671–2679, Oct. 2019.

[9] S. M. Noghabaei, R. L. Radin, Y. Savaria and M. Sawan, "A High-Sensitivity Wide Input-Power-Range Ultra-Low-Power RF Energy Harvester for IoT Applications," in *IEEE Transactions on Circuits and Systems I: Regular Papers (Early Access)*, doi: 10.1109/TCSI.2021.3099011.

[10] K. Yoon, S. Hong and G. Cho, "Double Pile-Up Resonance Energy Harvesting Circuit for Piezoelectric and Thermoelectric Materials," in *IEEE Journal of Solid-State Circuits*, vol. 53, no. 4, pp. 1049-1060, April 2018.

[11] W. Gomes et al., "8.1 lakefield and mobility compute: A 3D stacked 10 nm and 22FFL hybrid processor system in 12×12 mm$^2$, 1 mm package-on-package," in *IEEE Int. Solid-State Circuits Conf. (ISSCC) Dig. Tech. Papers*, Feb. 2020, pp. 144–146.

[12] S. Sankar, M. Goel, P. -H. Chen, V. R. Rao and M. S. Baghini, "Switched-Capacitor-Assisted Power Gating for Ultra-Low Standby Power in CMOS Digital ICs," in *IEEE Transactions on Circuits and Systems I: Regular Papers*, vol. 67, no. 12, pp. 4281-4294, Dec. 2020.

[13] S. Sankar, U. S. Kumar, M. Goel, M. S. Baghini and V. R. Rao, "Considerations for Static Energy Reduction in Digital CMOS ICs Using NEMS Power Gating," in *IEEE Transactions on Electron Devices*, vol. 64, no. 3, pp. 1399-1403, March 2017.

[14] S. Sankar, M. Goel, M. S. Baghini and V. R. Rao, "A Novel Method of Discrete-Time Signal Amplification Using NEMS Devices," in *IEEE Transactions on Electron Devices*, vol. 65, no. 11, pp. 5111-5117, Nov. 2018.

[15] S. Sankar, P. -H. Chen and M. S. Baghini, "An Efficient Inductive Rectifier Based Piezo-Energy Harvesting Using Recursive Pre-Charge and Accumulation Operation," *in IEEE Journal of Solid-State Circuits (Early Access)*, doi: 10.1109/JSSC.2022.3153590.

# Chapter 2

# Switched-Capacitor Assisted Power Gating

This chapter presents Switched-Capacitor assisted Power Gating (SwCap PG) for reducing the leakage currents of large digital circuits. For the first time, PG switch is biased in the super turn-off and the super turn-on mode during the off-state and the on-state, respectively. A simple switched-capacitor network reconfigures and biases the PG switch in four different possible states with low area and power overhead. During the super turn-off, voltage stress is avoided in the PG switch when the circuit load uses supply voltage equal to the nominal $V_{DD}$ in a given technology, and maximum possible leakage current reduction is achieved by the optimal biasing of the gate voltage. The proposed SwCap PG is experimentally validated in the 180nm CMOS technology. Measurement results of the CMOS SwCap PG are presented. In 180nm CMOS technology, the SwCap network and the control circuit occupies an area of 0.0065 mm$^2$ (14% of the PG switch area). However, for the same $R_{ON}$, the total area including the PG switch is reduced by 6.5%, due to the super turn-on implementation. An alternate solution for SwCap network using MEMS devices as the switching elements is implemented for additional benefits, and its measurement results are presented. Finally, the applicability of the SwCap PG in the nano-scale CMOS technologies is addressed. Parts of this chapter have been published in [44].

## 2.1 Introduction to Super Cut-off Power Gating

Leakage currents dominate the overall energy consumption of circuits with long idle times [1]. Portable devices like smartphones, tablets, wearable devices, medical implants, and sensor nodes have tremendous constraints on their energy consumption. Energy harvesting from ambient energy is one of the possible solutions to prolong battery life. However, it is not sufficient enough. Hence, significant efforts are invested in lowering idle time energy consumption. Power gating (PG) is one of the key techniques in reducing the leakage currents during the idle time [1]-[7], [12]-[27], and [39]. Fig. 2.1 (a) shows the conventional PG using a PMOS switch, which disconnects the circuit load from $V_{DD}$ during idle time.

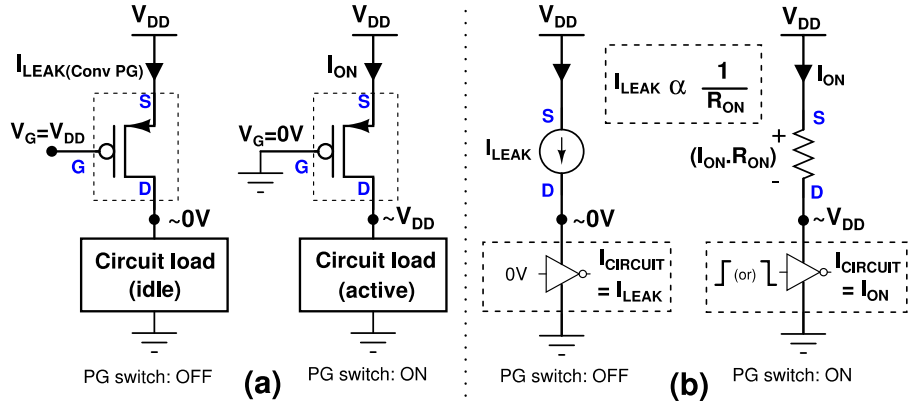

Fig. 2.1. (a) Conventional PG, and (b) equivalent model of the PG switch.

The equivalent model of the PG switch during the off-state and the on-state is given in Fig. 2.1 (b). Due to the finite off-state leakage current in the PG switch, the leakage current is still drawn from $V_{DD}$, but is now lower when compared to the case without power gating (since the width of the PG switch $\ll$ total width of transistors in the circuit load). The leakage current for a PMOS device is given by (2.1). The PG switch acts as a resistor in the on-state, and its on-resistance ($R_{ON}$) is given by (2.2).

$$I_{LEAK} = \mu_p C_{OX} \frac{W}{L}(n-1)V_t^2 e^{\frac{(V_{SGP}-|V_{THP}|)}{nV_t}} \qquad (2.1)$$

$$R_{ON} = \frac{1}{\mu_p C_{OX}\frac{W}{L}(V_{SGP}-|V_{THP}|)} \qquad (2.2)$$

where $\mu_p$ is the hole mobility, $C_{OX}$ is the oxide capacitance, $\frac{W}{L}$ is the aspect ratio of the device, $n$ is a constant (~1.5), $V_t$ is the thermal voltage, $V_{SGP}$ and $V_{THP}$ is the source to gate voltage and the threshold voltage of the PMOS device. In the case of the conventional PG, the leakage current and $R_{ON}$ is given by (2.3) and (2.4), respectively.

$$I_{LEAK} = \mu_p C_{OX} \frac{W}{L}(n-1)V_t^2 e^{\frac{-|V_{THP}|}{nV_t}} \qquad (2.3)$$

$$R_{ON} = \frac{1}{\mu_p C_{OX}\frac{W}{L}(V_{DD}-|V_{THP}|)} \qquad (2.4)$$

As seen from (2.3) and (2.4), reducing the IR drop across the switch in the on-state by using wider transistors increases the leakage currents in the off-state. Hence, achieving significant leakage reduction, along with lower IR drop across the switch, is one of the challenging tasks in power gating.

Dynamic body bias in the PG switch lowers $V_{THP}$ during the on-state, and vice-versa during the off-state. Hence, $R_{ON}$ and the leakage current can be reduced during the on-state and the off-state, respectively [7]. In the nano-scale CMOS technologies like FinFET or Gate all-around (GAA) FET devices, which are widely used in most of the SoC's, the body effect is negligible. This is due to the fact that the channel region in these devices are fully depleted [46]. Besides, GAAFETs have an inherently isolated channel from the substrate. Hence, a method for reducing the off-state leakage current and on-resistance of the PG switch that is applicable even for nano-scale CMOS technologies like FinFETs and GAA FETs is required. The works from [8]-[10] proposed using Nano-Electro-Mechanical Switches (NEMS) as a PG switch due to their near-infinite off-resistance. It leads to a significant area penalty for achieving $R_{ON}$ of few mΩ's and possess integration challenges with the CMOS process.

As observed from (2.1), for a PMOS device biased with a positive gate to source voltage ($V_{GSP}$), the leakage current exponentially reduces. This technique of leakage reduction in the PG switch is referred to as super cut-off (or super turn-off) PG and is illustrated in Fig. 2.2 (a). However, there exists an optimum value of $V_{GSP}$, referred to as $V_{GSP(opt)}$. This is because of the Gate-Induced Drain Leakage (GIDL) phenomenon, which starts to increase the leakage current for $V_{GSP} > V_{GSP(opt)}$ [11]. As shown in Fig. 2.2 (a), the gate voltage is raised above $V_{DD}$ in the super cut-off state, resulting in the gate to drain voltage ($V_{GDP}$) of the PMOS PG switch to exceed $V_{DD}$.

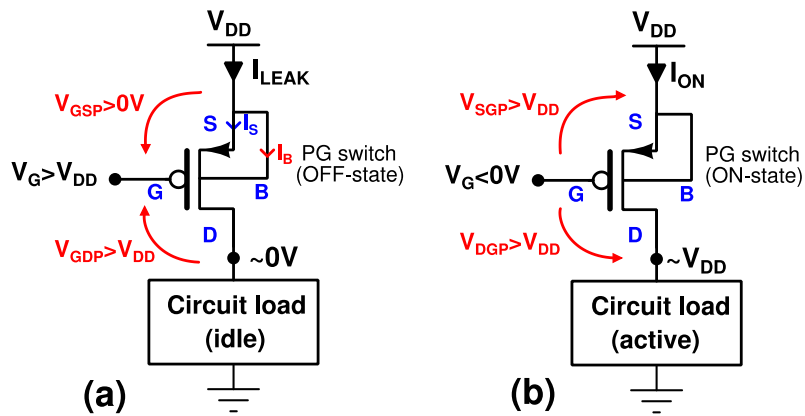

Fig. 2.2. (a) Super turn-off (cut-off) PMOS PG, and (b) super turn-on PMOS PG.

In [12]-[19], super cut-off PG is employed for near-threshold computing circuits, where the supply voltage ($V_{DD} \approx V_{TH}$) used by the circuit load is less than the nominal $V_{DD}$ of technology. Hence, the voltage stress on the PG switch during the super cut-off state was not a concern, and the core devices were used for the PG switch. In [20] and [21], core devices were used in the super cut-off PG switch for the circuit load whose supply voltage is equal to the nominal $V_{DD}$ of technology. Hence, the PG switch is voltage stressed in the super cut-off state, since $V_{GDP}$ exceeds the nominal allowed value of $V_{DD}$. None of the existing works allow super cut-off PG for the circuit load with nominal $V_{DD}$, while avoiding voltage stress and achieving maximum leakage current reduction (by biasing at $V_{GSP(opt)}$).

The new contributions in the proposed switched-capacitor assisted power gating (SwCap PG) technique are as follows,

1. For $V_{DD}$ = nominal $V_{DD}$ of the technology, voltage stress is avoided in the super cut-off PG switch by using thick-oxide device. $R_{ON}$ of the thick-oxide PG switch is reduced by biasing it in the super turn-on state, as shown in Fig. 2.2 (b).
2. For the first time, both the super turn-off and the super turn-on modes are realized together on the PG switch.
3. The proposed circuit realization allows for the optimum biasing of the gate voltage in the thick-oxide PG switch during the super turn-off (for maximum leakage reduction) and the super turn-on state (for maximum $R_{ON}$ reduction). This results in the lowest possible leakage current for the same $R_{ON}$ among the various PG switch options.
4. A simple switched-capacitor network reconfigures and biases the PG switch in four possible states with low power and area overhead. Two different implementations for the switched-capacitor network are realized.
    a) **CMOS SwCap PG** allows for the complete on-chip integration. The CMOS SwCap network is implemented without any additional circuit complexity, despite the node voltages exceeding the supply rails.
    b) **MEMS SwCap PG** is used for the process technology without an isolated NMOS device (to handle negative voltage). MEMS devices are used as the switching element in the switch network. This also provides an additional advantage of avoiding the switch leakage and the voltage error present in the CMOS SwCap network.

Due to the constraints on the energy consumption, commercial microprocessors in [2] and [22]-[24] uses thick-oxide NMOS device as the PG switch owing to the higher priority on minimizing the leakage currents rather than the area overhead. For reducing $R_{ON}$, the PG switch is super turned-on by applying a higher gate voltage ($>V_{DD}$) in the on-state. In [24], the benefits of combining the super turn-off and the super turn-on modes on a PG switch is briefly discussed. However, no circuit realization of the concept has been shown.

## 2.2 Existing Super Cut-off Power Gating Methods

### 2.2.1 Optimum bias point in super cut-off power gating

In Fig. 2.3 (a), leakage current of the PMOS PG switch shown in Fig. 2.2 (a) is plotted with respect to $V_{GSP}$ in 65nm CMOS technology. The leakage current reduces until the point where GIDL effect becomes dominant [11]. Hence, maximum leakage current reduction ratio ($I_{LEAK}(0)/I_{LEAK}(V_{GSP(opt)})$) is obtained for $V_{GSP} = V_{GSP(opt)}$. The value of $V_{GSP(opt)}$ increases with temperature and is constant with process corners, as shown in Fig. 2.3 (b) and 2.3 (c), respectively. The dependence of $V_{GSP(opt)}$ with temperature arises from the fact that band-to-band tunneling (BTBT) current, which causes GIDL, is insensitive to temperature [11]. In contrast, the channel leakage current is sensitive to temperature. However, both the current components are equally affected by the process conditions, which makes $V_{GSP(opt)}$ constant across process corners. The variation of maximum leakage current reduction ratio achievable using super cut-off PG with respect to temperature and process corners is plotted in Fig. 2.4 (a) and (b), respectively.

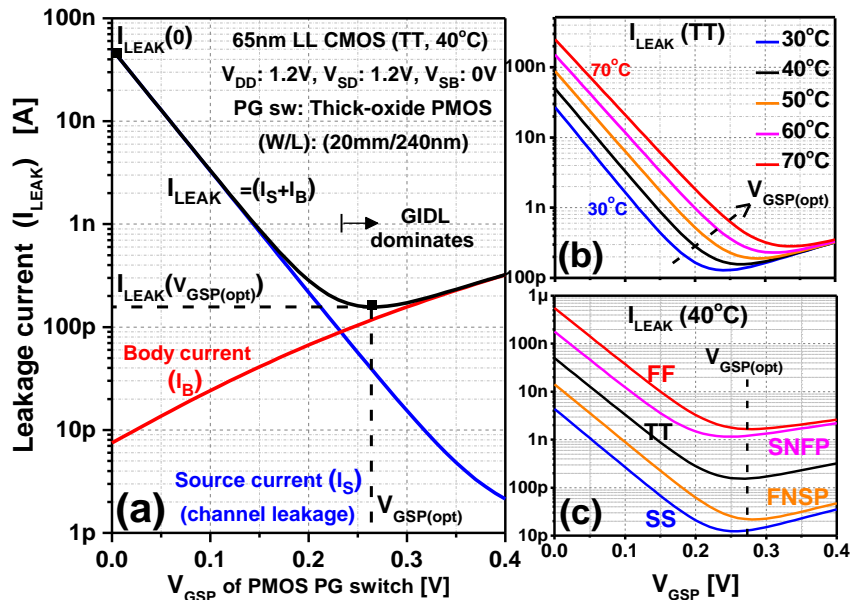

Fig. 2.3. (a) Simulated leakage current vs. $V_{GSP}$, (b) variation of $V_{GSP(opt)}$ with temperature, and (c) invariance of $V_{GSP(opt)}$ across process corners.

### 2.2.2 Prior works in super cut-off power gating

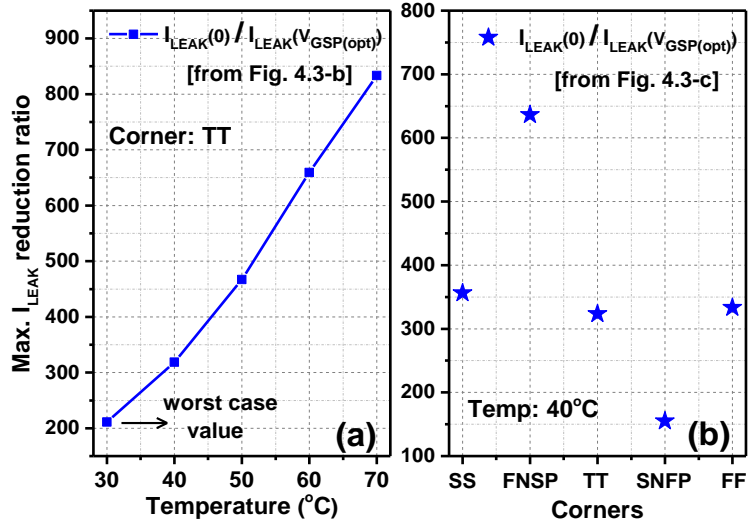

Fig. 2.4. Simulated maximum leakage current reduction using super cut-off PG with respect to (a) temperature, and (b) process corners in 65nm technology.

In [12], super cut-off PG was first proposed to reduce the leakage current of the PMOS PG switch by biasing it with a positive $V_{GSP}$ during the super cut-off state. The sleep signal is level shifted to a higher voltage and is provided to the gate. The charge pump based high voltage generator consumes significant power due to the higher clock frequency. Zig-zag super cut-off PG proposed in [13], groups the logic gates in the circuit load alternatively into two groups, and power gate them separately; one with PMOS and the other with NMOS PG switch. The alternate clustering of logic gates and individual super cut-off biasing in PG switch increases the design complexity.

The work in [14], biases the $V_{GSP}$ of the PMOS PG switch to a positive value during the off-state, using an already existing on-chip high voltage supply. $V_{GSP}$ is precisely set using an analog feedback circuit and results in a 75% reduction of the leakage power. However, the power overhead for gate biasing remains high due to the quiescent current consumption of the biasing circuits. Self-super cut-off PG proposed in [15] doesn't require any additional gate voltage boosting or charge pump. Even though 95× leakage savings are achieved, having two series PMOS PG switches for self-super cut-off biasing becomes ineffective for the circuits with high on-state current, which demand lesser IR drop across the PG switch.

In [16], super cut-off NMOS PG with a low frequency modified Dickson charge pump is proposed, which achieves 19.3× standby power reduction. In [17], dynamic standby controller (DSC) is used to bias the PMOS PG switch with positive $V_{GSP}$ using a programmable duty cycle and clock frequency. The leakage power of the PG switch is reduced by 89%.

However, the gate bias generator consumes more power than the leakage power of the PG switch. Super cut-off PG is realized in [18] and [19] by applying the PG signal to the gate of the PMOS PG switch from a higher voltage domain ($V_{DD,H}$). $V_{DD,H}$ is generated by the existing on-chip DC-DC converter.

The works from [12]-[19] were intended for near-threshold application (supply voltage of the circuit load ($\approx V_{TH}$) $\ll$ nominal $V_{DD}$ of given technology), and hence the voltage stress on the super cut-off PG switch was not a concern as discussed in Section I. In [20] and [21], super cut-off PG is used for the circuit load with supply voltage equal to the nominal $V_{DD}$ of the technology. A process and temperature insensitive control circuit is proposed in [20], but the power reduction is limited to a factor of 2.7×. Moreover, the voltage stress on the regular core device PG switch in the super cut-off state is not addressed.

In [21], super cut-off PMOS PG switch is implemented with regular core devices, and achieves maximum leakage current reduction by biasing at $V_{GSP(opt)}$. However, the PG switch is voltage stressed in the super cut-off state. The charge pump used to generate the gate voltage in [21] lowers its output voltage by sensing the dielectric aging of the PG switch. This reduces the benefits of super cut-off PG over time. Moreover, allowing the dielectric aging to occur in the PG switch increases $R_{ON}$ and causes larger IR drop during the on-state.

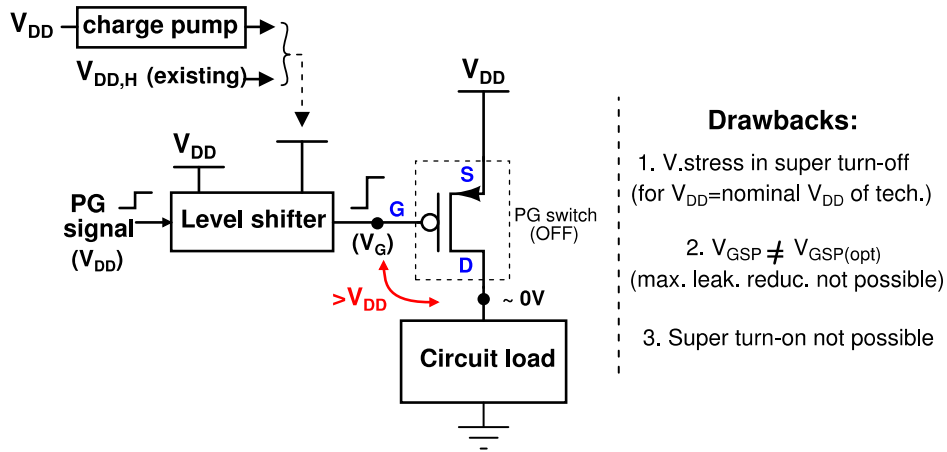

Fig. 2.5. Typical super cut-off PG method adopted by prior works.

A finer level of super cut-off PG at the standard cell level is proposed in [25]-[27]. It reduces the leakage power consumption to a few fW's/gate. However, the self cut-off PG mechanism in the individual logic gate limits the circuit load's maximum operating frequency to kHz range. Fig. 2.5 summarizes the typical super cut-off PG method followed in [12]-[21].

Stacking two core devices avoids the voltage stress problem in the super cut-off PG switch [12], as shown in Fig. 2.6 (a). However, it requires two separate charge pumps to generate the voltages $V_{DD}+V_B$ and $0.5V_{DD}+V_B$ as shown in Fig. 2.6 (b), which increases the power consumption due to the higher clock frequency required. Moreover, it doesn't always ensure that the PG switch is biased at $V_{GSP(opt)}$ due to its non-tunable gate voltage.

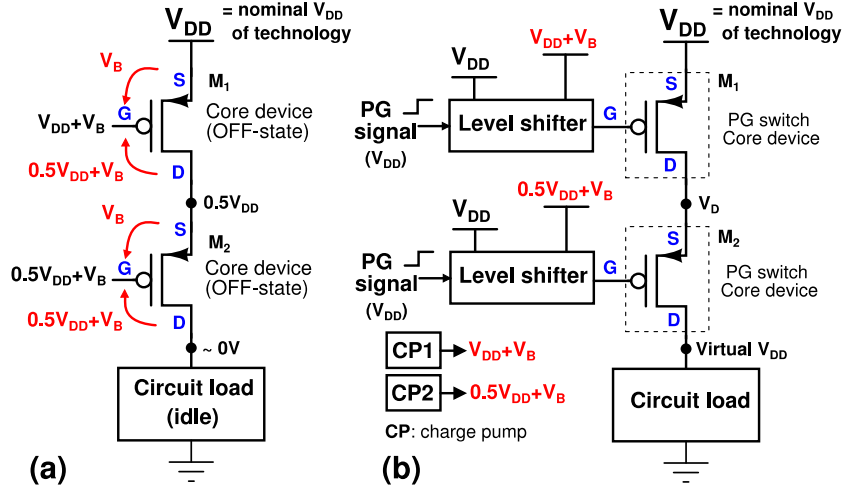

Fig. 2.6. (a) Avoiding voltage stress in the super cut-off state using stacking of thin-oxide devices, and (b) implementation of stacked super cut-off PG.

## 2.3 Switched-Capacitor Assisted Power Gating

### 2.3.1 Principle of operation

In the proposed SwCap PG, capacitor $C_X$ holding an appropriate voltage is connected across the source and gate terminals of the PG switch, as shown in Fig. 2.7. The gate voltage is equal to $V_{DD}+V_{B(off)}$ and $-V_{B(on)}$ during the super turn-off and the super turn-on state, respectively. As discussed in Section I, thick-oxide device is used as the PG switch for avoiding the voltage stress on them. From Section 2.2.1, the optimum value for $V_{B(off)}$ (referred as $V_{B(off)-opt}$) to achieve maximum leakage reduction is equal to $V_{GSP(opt)}$. From (2.1), the leakage current of the SwCap PG switch shown in Fig. 2.7 (a) is given by (2.5).

$$I_{LEAK(SwCap\,PG)} = \left(\mu_p C_{OX} \frac{W}{L}(n-1)V_t^2 e^{\frac{-|V_{THP}|}{nV_t}}\right) e^{\frac{-V_{B(off)-opt}}{nV_t}}$$

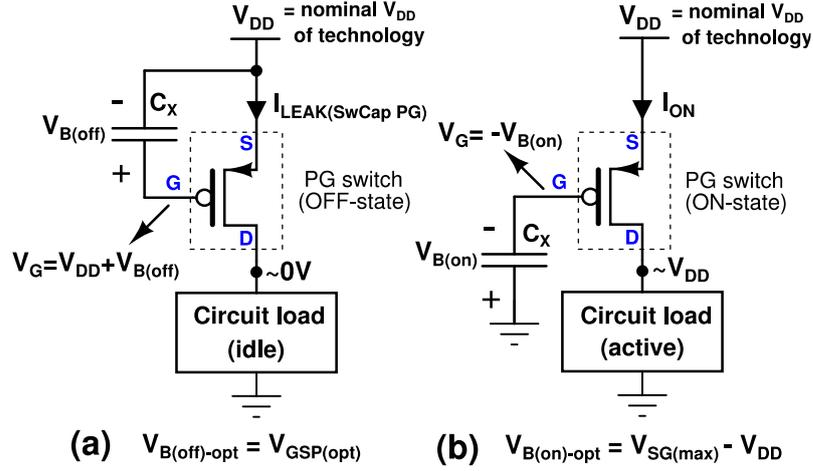

Fig. 2.7. Proposed switched-capacitor assisted power gating (SwCap PG) during (a) super turn-off state, and (b) super turn-on state.

$$= (I_{LEAK(Conv\ PG)}) e^{\frac{-V_{GSP(opt)}}{nV_t}} \qquad (2.5)$$

Biasing the gate voltage to $-V_{B(on)}$ in the super turn-on state, as shown in Fig. 2.7 (b), lowers $R_{ON}$ of the thick-oxide device. The optimum value for $V_{B(on)}$ (referred as $V_{B(on)-opt}$) to achieve the lowest possible $R_{ON}$ is equal to $V_{SG(max)}-V_{DD}$. $V_{SG(max)}$ is the maximum allowed source-to-gate voltage of the thick oxide device in the given technology. From (2.2), $R_{ON}$ of the SwCap PG switch shown in Fig. 2.7 (b) is given by (2.6).

$$R_{ON(SwCap\ PG)} = \frac{1}{\mu_p C_{OX} \frac{W}{L} (\{V_{DD} + V_{B(on)-opt}\} - |V_{THP}|)}$$

$$= \frac{1}{\mu_p C_{OX} \frac{W}{L} (\{V_{SG(max)}\} - |V_{THP}|)} \qquad (2.6)$$

For the same $R_{ON}$, the leakage current and the device area of various PG switch options are given in Table 3.1 for 65nm CMOS technology. As seen in Table 2.1 and 2.2, the proposed SwCap PG technique achieves the lowest possible leakage current for the same $R_{ON}$. This is due to the optimum biasing of the gate voltage in the thick-oxide PG switch during the super turn-off and the super turn-on state. However, the SwCap PG results in an increase in the PG switch area. Nevertheless, the utmost priority of extending the battery life in the energy-constrained applications overwhelms the increase in the PG switch area. The effect of GIDL (discussed in Section 2.2.1) is not modelled in the 180nm process design kit (PDK), and hence, 65 nm CMOS technology is used in the SPICE simulations for the comparison of the SwCap PG with other PG switch options.

Table 2.1: Comparison of various PG switch options in 65nm CMOS technology
($V_{DD}$: 1.2V, corner: TT, temperature: 40°C)

| PMOS PG switch type | (W/L) for $R_{ON}=0.15\Omega$ (mm/nm) | Leakage current (A) | Device area (μm$^2$) |
|---|---|---|---|
| LVT core device [&] | 8.08 / 60 | 18.378 μ | 2,250 |
| RVT core device [&] | 9.21 / 60 | 3.413 μ | 2,560 |
| HVT core device [&] | 11.49 / 60 | 232.8 n | 3,200 |
| Thick-oxide device [&] | 43.88 / 240 | 109.5 n | 28,478 |
| Proposed SwCap PG * $V_{B(on)-opt}$: 1.3V, $V_{B(off)-opt}$: 0.27V | 20 / 240 | 155 p | 12,980 |

[&]Conv. PG: $V_{SGP} = V_{DD}$ (on), 0V (off).  *Thick-oxide device $V_{SG(max)}$: 2.5V

Table 2.2: Cost-benefit trade-offs for PG switch in 65nm CMOS technology

| PMOS PG switch type | Reduction in leakage w.r.t LVT device | Increase in area w.r.t LVT device |
|---|---|---|
| RVT core device | 5.4 × | 1.14 × |
| HVT core device | 78.9 × | 1.42 × |
| Thick-oxide device | 167.8 × | 12.6 × |
| Proposed SwCap PG | 118,567 × | 5.8 × |

### 2.3.2 Method to overcome charge loss in the capacitor

The capacitor $C_X$ samples the voltage, and is then configured across the PG switch for biasing during the super turn-off and the super turn-on state, as shown in Fig. 2.8 (a). The critical observation made in SwCap PG is the fact that the gate of a MOS transistor is a high impedance node, which does not load $C_X$. In the actual implementation, gate leakage current, switch network leakage current, and other parasitic capacitances cause charge loss in $C_X$. The various elements that cause charge loss in $C_X$ are highlighted in Fig. 2.8 (b) and (c) during the super turn-off and the super turn-on state, respectively. $C_{GS}$ and $C_{GD}$ represent the gate-to-source and gate-to-drain capacitance of the PG switch. $C_P$ represents the parasitic capacitance of the switch network. $I_G$ and $I_{SW}$ represent the gate leakage current of the PG switch and the switch network leakage current. $I_{dis}$ (~$I_G+I_{SW}$) is the total discharge current flowing through $C_X$. Because of the charge loss in $C_X$, the sampled voltage held across $C_X$ ($V_{CX}(t)$ in Fig. 2.8 (b) and (c)) decays over time.

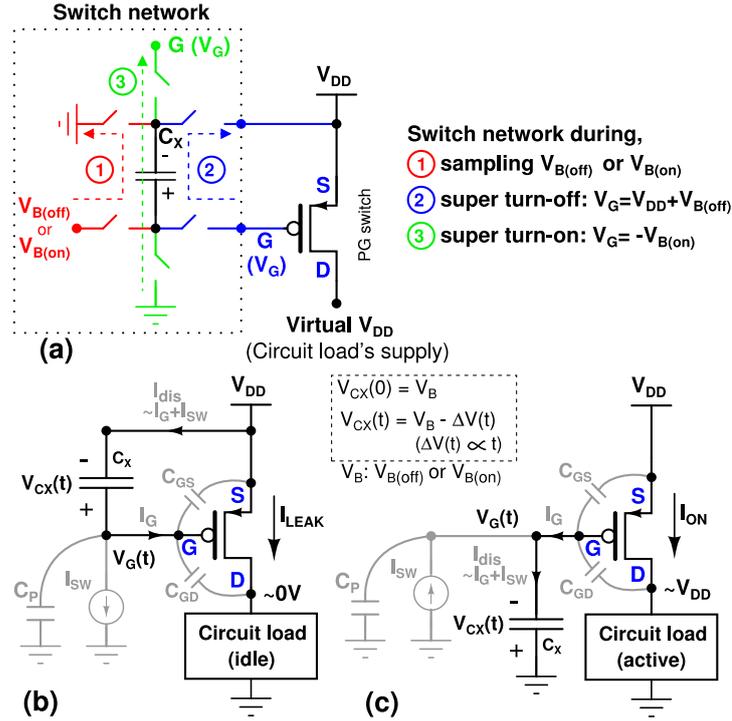

Fig. 2.8. (a) Switch network for biasing the PG switch, (b) simplified representation of the parasitic elements in the SwCap PG during the super turn-off state, and (c) during the super turn-on state of the PG switch.

Hence, to avoid complete decay of the voltage across $C_X$, two identical switched-capacitor networks are employed, and are periodically refreshed in a time-interleaved manner. Fig. 2.9 shows the periodic refreshment of $V_{B(off)}$ during the super turn-off state. When capacitor $C_2$ (in SwCap network-2) is connected between the source and the gate of the PG switch (to keep $V_G = V_{DD} + V_{B(off)}$), capacitor $C_1$ (in SwCap network-1) samples the voltage $V_{B(off)}$, and vice versa. Thus refreshment of voltage occurs every half clock cycle. Similarly, the periodic refreshment of $V_{B(on)}$ during the super turn-on state is done to maintain the gate voltage close to $-V_{B(on)}$, as shown in Fig. 2.10.

For simplifying the analysis without losing key information, the discharge current ($I_{dis}$) indicated in Fig. 2.8 (b) and (c) is approximated to be constant. Hence, the gate voltage $V_G(t)$ changes linearly with time until the next refresh instant. The approximated waveform of $V_G(t)$ during the super turn-off and the super turn-on state is shown in Fig. 2.11 (a) and (b), respectively. From Fig. 2.11 (a), the average value of $V_G$ during the super turn-off state ($V_{G,off}(avg)$) is given by (2.7). Substituting the terms $V_{G,off}(0)$ and $V_{G,off}(T/2)$ from Fig. 2.11 (a) into (2.7), the final expression for $V_{G,off}(avg)$ is given by (2.8).

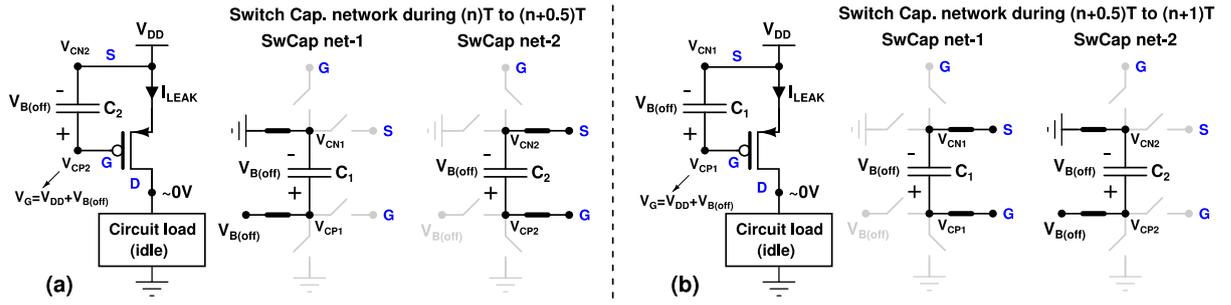

Fig. 2.9. Periodic refreshment of $V_{B(off)}$ in the capacitors $C_1$ and $C_2$ during the super turn-off state between (a) nT to (n+0.5)T and (b) (n+0.5)T to (n+1)T. T (=$1/f_{clk}$) is the period of the refreshing clock. The value of capacitance, $C_1=C_2=C_X$.

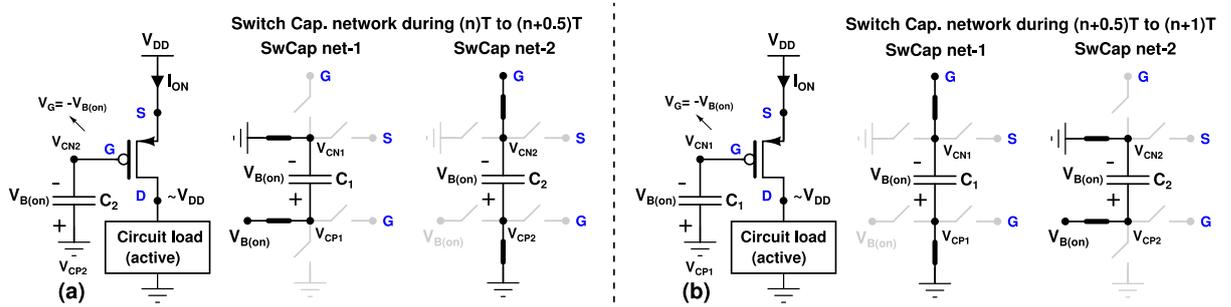

Fig. 2.10. Periodic refreshment of $V_{B(on)}$ in the capacitors $C_1$ and $C_2$ during the super turn-on state between (a) nT to (n+0.5)T and (b) (n+0.5)T to (n+1)T. T (=$1/f_{clk}$) is the period of the refreshing clock. The value of capacitance, $C_1=C_2=C_X$.

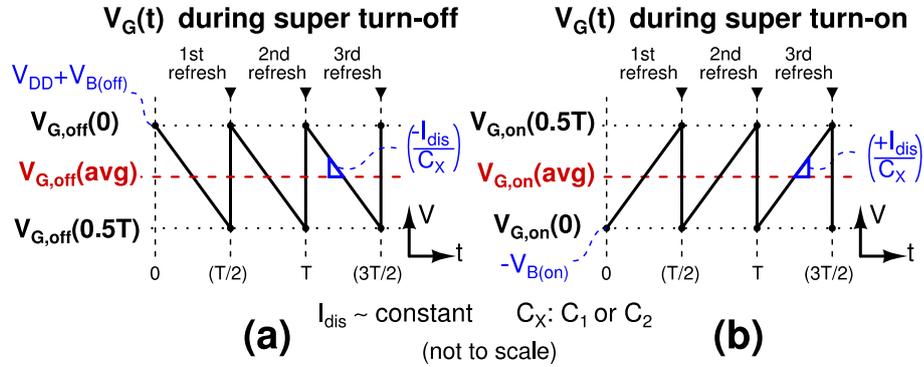

Fig. 2.11. Approximated waveform of $V_G(t)$ after refreshing $C_X$ ($C_1$ or $C_2$), during the (a) super turn-off state, and (b) super turn-on state. T=($1/f_{clk}$).

$$V_{G,off}(avg) = \left(\frac{V_{G,off}(0) + V_{G,off}(T/2)}{2}\right) \qquad (2.7)$$

$$= V_{DD} + V_{B(off)} - \left(\frac{I_{dis} \cdot T}{4C_X}\right) \qquad (2.8)$$

From Fig. 2.11 (b), the average value of $V_G$ during the super turn-on state ($V_{G,on}(avg)$) is given by (2.9). Substituting the terms $V_{G,on}(0)$ and $V_{G,on}(T/2)$ from Fig. 2.11 (b) into (2.9), the final expression for $V_{G,on}(avg)$ is given by (2.10).

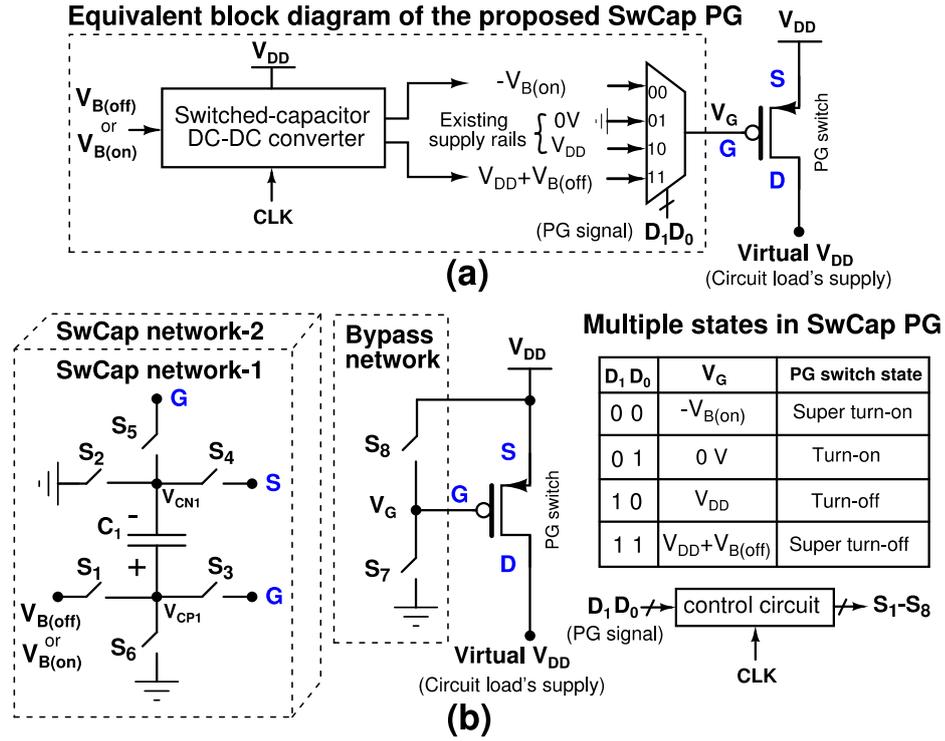

Fig. 2.12. (a) Equivalent block diagram of the proposed SwCap PG technique, and (b) corresponding implementation of the SwCap PG technique.

$$V_{G,on}(avg) = \left(\frac{V_{G,on}(0) + V_{G,on}(T/2)}{2}\right) \quad (2.9)$$

$$= -V_{B(on)} + \left(\frac{I_{dis} \cdot T}{4C_X}\right) \quad (2.10)$$

As evident from (2.8) and (2.10), for an increased amount of discharge current, either period of refresh has to be decreased or the value of $C_X$ has to be increased to maintain the average value of $V_G$ close to its ideal one. Thus, the period of refresh is dependent on the leakage time constant associated with $C_X$. The usage of a thick-oxide device as the PG switch helps in reducing the gate leakage current. The switch network leakage current is reduced by using thick-oxide devices with longer channel lengths as the switching elements. Hence, frequencies on the order of sub-100 Hz are sufficient enough to maintain the average gate voltage close to its ideal value.

As an example, consider the expression for the gate voltage in the super turn-off state given by (2.8). For a 5% error in the gate voltage, the frequency of the refresh clock ($f_{CLK}$) must be as given by (2.11)

$$f_{CLK} = \frac{1}{T} = \frac{5I_{dis}}{(V_{DD}+V_B)C_X} \quad (2.11)$$

In equation (2.11), the discharge current ($I_{dis}$) depends on the gate leakage current ($I_G$) and the switch network leakage current ($I_{SW}$). However, due to larger PG switch size, the term $I_G$ will be dominant. Hence, $f_{CLK}$ can be approximated as given by equation (2.12).

$$f_{CLK} \approx \frac{5 I_G}{(V_{DD}+V_B)C_X} \qquad (2.12)$$

For a thick-oxide device in 22nm FinFET technology, the gate leakage current is estimated to be 10fA/μm [35]. Hence, for a 10mm PG switch, the total gate leakage current will be 100pA. For a $V_{DD}$=0.7V $V_B$=0.5V and $C_X$=5pF, the refresh frequency required is 83 Hz.

However, at higher temperatures, the gate leakage current is estimated to increase to a value of 100fA/μm [45]. Hence, the required refresh frequency of the switched-capacitor network will increase to 830 Hz. On-chip generation of 830 Hz clock frequency using an ultra-low power relaxation oscillator proposed in [33] consumes around 1.8 nW of power. An optimum trade-off between the required refresh frequency and capacitor area helps in achieving lower dynamic power consumption and silicon area.

Apart from the super turn-off and the super turn-on state, the proposed PG technique also allows to bias the gate of the PG switch at $V_{DD}$ or 0 V as in the case of the conventional PG. Fig. 2.12 (a) shows the equivalent block diagram of the proposed SwCap PG technique. The corresponding implementation of the proposed SwCap PG technique is depicted in Fig. 2.12 (b). Multiple states of the PG switch allow the circuit load to reside in a wide range of operating states common in energy-efficient microprocessors [3] and [5].

## 2.4 Circuit Implementation of CMOS SwCap PG

### 2.4.1 Challenges in switch network when the voltages exceed supply rails

The node voltages $V_{CP1}$ (or $V_{CP2}$ in SwCap network-2) and $V_G$ indicated in Fig. 2.9 and Fig. 2.12 (b) swing to $V_{DD}+V_{B(off)}$ during the super turn-off state of the PG switch. Similarly, the node voltages $V_{CN1}$ (or $V_{CN2}$ in SwCap network-2) and $V_G$ indicated in Fig. 2.10 and Fig. 2.12 (b) swing to -$V_{B(on)}$ during the super turn-on state of the PG switch. When the node voltages swing beyond the supply rails ($V_{DD}$ and GND), it is impossible to turn off the transistors with standard digital signals. Moreover, parasitic diodes in the transistor gets forward biased.

A common approach to turn-off the transistors under these conditions is to provide the appropriate level shifted signals (shifted to $V_{DD}+V_{B(off)}$ or $-V_{B(on)}$) to the gate [28]. Similarly, to ensure that the parasitic diodes are always reverse biased, dynamic body-switching can be done [29]. However, both the solution increases the circuit complexity. Instead, a simpler approach of using the automatic self cut-off operation of the diode-connected transistor is utilized in this work.

*A). Turning off NMOS device when node voltages go below 0 V*

The NMOS device $M_{N1}$ and the diode shown in Fig. 2.13 (a) is on, due to the source voltage being negative. To solve this problem, a diode-connected transistor $M_{N2}$ is inserted, as shown in Fig. 2.13 (b). The node voltage $V_{N1}$ lies between 0 V and $-V_{B(on)}$, which forces $V_{GS}$ of $M_{N2}$ to be 0 V, as indicated in Fig. 2.13 (b). Since $M_{N2}$ is off, the path is effectively disconnected. Switch $S_2$ shown in Fig. 2.12 (b) needs to block $-V_{B(on)}$ during its off-state, and does not experience voltage swings higher than $V_{DD}$ at its terminals. Hence, the switch architecture described in Fig. 2.13 (b) is used to realize $S_2$.

Switches $S_5$ and $S_7$ shown in Fig. 2.12 (b) needs to block $-V_{B(on)}$ as well as $V_{DD}+V_{B(off)}$ during its off-state. The switch architecture described in Fig. 2.13 (c) is used to realize $S_5$ and $S_7$. The transistor $M_{N3}$ is on, and the node $V_{N2}$ is equal to $-V_{B(on)}$. This forces $V_{GS}$ of $M_{N2}$ to be 0 V, as shown in Fig. 2.13 (c). Hence the diode-connected transistor $M_{N2}$ is off, which ensures that the path is effectively disconnected. The switch architecture for $S_5$ and $S_7$ is derived from $S_2$ by interchanging the position of the diode-connected transistor. This avoids the P-body to N-well diode from being forward biased for voltages above $V_{DD}$.

*B). Turning off PMOS device when node voltages exceed $V_{DD}$*

The PMOS device $M_{P1}$ and the diode shown in Fig. 2.13 (d) is on, due to the source voltage exceeding $V_{DD}$. Using the similar self cut-off feature of the diode-connected transistor as exploited earlier, $M_{P2}$ is inserted to effectively disconnect the path, as shown in Fig. 2.13 (e). Switches $S_3$ and $S_8$ shown in Fig. 2.12 (b) needs to block $V_{DD}+V_{B(off)}$ as well as $-V_{B(on)}$ during its off-state. The switch architecture described in Fig. 2.13 (f) is used to realize $S_3$ and $S_8$. The transistor $M_{P1}$ is on, and the node voltage $V_{N4}$ equals $V_{DD}+V_{B(off)}$. This forces $V_{SG}$ of $M_{P2}$ to be

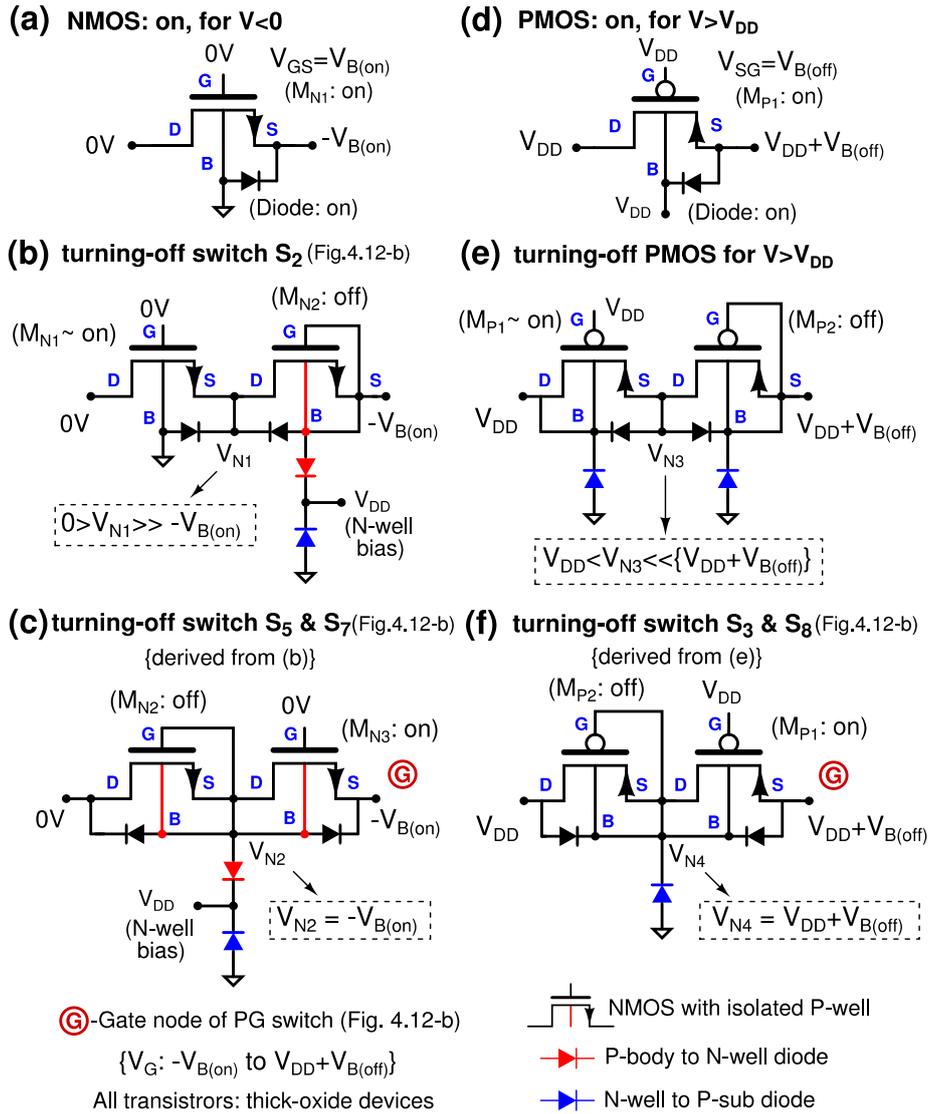

Fig. 2.13. Implementation of switch network. (a) Problem with NMOS due to negative voltages, (b) realization of switch $S_2$, (c) realization of switch $S_5$ & $S_7$, (d) problem with PMOS due to voltages higher than $V_{DD}$, (e) intermediate stage in the realization of switches $S_3$ & $S_8$, and (f) final realization of $S_3$ & $S_8$.

0 V, as shown in Fig. 2.13 (f). Hence the diode-connected transistor $M_{P2}$ is off, which ensures that the path is effectively disconnected. The switch architecture for $S_3$ and $S_8$ is derived from the one shown in Fig. 2.13 (e), by interchanging the position of the diode-connected transistor. This avoids the N-well to P-sub diode from being forward biased for voltages below 0V.

As observed in Fig. 2.13 (b), (c), and (f), the parasitic diodes always appear in a back-to-back connected manner. Hence, the transistors and the parasitic diodes getting turned on unwantedly due to the node voltages exceeding the supply rails are avoided by just using the simple self cut-off feature of the diode-connected transistor as outlined in Fig. 2.13. This comes at the cost of voltage drop across the diode-connected devices when the switch is on, and is analyzed in Section 2.4.2.

The complete transistor level implementation of the CMOS SwCap network is shown in Fig. 2.14. The capacitors $C_1$ and $C_2$ are implemented using the P-MOSCAP, as shown in Fig. 2.15, due to the availability of PMOS devices in any CMOS process (unlike the case of MIM capacitors). The nominal value of $C_1$ and $C_2$ is approximately 5pF. A 15 mm wide thick-oxide low $V_{TH}$ PMOS device is used as the PG switch. The low $V_{TH}$ flavor of the thick-oxide device is used for allowing the leakage current of the PG switch to be in a measurable quantity, since 180 nm process is used for implementation. The presence of low $V_{TH}$ flavor thick-oxide device option in the process technology additionally helps in reducing $R_{ON}$ of the thick-oxide device, and is not necessary for the SwCap PG technique.

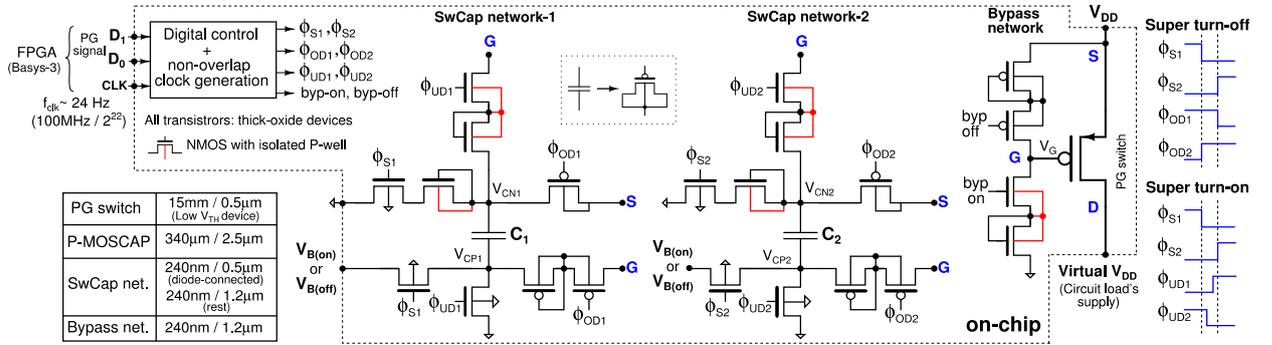

Fig. 2.14. Circuit implementation of the CMOS SwCap PG.

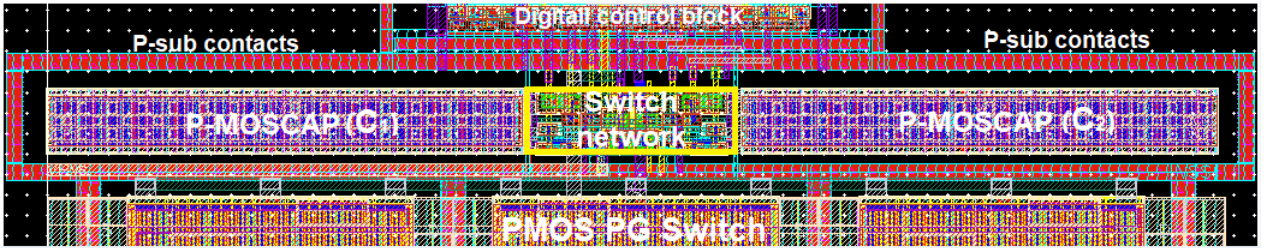

Fig. 2.15. Layout showing the implementation of $C_1$ & $C_2$ realized using P-MOSCAP. The layout of the entire PMOS PG switch is not shown in the figure.

### 2.4.2 Penalty due to diode-connected devices in the switch network

The effect of diode-connected devices in the CMOS SwCap network (Fig. 2.14) is depicted in Fig. 2.16. The on-resistances of the other switches in the network are not shown, since the period of the refresh is larger than the RC time constants associated with them. In the steady-state, the charge lost by the capacitor $C_X$ is exactly replenished by refreshing it in the next half-cycle. Hence in the steady-state, $I_{charge} = I_{dis} = I_{sub}$. For brevity, consider $\mu_n = \mu_p =$ $\mu$, $V_{THN} = |V_{THP}| = V_{TH}$, and $V_{GSn} = V_{SGp} = V_{GS}$. From Fig. 2.16 (a), the average value of $V_G$ during the super turn-off state ($V_{G,off}(avg)$) is given by (2.13). Substituting for $V_{CX}$ from Fig. 2.16 (a) into (2.13), the final expression for $V_{G,off}(avg)$ is given by (2.14).

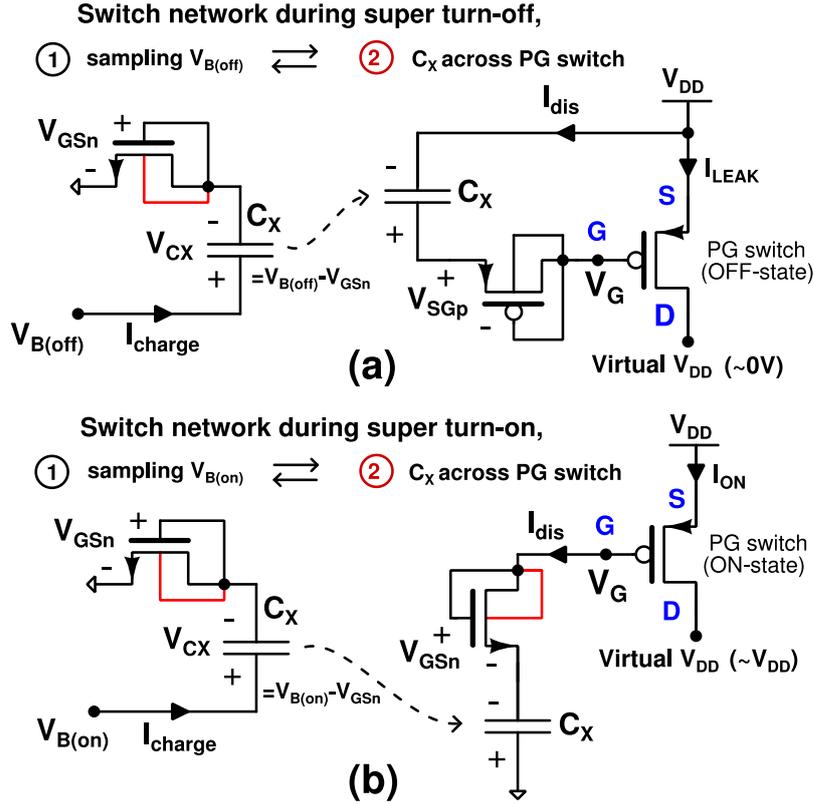

Fig. 2.16. Effect of diode-connected devices in CMOS SwCap network during (a) super turn-off state and (b) super turn-on state. $C_X$ denotes $C_1$ or $C_2$.

$$V_{G,off}(avg) = V_{DD} + V_{CX} - V_{SGp} \qquad (2.13)$$

$$= V_{DD} + V_{B(off)} - (2V_{GS}) \qquad (2.14)$$

From Fig. 2.16 (b), the average value of $V_G$ during the super turn-on state ($V_{G,on}(avg)$) is given by (2.15). Substituting for $V_{CX}$ from Fig. 2.16 (b) into (2.13), the final expression for $V_{G,on}(avg)$ is given by (2.16).

$$V_{G,on}(avg) = -V_{CX} + V_{SGn} \qquad (2.15)$$

$$= -V_{B(on)} + (2V_{GS}) \qquad (2.16)$$

In steady state, $I_{charge}$ and $I_{dis}$ shown in Fig. 2.16 are equal ($=I_{sub}$), and low enough ($\ll$ nA) in value. Hence, the diode-connected devices are in the sub-threshold region. Based on (2.1), $V_{GS}$ for a sub-threshold transistor is written as given by (2.17), and the inequality shown in (2.18) holds true. The value of $\alpha$ in (2.17) is always less than one and is given by (2.19).

$$V_{GS} = nV_t \ln\left(\frac{I_{sub}}{\mu C_{ox}\frac{W}{L}(n-1)V_t^2}\right) + V_{TH} = \alpha V_{TH} \tag{2.17}$$

$$I_{sub} \ll \mu C_{ox}\frac{W}{L}(n-1)V_t^2 \tag{2.18}$$

$$\alpha = 1 - \frac{nV_t}{V_{TH}}\ln\left(\frac{\mu C_{ox}\frac{W}{L}(n-1)V_t^2}{I_{sub}}\right) \ll 1 \tag{2.19}$$

By substituting (2.17) into (2.14) and (2.16), the new expressions for the average value of $V_G$ during the super turn-off and the super turn-on state is given by (2.20) and (2.21), respectively.

$$V_{G,off}(avg) = V_{DD} + V_{B(off)} - (2\alpha V_{TH}) \tag{2.20}$$

$$V_{G,on}(avg) = -V_{B(on)} + (2\alpha V_{TH}) \tag{2.21}$$

After including the effect of the charge loss in $C_X$ and the periodic refreshment of voltage as discussed in Section 2.3.2, the average value of $V_G$ during the super turn-off and the super turn-on state is given by (2.22) and (2.23), respectively. The average voltage error in $V_G$ ($V_{error,avg}$) during the super turn-off and the super turn-on state is provided by (2.24).

$$V_{G,off}(avg) = V_{DD} + V_{B(off)} - \left(2\alpha V_{TH} + \frac{I_{dis} \cdot T}{4C_X}\right) \tag{2.22}$$

$$V_{G,on}(avg) = -V_{B(on)} + \left(2\alpha V_{TH} + \frac{I_{dis} \cdot T}{4C_X}\right) \tag{2.23}$$

$$V_{error,avg} = 2\alpha V_{TH} + \left(\frac{I_{dis} \cdot T}{4C_X}\right) \tag{2.24}$$

As observed from (2.24), the error term is made of two components. The former one is due to the effect of diode-connected devices. The latter one is due to the effect of periodic refresh in $C_X$. For a quantitative understanding on $V_{error,avg}$, consider the following values corresponding to the CMOS SwCap PG implementation (Fig. 2.14) in 180 nm technology: $\mu C_{OX}\frac{W}{L} = \frac{60\mu A}{V^2}$, $n = 1.5$, $V_{TH} = 0.5V$ and $I_{sub}$ (or $I_{dis}$) = 10 pA. The value of $C_X$ ($C_1$ or $C_2$) used in this work is 5pF. Based on simulations, the refresh clock ($f_{clk}$) used in this work is 24 Hz. The calculated value of $\alpha$ and $V_{error,avg}$ is 0.434 and 0.45 V, respectively. $V_{error,avg}$ is dominated by the voltage drop across the diode-connected devices (96% of $V_{error,avg}$). Despite the various simplifications made in arriving at equations (2.22)-(2.24), the effect of non-idealities arising from the circuit implementation of the CMOS SwCap PG are well captured.

### 2.4.3 Optimum $V_{B(off)}$ for maximum leakage reduction

As seen from (2.22) and (2.23), for maintaining the gate voltage of the PG switch at $V_{DD}$ or 0 V (as in the case of conventional PG), the required value of $V_{B(off)}$ or $V_{B(on)}$ in the CMOS SwCap PG is equal to the voltage error given by (2.24). $V_{B(off)-opt}$ for obtaining the maximum leakage reduction is given by (2.25).

$$V_{B(off)-opt} = V_{GSP(opt)} + V_{error,avg}$$

$$= V_{GSP(opt)} + 2\alpha V_{TH} + \left(\frac{I_{dis} \cdot T}{4C_X}\right) \quad (2.25)$$

In (2.25), $V_{GSP(opt)}$ is independent of process corners as described in Section 2.2.1. But the terms arising due to the voltage error in $V_G$ leads to the dependency of $V_{B(off)-opt}$ on the process corners. $V_{GSP(opt)}$ varies with temperature, as shown in Fig. 2.3 (b). Also, the voltage error in $V_G$ has a dependence on temperature. Hence in given CMOS technology, the value of $V_{B(off)-opt}$ varies with the temperature and the process corners.

## 2.5 MEMS Switched-Capacitor Assisted Power Gating

Some of the pure digital CMOS technology may not offer an NMOS device with an isolated-well option. Hence, super turn-on implementation for the PG switch is not possible in those cases. MEMS switches are made of metallic beams and can handle voltages beyond the supply rails without any additional circuit techniques. They possess near-infinite off-resistance, thereby avoiding switch leakage in the off-state. The works from [8]-[10] proposed to use the MEMS switches for power gating. But for achieving very low $R_{ON}$ in the PG switch (on the order of mΩ's), several 100-1000's of MEMS switches have to be used in parallel, which consumes significant area.

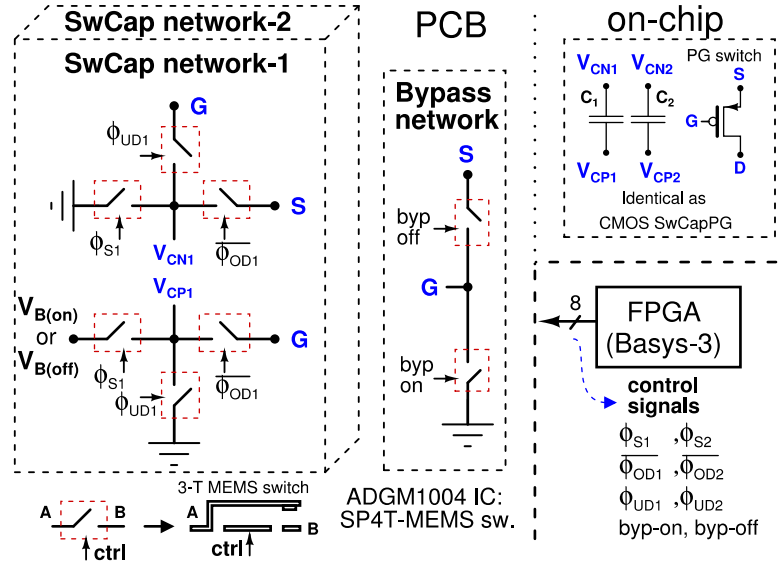

Fig. 2.17. Implementation of MEMS SwCap PG using external MEMS switches.

Therefore, MEMS SwCap PG is proposed for the process without an isolated NMOS device. MEMS switches are used as the switch elements in the SwCap network and the bypass network for biasing the PMOS PG switch. Since the period of refresh in the SwCap PG is on the order of 10's of ms, MEMS switches with even kΩ's of on-resistance are sufficient [30].

MEMS switches provide an additional advantage of avoiding the voltage drop due to the diode-connected devices and the switch network leakage current present in the CMOS SwCap network (Fig. 2.14). However, the gate leakage current of the PMOS PG switch still exists and periodic refresh of $V_{B(off)}$ or $V_{B(on)}$ is performed to keep the gate voltage close to its ideal value. Nevertheless, the voltage error ($V_{error,avg}$) term which is mainly dominated by the error due to the diode-connected device is approximately zero in (2.24)-(2.24) for the case of MEMS SwCap PG. Hence, the average value of $V_G$ is close to its ideal one in the super turn-off and the super turn-on state as given by (2.26) and (2.27), respectively. $V_{B(off)-opt}$ for obtaining the maximum leakage reduction is close to the ideal value of $V_{GSP(opt)}$.

$$V_{G,off}(avg) = V_{DD} + V_{B(off)} - \left(\frac{I_G \cdot T}{4C_X}\right) \approx V_{DD} + V_{B(off)} \qquad (2.26)$$

$$V_{G,on}(avg) = -V_{B(on)} + \left(\frac{I_G \cdot T}{4C_X}\right) \approx -V_{B(on)} \qquad (2.27)$$

The proposed MEMS SwCap PG requires only 14 MEMS switches for implementing the SwCap and the bypass network, as shown in Fig. 2.17. The PMOS PG switch and the capacitors are implemented on-chip and are identical from the CMOS SwCap PG (Fig. 2.14). For demonstration, MEMS SwCap PG is implemented by interfacing the commercial MEMS switch ICs (ADGM1004 [31]) with the on-chip elements, as shown in Fig. 2.17. However, the development of miniaturized CMOS compatible MEMS switches allows for better integration of switch network with the on-chip elements [30], [32], and [43].

## 2.6 Chip Measurement Results

The proposed SwCap PG is implemented in the 180 nm CMOS technology. The chip micrograph from the two different process runs is shown in Fig. 2.18 (a) and (b), respectively. The test circuits corresponding to the CMOS SwCap PG and the on-chip elements for the MEMS SwCap PG were included in the 1[st] process run. Only the standalone CMOS SwCap PG switch was included in the 2[nd] process run. All the measurements in this work are carried out at room temperature, which provides the pessimistic value of the leakage current reduction ratio.

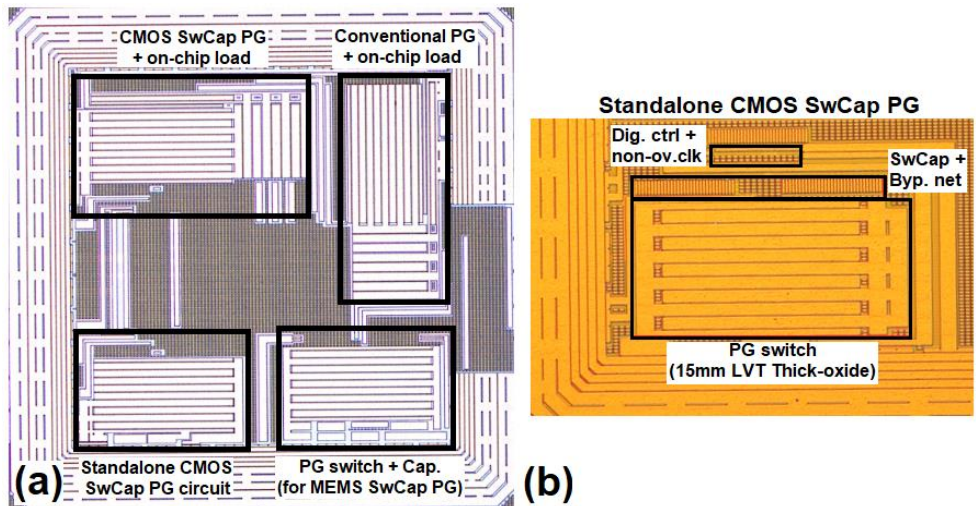

Fig. 2.18. Chip micro graph. (a) Four different test circuits in the 1[st] process run, and (b) standalone CMOS SwCap PG switch in the 2[nd] process run.

The accurate measurement of leakage currents and low on-resistance provide several challenges. The test setup for the measurement of leakage current and low on-resistance is shown in Fig. 2.19 (a) and (b), respectively. The leakage current is measured using rear triax cables in Keithley-2450 Sourcemeter. The leakage current in the PCB without test chip at the appropriate test point is de-embedded. $R_{ON}$ is measured using the 4-wire sense method as shown in Fig. 2.19 (b). The parasitic series resistance of the IC package, chip socket and the PCB tracks are de-embedded from the measured value. The measured value of the leakage current and $R_{ON}$ from the Sourcemeter was internally averaged using its digital filter, after which several readings were averaged externally to obtain a single data point. The actual experimental measurement setup of the test chip for leakage current and $R_{ON}$ measurement is shown in Fig. 2.20 (a) and (b), respectively.

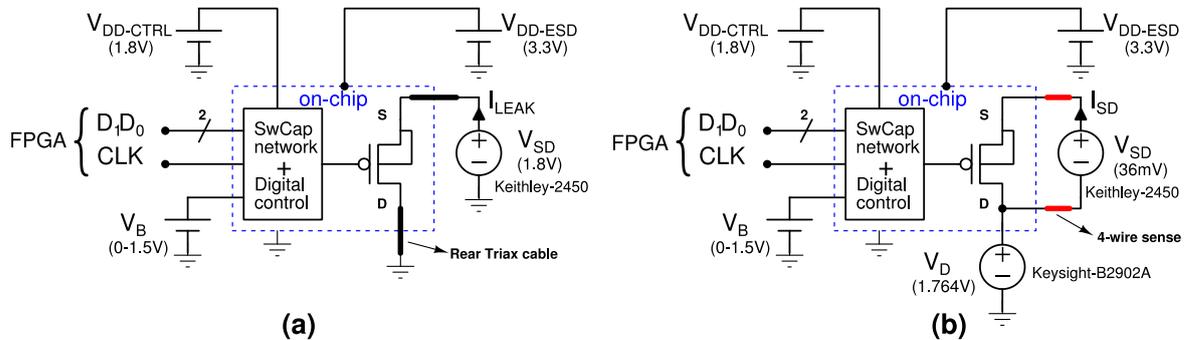

Fig. 2.19. (a) Test setup for the measurement of leakage current, and (b) test setup for the measurement of on-resistance of the PG switch.

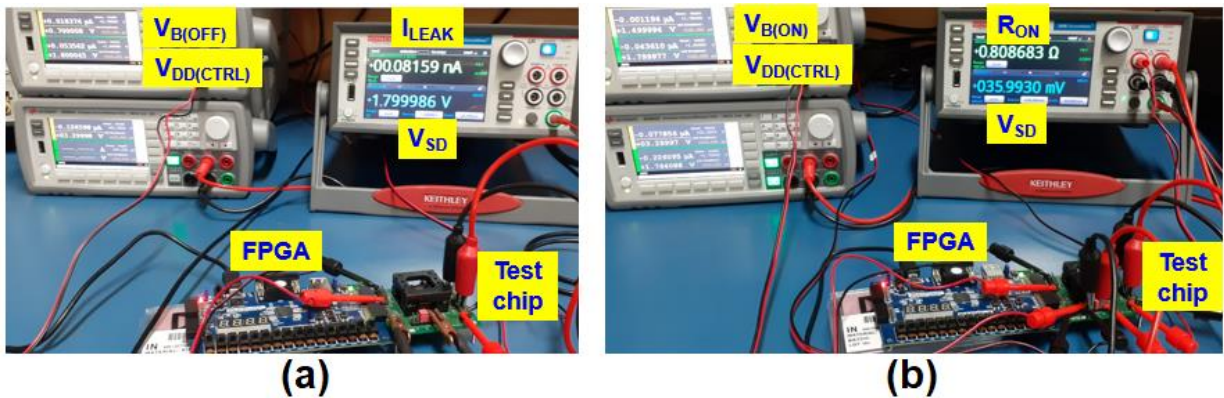

Fig. 2.20. Experimental characterization of SwCap PG test chip in the lab. (a) Measurement of leakage current, and (b) measurement of $R_{ON}$.

## 2.6.1 Measurement results of the CMOS SwCap PG

The test circuits corresponding to the CMOS SwCap PG and the conventional PG with on-chip load are shown in Fig. 2.21 (a) and (b), respectively. To estimate the effectiveness of the super turn-off and the super turn-on state in the SwCap PG, the conventional PG is made of the same PG switch (15mm thick-oxide LVT device), as indicated in Fig. 2.21. Leakage current source ($M_L$) in parallel with a capacitor ($C_L$) is used as an on-chip load for both the cases. When the individual PG switches are on, the current drawn by the on-chip load is approximately 100μA. Measurement of the leakage current after power gating is plotted in Fig. 2.22. For the conventional PG, the leakage current after power gating is 28.457 nA. Due to the super turn-off implementation in the CMOS SwCap PG, the lowest average value of the leakage current achieved is 153 pA at a $V_{B(off)-opt} = 0.8$ V. This translates to a leakage reduction of 186× when compared to the conventional PG technique.

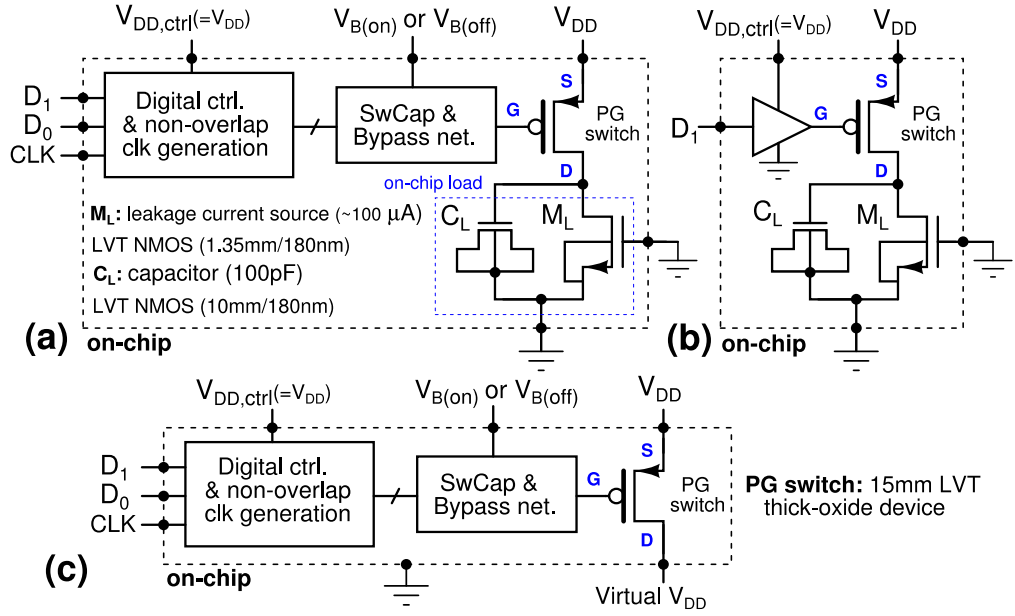

Fig. 2.21. (a) CMOS SwCap PG with on-chip load, (b) conventional PG with on-chip load, and (c) standalone CMOS SwCap PG switch.

From Fig. 2.22, it can be observed that for a $V_{B(off)} = 0.5$ V, the leakage current of the CMOS SwCap PG and the conventional PG is the same. Hence to maintain the gate voltage of the CMOS SwCap PG switch at $V_{DD}$, the required value of $V_{B(off)}$ is 0.5 V. This is due to the voltage error ($V_{error,avg}$) arising from the non-idealities in the circuit implementation, as discussed in Section 2.4.3. From Fig. 2.22, $V_{GSP(opt)}$ is obtained to be 0.3 V.

The leakage current measurements of the standalone CMOS SwCap PG switch (Fig. 2.21 (c)) from the two process runs are plotted in Fig. 2.23 (a) and (b), respectively. The lowest average value of the leakage current obtained across the two process runs is 161 pA and 87 pA, respectively. The corresponding value of $V_{B(off)-opt}$ is 0.8 V and 0.7 V. The difference in $V_{B(off)-opt}$ across different process runs is expected, as discussed in Section 2.4.3. Since the value of $V_{GSP(opt)}$ is constant with process, as discussed in Section 2.2.1, the value of $V_{error,avg}$ in the 2$^{nd}$ process run is obtained as 0.4 V. Hence, the leakage reduction factor in the 2$^{nd}$ process run is obtained to be 226×, as shown in Fig. 2.23 (b).

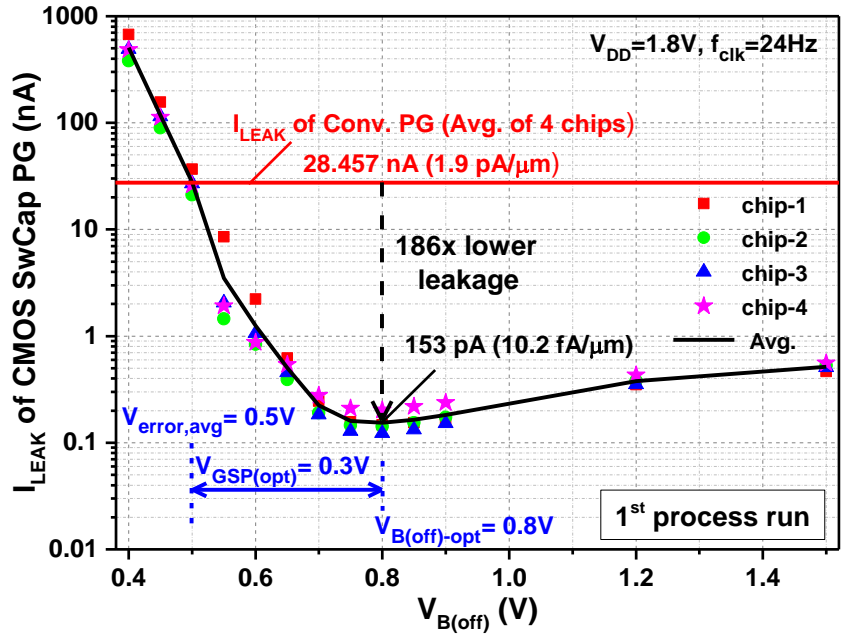

Fig. 2.22. Leakage current measurement of CMOS SwCap PG (Fig. 2.21-a) and conventional PG (Fig. 2.21-b) with on-chip load after power gating.

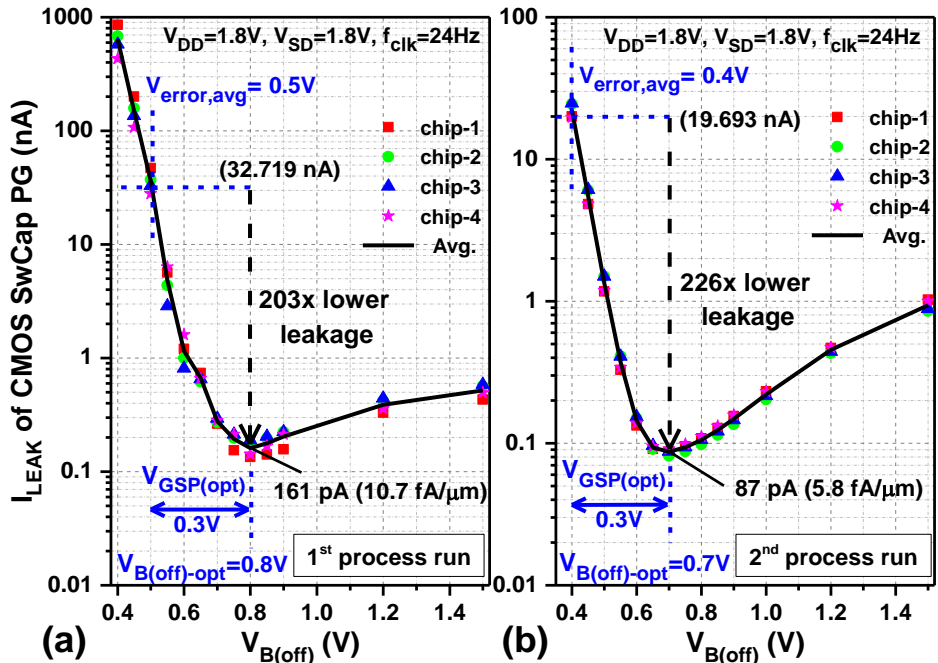

Fig. 2.23. Leakage current measurement of standalone CMOS SwCap PG (Fig. 2.21-c) during super turn-off state in (a) 1$^{st}$ process run, and (b) 2$^{nd}$ process run.

$R_{ON}$ measurement of the standalone CMOS SwCap PG switch (Fig. 2.21 (c)) is plotted in Fig. 24. Due to super turn-on, for a $V_{B(on)} = 1.5$ V, $R_{ON}$ actually reduces by 26% compared to $V_{B(on)} = 0$ V. But due to the voltage error ($V_{error,avg}$) indicated in (2.24), the CMOS SwCap PG achieves only 18% reduction in $R_{ON}$ compared with the conventional PG, as shown in Fig. 2.24. The difference in the value of $V_{error,avg}$ during the super turn-off and the super turn-on state is not shown by (2.22) and (2.23), due to the simplifications made in arriving at the same. $R_{ON}$ plotted in Fig. 2.24 includes the parasitic series resistance of on-chip metal interconnects in the PG switch.

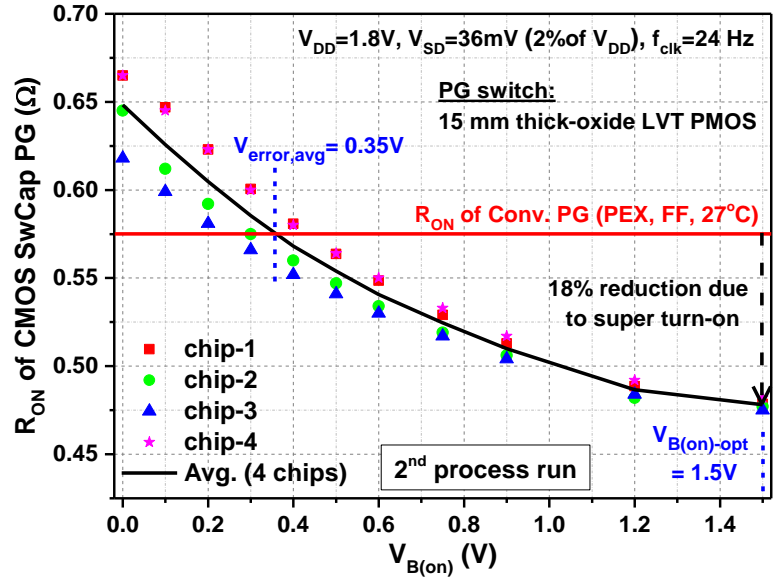

Fig. 2.24. $R_{ON}$ measurement of the standalone CMOS SwCap PG (Fig. 2.21-c) in the super turn-on state.

### 2.6.2 Area and power overhead in the CMOS SwCap PG

The PG switch occupies an area of 46,400 µm². The combined area of the switch network and the digital control block is 6,500 µm² (14% of the PG switch area). However, for the same $R_{ON}$ (0.475Ω, corresponding to $V_{B(on)}$=1.5V in Fig. 2.24), the total area of the CMOS SwCap PG (52,900 µm²) is 6.5% lower than the conventional PG technique (56,585 µm²), due to the super turn-on implementation.

In this work, silicon verification of the proposed SwCap PG was of prime concern. For a $R_{ON}$ of 0.475Ω, the leakage current and the area of the conventional PG switch made using RVT core device in the 180nm process is 58.884 nA (TT corner, 27°C) and 7,820 µm², respectively. Hence, for the same $R_{ON}$, CMOS SwCap PG reduces the leakage current by 677×, compared to the conventional PG made using RVT core device. However, this leakage reduction comes with an increase in the total PG switch area by a factor of 6.7×.

The measured power consumption in the digital control block and the SwCap network ($V_{B(off)}$=1.5V, $f_{clk}$=24Hz) is 50 pW and 15 pW, respectively. The power dissipation due to the periodic refresh in the SwCap network scales linearly with the PG switch size. For flexibility in testing, the refresh clock and the voltage ($V_{B(off)}$ or $V_{B(on)}$) is provided to the chip. However, the total system power overhead should also be lower. From [33], the power consumption for generating a 24 Hz clock in the 180 nm technology is estimated to be 51 pW. The overhead for generating multiple discrete values of voltage using the series-stack of diode-connected transistors in 180 nm technology is estimated to be 100 pW to few nW's of power [34]. This power overhead is shared by all the multiple SwCap PG units in the chip. For the programmability of $V_{B(on)}$ or $V_{B(off)}$, the appropriate voltage is then selected among the multiple discrete values using a multiplexer.

### 2.6.3 Measurement results of the MEMS SwCap PG

Fig. 2.25 shows the PCB containing MEMS switch ICs interfaced with the chip. The leakage current measurement of the MEMS SwCap PG is shown in Fig. 2.26. By utilizing the MEMS devices as switch elements, the gate voltage of the PG switch is close to its ideal value ($V_{error,avg} \approx 0V$) as discussed in Section 2.5. Hence, for a $V_{B(off)} = 0V$, the leakage current corresponds to the conventional PG. In the super turn-off state, the MEMS SwCap PG achieves a maximum leakage reduction of 172× at a $V_{B(off)-opt} = 0.3V$, compared with the conventional PG. Due to the super turn-on, $R_{ON}$ reduction of 26% is achieved with respect to the conventional PG, as indicated in Fig. 2.27. For a given PG signal, the gate voltage of the MEMS SwCap PG, indicating the four different possible states in the PG switch, is shown in Fig. 2.28.

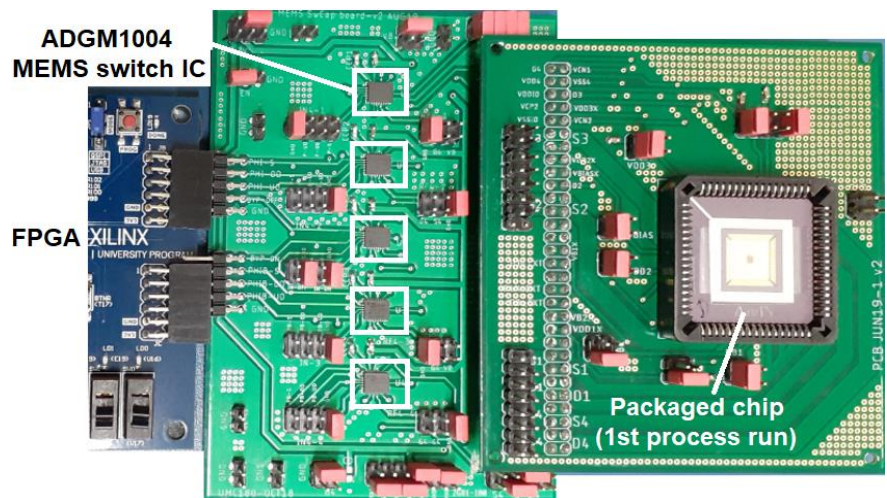

Fig. 2.25. PCB showing the realization of MEMS SwCap PG (Fig. 2.17). Five ADGM1004 (SP4T) ICs are used to realize the SwCap and bypass network.

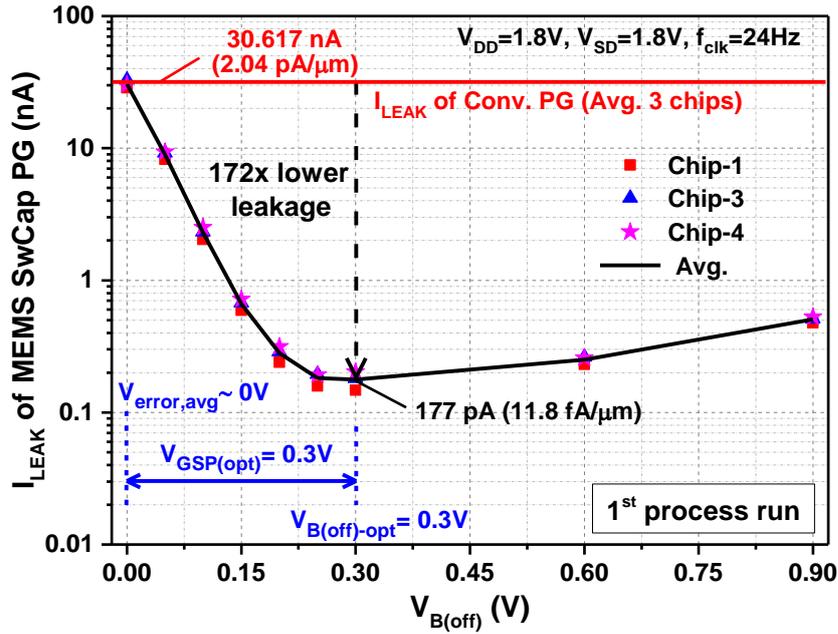

Fig. 2.26. Leakage current measurement of the MEMS SwCap PG (Fig. 2.17) in the super turn-off state.

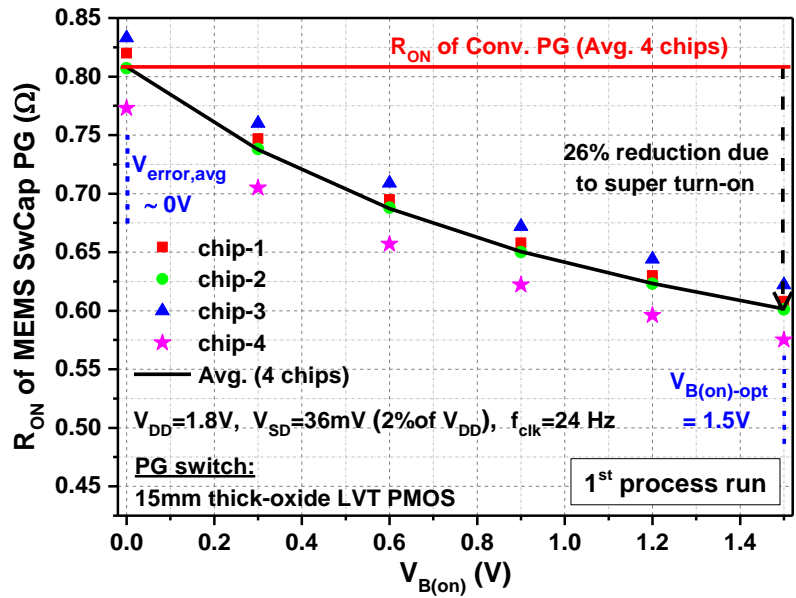

Fig. 2.27. $R_{ON}$ measurement of the MEMS SwCap PG (Fig. 2.17) in the super turn-on state.

### 2.6.4 Comparison with state of the art super cut-off PG implementations

Table 2.3 provides the comparison of the proposed SwCap PG with the other super cut-off PG methods. As seen in Table 2.3, the SwCap PG implements the super turn-off and the super turn-on state together for the first time. Maximum leakage current reduction (by biasing at $V_{GSP(opt)}$) in the super turn-off state is achieved while avoiding voltage stress on the PG switch. The proposed gate biasing method incurs the least additional power overhead among the various other reported works.

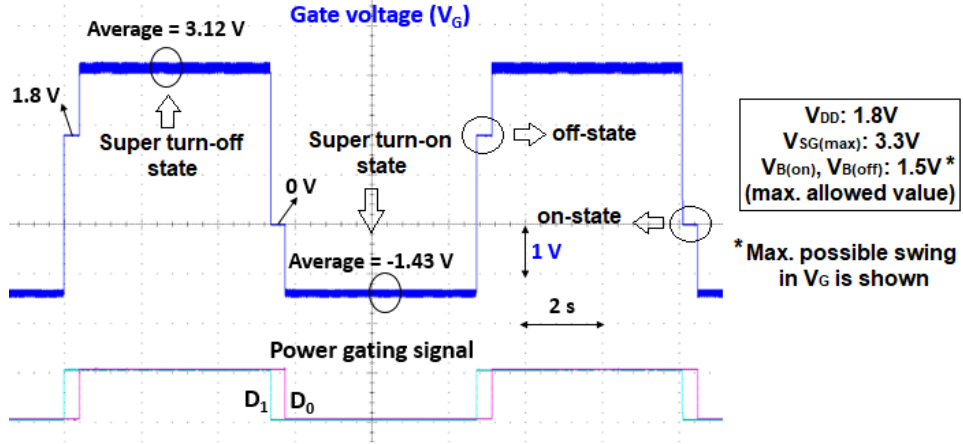

Fig. 2.28. Measured $V_G$ in MEMS SwCap PG indicating four different states of the PG switch. Due to the loading of the oscilloscope probe (1MΩ) at the gate node, 30nF of external capacitors were used in parallel to the $C_1$ and $C_2$.

Table 2.3: Comparison with the other super cut-off power gating implementations

| References | JSSC 2000, [12] | JSSC 2008, [21] | ISSCC 2013, [15] | TCAS-II 2013, [16] | JSSC 2014, [17] | JSSC 2018, [18] | TCAS-I 2018, [19] | This work 2020 |
|---|---|---|---|---|---|---|---|---|
| CMOS process | 0.3 µm | 65 nm | 130 nm | 180 nm | 90 nm | 28 nm | 28 nm | 180 nm |
| Nominal $V_{DD}$ | 3.3 V | 1.2 V | 1.2 V | 1.8 V | 1.0 V | 1.05 V | 1.05 V | 1.8 V |
| $V_{DD}$ used | 0.5-0.8 V | 1.2 V | 0.3 V | 0.5 V | 0.5 V | 0.55 V | 0.25 V | 1.8 V |
| PG switch type | Core LVT PMOS | Core RVT PMOS | Core LVT PMOS | Core RVT NMOS | Core RVT PMOS | Core HVT PMOS | Core RVT PMOS | Thick-oxide LVT PMOS |
| Voltage stress | No | Yes | No | No | No | No | No | No |
| Width of PG switch | 10 µm | 1 mm | 2 mm | 17.16 µm | N/R | N/R | N/R | 15 mm |
| Leak. reduction ratio in super cut-off | <1pA per logic gate | ~1000× | 95× | 19.3× | 9.8× | 35× | 31.5× | [a]186 - [b]226× (CMOS) [c]172× (MEMS) |
| $V_{GSP(opt)}$ in super cut-off | No | Yes | No | No | No | No | No | Yes |
| $f_{clk}$ (for gate bias) | 10 kHz | 100 kHz | Self super cut-off | 10 Hz | 10 kHz | $V_{DD,H}$ available | $V_{DD,H}$ available | 24 Hz |
| $P_{OVERHEAD}$ (∝ PG switch size) | 50 nW | 45 nW | Nil | 0.5 pW | 19.5 nW | Nil | Nil | 65 pW [*] 160 pW [&] |
| $P_{OVERHEAD}$ / $P_{LEAK\ (SUPER\ CUT-OFF)}$ | 990 | 7.50 | Nil | 1.66 | 2.19 | Nil | Nil | 1.02 |
| Super turn-on implementation ($R_{ON}$ reduction) | No | No | No | No | No | No | No | Yes 18% (CMOS) 26% (MEMS) |

N/R: Not-reported, [*]Measured value, [&]Total system power including the estimated value for 24 Hz clock and voltage ($V_{B(off)}$ or $V_{B(on)}$) generation
[a] & [b] value for CMOS SwCap PG in 1st & 2nd process run, [c] value for MEMS SwCap PG in 1st process run

## 2.7 Applicability of SwCap PG in the Nano-Scale CMOS Technologies

Due to cost considerations and affordability reasons, we have experimentally demonstrated our proposed SwCap PG in a relatively older 180nm CMOS technology. However, leakage currents of large digital circuits in the nano-scale CMOS technologies are of prime concern in today's era. As an example, the leakage currents of high performance (LVT) and standard performance (RVT) core device is 10nA/µm and 0.1nA/µm, respectively in 22nm FinFET technology [35]. For the 14nm FinFET technology, the leakage currents of LVT and RVT core device amounts to 100nA/µm and 1nA/µm, respectively [36]. This results in an enormous amount of net leakage current in the circuit load when packing billions of transistors together on a chip. The proposed SwCap PG achieves the lowest possible leakage current for the same $R_{ON}$ among the various PG switch options, as discussed in Section 2.3.1. The requirements for the applicability of the proposed SwCap PG technique in any given nano-scale CMOS technology are outlined as follows.

*1) Presence of thick-oxide devices* for avoiding voltage stress, reducing the gate leakage current of the PG switch and the CMOS switch network leakage current. FinFET is one of the main technology for implementing microprocessors. Thick-oxide devices used for I/O signaling or power management functions are available in the FinFET technology [35]-[37]. The increased lateral silicon area due to the thick-oxide device as the PG switch is not a concern if the PMOS PG switches are integrated vertically (3D-IC) as demonstrated in [38] and [39].

*2) Availability of NMOS devices with isolated well* for handling negative voltages in the CMOS super turn-on implementation. In a System-on-Chip (SoC), logic, memory, DSP, analog, mixed-signal, RF, and power management circuits are integrated onto the same silicon die. This leads to the development of process technology for integrating the aforementioned functional units. Hence, NMOS devices with the isolated well continues to be available in the FinFET technology due to its requirements in the SoC [35]-[37]. Gate-All-Around (GAA) nanowire FETs have their channel completely isolated from the substrate due to the gate material wrapped around them [40]-[41]. This inherent isolation of the depleted channel from the substrate avoids the formation of parasitic diodes between the source (or drain) and the substrate. The Fully-Depleted Silicon-On-Insulator (FD-SOI) technology has a similar feature of the inherently isolated channel [42]. Hence, isolated NMOS device is inherently present in GAAFETs and FDSOI-FETs.

*3) Long channel length devices* for reducing the leakage current in the CMOS switch network. In FinFETs or GAAFETs, the length of a transistor cannot be arbitrarily controlled, due to the restriction from the fabrication process [37]. Hence, the leakage currents in the CMOS switch network can be reduced by increasing the effective channel length using the series stacked configuration of transistors [37].

The proposed SwCap PG technique does not rely on dynamic body bias effect which is absent in advanced CMOS nodes, and duly considers the effect of GIDL which is highly prevalent in nano-scale CMOS devices. The physical mechanism underlying the proposed SwCap PG for maximum leakage reduction ($V_{GSP}=V_{GSP(opt)}$) in super turn-off, and for reducing $R_{ON}$ ($V_{SGP}>V_{DD}$) in super turn-on state, is valid in any process technology including the emerging beyond-CMOS devices.

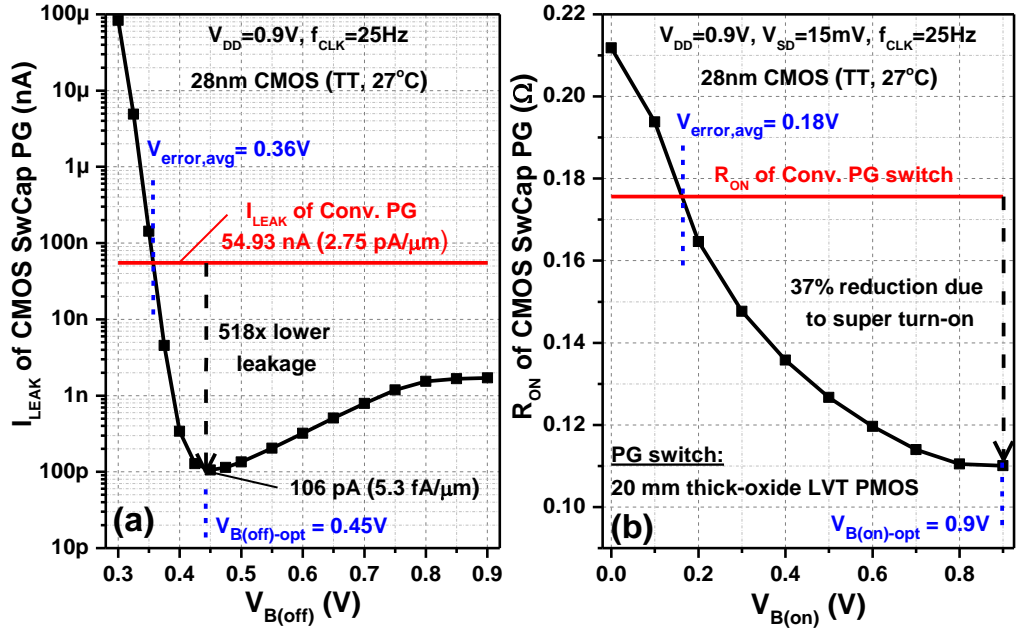

Fig. 2.29. (a) Leakage current during super turn-off state, and (b) $R_{ON}$ during super turn-on state of the proposed CMOS SwCap PG in 28nm CMOS technology.

Table 2.4: Comparison of SwCap PG in 180nm and 28nm CMOS technology

| CMOS process | 180 nm (measured) | 28 nm (simulation) |
|---|---|---|
| Nominal $V_{DD}$ | 1.8 V | 0.9 V |
| $V_{DD}$ used | 1.8 V | 0.9 V |
| PG switch type | Thick-oxide LVT PMOS | Thick-oxide LVT PMOS |
| Voltage stress | No | No |
| Width of PG switch | 15 mm | 20 mm |
| Leak. reduction ratio in super cut-off | 186 - 226× | 518 × |
| Always $V_{GSP(OPT)}$ in super cut-off | Yes | Yes |
| $f_{CLK}$ (for gate bias) | 24 Hz | 24 Hz |
| $P_{OVERHEAD}$ ($\propto$ PG switch size) | 160 pW | 250 pW |
| $P_{OVERHEAD} / P_{LEAK\ (SUPER\ CUT-OFF)}$ | 1.02 | 2.62 |
| Super turn-on ($R_{ON}$ reduction) | Yes (18%) | Yes (37%) |

As a validation, Fig. 2.29 (a) and (b) shows the leakage current reduction and $R_{ON}$ reduction during the super turn-off and super turn-on state respectively in 28nm CMOS process. Table 2.4 summarises the performance comparison of the CMOS SwCap PG in 28nm technology with the chip measurements made with 180nm CMOS technology. As expected, the benefits of CMOS SwCap PG increases for scaled CMOS technology. Table 2.4 shows the comparison of benefits obtained using the proposed SwCap PG in 180nm and 28nm CMOS technology.

## 2.8 Conclusion

Switched-capacitor assisted PG is proposed to reduce the standby power of the digital circuits. For the first time, PG switch is biased in the super turn-off and the super turn-on mode during the off-state and the on-state, respectively. The voltage stress on the PG switch is avoided when the circuit load operates with the nominal $V_{DD}$ of the technology. Due to the optimal gate bias, maximum possible leakage current reduction is achieved. The proposed SwCap PG has four possible states, and requires low power and area overhead. Measurement results of CMOS SwCap PG in 180 nm CMOS technology show that the leakage current and $R_{ON}$ reduce by 186-226× and 18%, respectively, compared with the conventional PG technique. In other words, for the same $R_{ON}$, CMOS SwCap PG achieves an equivalent leakage current reduction of 227-275× compared with the conventional PG. Measurement results of MEMS SwCap PG show that the leakage current and $R_{ON}$ reduce by 172× and 26%, respectively compared to the conventional PG. Moreover, for same $R_{ON}$, CMOS SwCap PG lowers the leakage current even further (677× obtained from simulation) compared with the conventional PG made using RVT core device. The applicability of the SwCap PG in the nano-scale CMOS technologies is duly addressed. The integration of the proposed CMOS SwCap network as a hard macro block, which can be instantiated in the digital design flow, is shown in Fig. 2.30. A similar integration approach is applicable to the MEMS SwCap network. The MEMS switches are interfaced with the on-chip PMOS PG switch and the digital control block at the die level or packaged chip level [43].

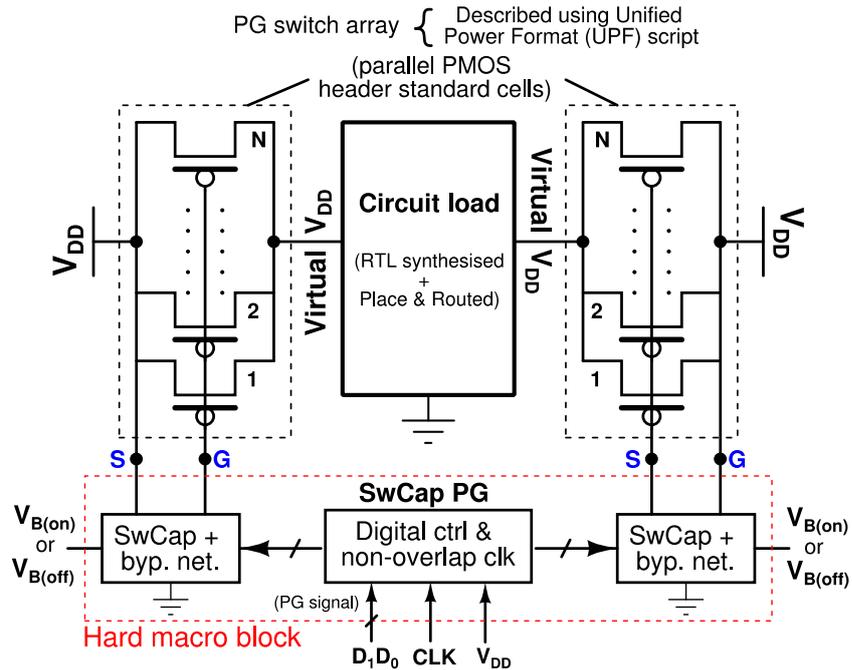

Fig. 2.30. Integration of proposed CMOS SwCap PG into the standard digital ASIC design.

## 2.9 References

[1] R. Salvador, A. Sanchez, X. Fan, and T. Gemmeke, " A cortex-M3 based MCU featuring AVS with 34nW static power, 15.3pJ/inst. active energy, and 16% power variation across process and temperature," in *Proc. IEEE ESSCIRC*, Sep. 2018, pp. 278–281.

[2] J. Myers, A. Savanth, R. Gaddh, D. Howard, P. Prabhat, and D. Flynn, "A subthreshold ARM cortex-M0+ subsystem in 65 nm CMOS for WSN applications with 14 power domains, 10T SRAM, and integrated voltage regulator," *IEEE J. Solid-State Circuits*, vol. 51, no. 1, pp. 31–44, Jan. 2016.

[3] G. Lallement, F. Abouzeid, M. Cochet, J. Daveau, P. Roche, and J. Autran, "A 2.7 pJ/cycle 16 MHz, 0.7 μW Deep Sleep Power ARM Cortex-M0+ Core SoC in 28 nm FD-SOI," *IEEE J. Solid-State Circuits*, vol. 53, no. 7, pp. 2088-2100, July 2018.

[4] Y. Shin *et al.*, "28 nm high-k metal-gate heterogeneous quad-core CPUs for high-performance and energy-efficient mobile application processor," in *IEEE Int. Solid-State Circuits Conf. (ISSCC) Dig. Tech. Papers*, Feb. 2013, pp. 154–155.

[5] H. Mair *et al.*, "A highly integrated smartphone SoC featuring a 2.5 GHz octa-core CPU with advanced high-performance and low power techniques," in *IEEE Int. Solid-State Circuits Conf. (ISSCC) Dig. Tech. Papers*, Feb. 2015, pp. 424–425.

[6] M. Cho *et al.*, "Postsilicon Voltage Guard-Band Reduction in a 22 nm Graphics Execution Core Using Adaptive Voltage Scaling and Dynamic Power Gating," *IEEE J. Solid-State Circuits*, vol. 52, no. 1, pp. 50-63, Jan. 2017.

[7] J. Tschanz, S. Narendra, Y. Ye, B. Bloechel, S. Borkar, and V. De, "Dynamic sleep transistor and body bias for active leakage power control of microprocessors," *IEEE J. Solid-State Circuits*, vol. 38, no. 11, pp. 1838–1845, Nov. 2003.

[8] H. Fariborzi *et al.*, "Analysis and demonstration of MEM-relay power gating," in *Proc. IEEE Custom Integr. Circuits Conf. (CICC)*, Sep. 2010, pp. 1–4.

[9] M. B. Henry and L. Nazhandali, "NEMS-based functional unit power– gating: Design, analysis, and optimization," *IEEE Trans. Circuits Syst. I, Reg. Papers*, vol. 60, no. 2, pp. 290–302, Feb. 2013.

[10] S. Sankar, U. S. Kumar, M. Goel, M. S. Baghini and V. R. Rao, "Considerations for Static Energy Reduction in Digital CMOS ICs Using NEMS Power Gating," *IEEE Trans. on Electron Devices*, vol. 64, no. 3, pp. 1399-1403, March 2017.

[11] Y. Tsividis and C. McAndrew, *Operation and Modeling of the MOS Transistor*, 3rd ed. New York: Oxford Univ. Pr., 2010.

[12] H. Kawaguchi, K. Nose, and T. Sakurai, "A super cut-off CMOS (SCCMOS) scheme for 0.5-V supply voltage with picoampere stand-by current," *IEEE J. Solid-State Circuits*, vol. 35, no. 10, pp. 1498-1501, Oct. 2000.

[13] K.-S. Min, H.-D. Choi, H.-Y. Choi, H. Kawaguchi, and T. Sakurai, "Leakage-suppressed clock-gating circuit with zigzag super cut-off CMOS (ZSCCMOS) for leakage-dominant sub-70-nm and sub-1-V-$V_{DD}$ LSIs," *IEEE Trans. Very Large Scale Integr. (VLSI) Syst.*, vol. 14, no. 4, pp. 430–435, Apr. 2006.

[14] M. Yuffe *et al.*, "A fully integrated multi-cpu, processor graphics, and memory controller 32-nm processor," *IEEE J. Solid-State Circuits*, vol. 47, no. 1, pp. 194–205, Jan. 2012.

[15] J. S. Chen, C. W. Yeh, and J. S. Wang, "Self-super-cutoff power gating with state retention on a 0.3 V 0.29 fJ/cycle/gate 32b RISC core in 0.13 μm CMOS," in *IEEE Int. Solid-State Circuits Conf. (ISSCC) Dig. Tech. Papers*, 2013, pp. 426–427.

[16] Y. Lee, M. Seok, S. Hanson, D. Sylvester and D. Blaauw, "Achieving Ultralow Standby Power With an Efficient SCCMOS Bias Generator," *IEEE Trans. Circuits Syst. II, Exp. Briefs*, vol. 60, no. 12, pp. 842-846, Dec. 2013.

[17] S.-Y. Hsu, Y. Ho, P.-Y. Chang, C. Su, and C.-Y. Lee, "A 48.6-to-105.2 µW machine learning assisted cardiac sensor SoC for mobile healthcare applications," *IEEE J. Solid-State Circuits*, vol. 49, no. 4, pp. 801–811, Apr. 2014.

[18] Y. Pu *et al.*, "A 9-mm$^2$ Ultra-Low-Power Highly Integrated 28-nm CMOS SoC for Internet of Things," *IEEE J. Solid-State Circuits*, vol. 53, no. 3, pp. 936-948, March 2018.

[19] Y. Chien and J. Wang, "A 0.2 V 32-Kb 10T SRAM With 41 nW Standby Power for IoT Applications," *IEEE Trans. Circuits Syst. I, Reg. Papers*, vol. 65, no. 8, pp. 2443-2454, Aug. 2018.

[20] M. Drazdziulis, P. Larsson-Edefors and L. Svensson, "Overdrive Power-Gating Techniques for Total Power Minimization," *IEEE Computer Society Annual Symposium on VLSI (ISVLSI)*, Mar. 2007, pp. 125-132.

[21] A. Valentian and E. Beigne, "Automatic Gate Biasing of an SCCMOS Power Switch Achieving Maximum Leakage Reduction and Lowering Leakage Current Variability," *IEEE J. Solid-State Circuits*, vol. 43, no. 7, pp. 1688-1698, July 2008.

[22] K. Kawasaki, T. Shiota, K. Nakayama and A. Inoue, "A Sub-µs Wake-Up Time Power Gating Technique With Bypass Power Line for Rush Current Support," *IEEE J. Solid-State Circuits*, vol. 44, no. 4, pp. 1178-1183, April 2009.

[23] M. Fujigaya *et al.*, "A 28nm High-κ metal-gate single-chip communications processor with 1.5GHz dual-core application processor and LTE/HSPA+-capable baseband processor," in *IEEE Int. Solid-State Circuits Conf. (ISSCC) Dig. Tech. Papers*, Feb. 2013, pp. 156-157.

[24] T. Inukai, M. Takamiya, K. Nose, H. Kawaguchi, T. Hiramoto, and T. Sakurai, "Boosted gate MOS (BGMOS): device/circuit cooperation scheme to achieve leakage-free giga-scale integration," in *Proc. IEEE Custom Integrated Circuits Conf. (CICC)*, May 2000, pp. 409-412.

[25] W. Lim, I. Lee, D. Sylvester, and D. Blaauw, "Batteryless sub-nW cortexM0+ processor with dynamic leakage-suppression logic," in *IEEE Int. Solid-State Circuits Conf. (ISSCC) Dig. Tech. Papers*, Feb. 2015, pp. 1–3.

[26] J. P. Cerqueira, J. Li, and M. Seok, "A fW- and kHz-Class Feedforward Leakage Self-Suppression Logic Requiring No External Sleep Signal to Enter the Leakage Suppression Mode," in *IEEE Solid-State Circuits Letters*, vol. 1, no. 6, pp. 150-153, June 2018.

[27] D. S. Truesdell, J. Breiholz, S. Kamineni, N. Liu, A. Magyar, and B. H. Calhoun, "A 6–140-nW 11 Hz–8.2-kHz DVFS RISC-V Microprocessor Using Scalable Dynamic

Leakage-Suppression Logic," in *IEEE Solid-State Circuits Letters*, vol. 2, no. 8, pp. 57-60, Aug. 2019.

[28] P.-H. Chen, C.-S. Wu, and K.-C. Lin, "A 50 nW-to-10 mW output power tri-mode digital buck converter with self-tracking zero current detection for photovoltaic energy harvesting," *IEEE J. Solid-State Circuits*, vol. 51, no. 2, pp. 523–532, Feb. 2016.

[29] C.-L. Lo, H.-C. Cheng, P.-C. Liao, Y.-L. Chen, and P.-H. Chen, "An 82.1%-power-efficiency single-inductor triple-source quad-mode energy harvesting interface with automatic source selection and reversely polarized energy recycling," in *Proc. IEEE Asian Solid-State Circuits Conf. (A-SSCC)*, Nov. 2018, pp. 165–168.

[30] M. Spencer *et al.*, "Demonstration of integrated micro-electro-mechanical relay circuits for VLSI applications," *IEEE J. Solid-State Circuits.*, vol. 46, no. 1, pp. 308–320, Jan. 2011.

[31] (2020). *Analog Devices ADGM1004 SP4T MEMS switch*. [Online]. Available: https://www.analog.com/en/products/adgm1004.html

[32] S. Saha, M. S. Baghini, M. Goel and V. R. Rao, "Sub-50-mV Nanoelectromechanical Switch Without Body Bias," *IEEE Trans. on Electron Devices*, vol. 67, no. 9, pp. 3894-3897, Sept. 2020.

[33] P. M. Nadeau, A. Paidimarri, and A.P. Chandrakasan, "Ultra low-energy relaxation oscillator with 230 fJ/cycle efficiency," *IEEE J. Solid-State Circuits*, vol. 51, no. 4, pp. 789–799, Apr. 2016.

[34] I. Lee, D. Sylvester, and D. Blaauw, "A subthreshold voltage reference with scalable output voltage for low-power IoT systems," *IEEE J. Solid State Circuits*, vol. 52, no. 5, pp. 1443–1449, May 2017.

[35] B. Sell *et al.*, "22FFL: A high performance and ultra low power FinFET technology for mobile and RF applications," in *IEDM Tech. Dig.*, 2017, pp. 29.4.1-29.4.4.

[36] C.-H. Jan *et al.*, "A 14 nm SoC platform technology featuring 2nd generation Tri-Gate transistors, 70 nm gate pitch, 52 nm metal pitch, and 0.0499 um$^2$ SRAM cells, optimized for low power, high performance and high density SoC products," in *Symp. VLSI Technology Dig.*, 2015, pp. T12-T13.

[37] A.L.S. Loke *et al.*, "Analog/mixed-signal design challenges in 7-nm CMOS and beyond," in *Proc. IEEE Custom Integrated Circuits Conf. (CICC)*, 2018, pp. 1-8.

[38] P.-Y. Hsieh *et al.*, "Monolithic 3D BEOL FinFET switch arrays using location-controlled-grain technique in voltage regulator with better FOM than 2D regulators," in *IEDM Tech. Dig.*, 2019, pp. 3.1.1-3.1.4.

[39] W. Gomes *et al.*, "Lakefield and Mobility Compute: A 3D Stacked 10nm and 22FFL Hybrid Processor System in 12×12mm$^2$, 1mm Package-on-Package," in *IEEE Int. Solid-State Circuits Conf. (ISSCC) Dig. Tech. Papers*, Feb. 2020, pp. 144-146.

[40] G. Bae *et al.*, "3nm GAA Technology featuring Multi-Bridge-Channel FET for Low Power and High Performance Applications," in *IEDM Tech. Dig.*, 2018, pp. 28.7.1-28.7.4.

[41] J. Ryckaert *et al.*, "Enabling Sub-5nm CMOS Technology Scaling Thinner and Taller!," in *IEDM Tech. Dig.*, 2019, pp. 29.4.1-29.4.4.

[42] J.L. Coz, B.P. Prayer, B. Giraud, F. Giner and P. Flatresse, "DTMOS power switch in 28 nm UTBB FD-SOI technology," in *IEEE SOI-3D-Subthreshold Microelectronics Technology Unified Conference (S3S)*, 2013, pp. 1-2.

[43] A. C. Fischer *et al.*, "Integrating MEMS and ICs", *Nature Microsyst. Nanoeng.*, vol. 1, no. 15005, pp. 1-16, May 2015.

[44] S. Sankar, M. Goel, P. -H. Chen, V. R. Rao, and M. S. Baghini, "Switched-Capacitor-Assisted Power Gating for Ultra-Low Standby Power in CMOS Digital ICs," *IEEE Trans. Circuits Syst. I, Reg. Papers*, vol. 67, no. 12, pp. 4281-4294, Dec. 2020.

[45] G. Joshi, D. Singh, and S. Thangjam, "Effect of temperature variation on gate tunneling currents in nanoscale MOSFETs," *in Proc. IEEE Conf. Nanotechnology*, Aug. 2008, pp. 37–41.

[46] C. Auth, "22-nm fully-depleted tri-gate CMOS transistors," *in Proceedings of IEEE Custom Integrated Circuits Conference (CICC)*, Sep. 2012, pp. 1–6.

# Chapter 3

# Static Energy Reduction in Digital CMOS ICs using NEMS Power Gating

Due to its infinite off-resistance, NEMS switches have been recently proposed to reduce leakage currents during the standby mode in large-scale digital ICs at the nano-scale regime. However, a detailed analysis of the conditions at which the NEMS devices will have an impact is missing. In this chapter, the technique of power gating is analyzed with a NEMS switch using detailed circuit-level simulations to obtain the conditions under which one can obtain net energy savings as compared to FinFET based power gating. Finally, applicability in the energy reduction on a futuristic System-on-Chip (SoC) for the mobile platform made using 14 nm gate length FinFET device is evaluated. Minimum estimated area required for implementing the NEMS PG switch for power gating the test logic block considered in this work in 14 nm FinFET technology is 1725 $\mu m^2$ (81.9% of the logic block area). The inclusion of NEMS PG switches during the ASIC design phase and their integration with CMOS devices is addressed. Parts of this chapter have been published in [12].

## 3.1 Motivation for NEMS Power Gating

The leakage current has become comparable to the average switching current in nano-scale CMOS technologies, and hence the static energy dissipation amounts to a significant percentage of the total energy consumed in a digital chip [1]. Although leakage current per unit width of a FinFET device is lower when compared to its planar counterpart, packing billions of these devices in IC's makes the overall leakage current still a concern. Continuous innovation in the device is needed to address the concerns of static energy consumption in the future.

Power gating is a common technique used to reduce the leakage current by inserting an additional switch in series with the digital block and is turned off during the idle time, as shown in Fig. 3.1. The leakage path between $V_{DD}$ and ground will be completely eliminated only with an ideal switch. One of the drawbacks of power gating is the reduced effective supply voltage across the digital block due to "IR" drop across the switch resistance, which in turn increases the delay of the logic block.

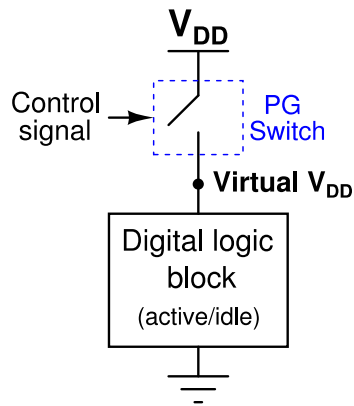

Fig. 3.1. Concept of power gating technique.

The power gating entails a tradeoff between the incurred delay and saving in the leakage power. The former needs a larger width transistor in the on-state (to reduce "IR" drop and subsequently the delay), while the later needs a shorter width transistor in the off-state (so as to have a low leakage current). With the conventional transistor, it is easy to reduce the on-state resistance by adding multiple units in parallel, but the off-state leakage in turn increases. Nano-electro mechanical switches (NEMS) in principal can offer close to infinite resistance in the off-state. So with NEMS power gating switches, the on-state performance can be improved without increasing the leakage current during the off-state, which eventually breaks the tradeoff mentioned earlier.

Some of the proposed solutions for power gating using NEMS switches include [2]-[4]. However, there is no generic analysis of the conditions to predict the effectiveness of NEMS power gating. The work in [5] investigates various parameters of a highly scaled NEMS switch intended for low power circuits. Hence in this work, the cantilever based NEMS switch from [5] is being considered for the analysis of NEMS power gating.

In this chapter, we identify the conditions under which one can obtain saving in energy consumption with NEMS power gating as compared to the FinFET based gating for the same on-state performance. This study is important since energy-saving using power gating depends on the standby scheme in the target digital block. Accordingly, for a mobile system-on-chip (SoC) using 14nm gate length FinFET, the possible net energy reduction using NEMS power gating is evaluated. The method for including the electrical characteristics of the NEMS PG switches during the digital ASIC design phase is discussed. The integration of NEMS PG switches along with the digital CMOS ICs is also addressed.

## 3.2 Description of NEMS Switch and its Equivalent Circuit Model

The structure of the NEMS switch [5] and its corresponding dimensions are shown in Fig. 3.2. The cantilever beam forms the source terminal, which makes electrical contact with the drain through the contact bump at the edge. If the electrostatic force between the gate and source terminal is higher than the spring restoring force of the beam, then the source and drain terminals are shorted, and the switch is set in on-state; else, it is off. For the on-condition, $|V_{GS}|>V_{PI}$ (pull-in voltage).

The Verilog-A model of the NEMS switch, which is used for the purpose of co-simulation with the CMOS circuits, leads to time-consuming simulations and has convergence issues when it is used with large scale circuits [6]. Hence, a simpler model of the switch is obtained by evaluating the parameters under the pessimistic condition for the purpose of fast and still accurate first-order estimation of energy consumption. Important parameters of the switch considered are: pull-in voltage, actuation time delay (mechanical parameters), on-resistance, and parasitic capacitances (electrical parameters). The parameters of the simplified NEMS switch model are illustrated in Fig. 3.3.

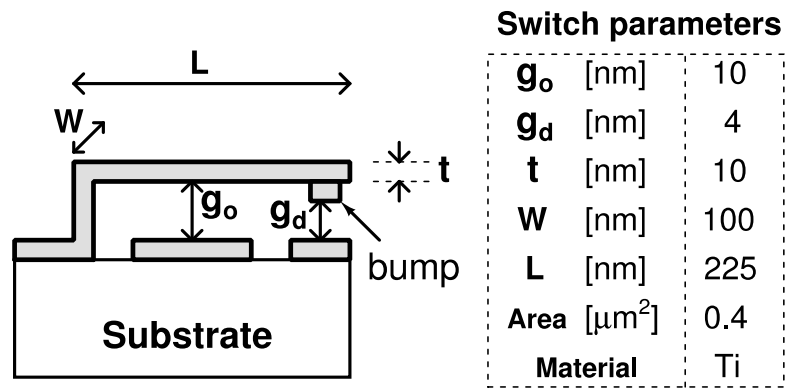

Fig. 3.2. Side view of the 3-terminal NEMS switch.

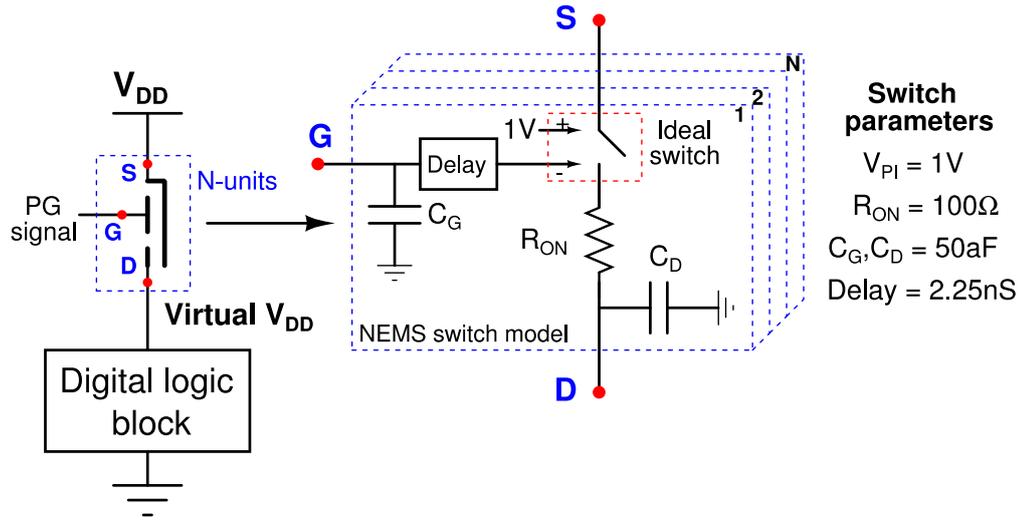

Fig. 3.3. Power gating using NEMS switch and its simplified circuit model.

## 3.3 Definition of Energy Gain

While evaluating the total energy consumption, the following assumptions have been made to make the analysis tractable without losing any key information. The power gating signal is assumed to be periodic, and the internal $V_{DD}$ is considered to be the same as the supply $V_{DD}$ (the assumption is justified in the next section). The "net energy gain ($E_G$)" obtained using NEMS power gating over the FinFET based gating is defined by equation (3.1).

$$E_G = \frac{E_F}{E_N} = \left( \frac{\overbrace{C_{G,F}V_{DD}^2 f_{PG}\left(1+\frac{T_{ON}}{T_{OFF}}\right)}^{\text{dynamic power to turn on PG switch}} + \overbrace{P_{Active}\left(\frac{T_{ON}}{T_{OFF}}\right)}^{\text{on-state active power}} + \overbrace{V_{DD}I_{Leakage}}^{\text{off-state static power}}}{C_{G,N}V_{PG}^2 f_{PG}\left(1+\frac{T_{ON}}{T_{OFF}}\right)+P_{Active}\left(\frac{T_{ON}}{T_{OFF}}\right)} \right) \qquad (3.1)$$

For the NEMS power gating to achieve net energy saving, the value of "$E_G$" must be > 1. The term $P_{Active}$ which is the same for FinFET and NEMS power gating is given by the equation (3.2). The variables used in the equation (3.1) are explained in Table 3.1.

$$P_{Active} = V_{DD}\,I_{Average} = V_{DD}[\,\alpha C_L V_{DD} f_{Logic} + I_{Static}] \qquad (3.2)$$

Table 3.1: Notations used in the energy gain expression

| | |
|---|---|
| $E_F$ | Total energy consumed by the circuit with FinFET power gating during $T_{ON}+T_{OFF}$ |
| $E_N$ | Total energy consumed by the circuit with NEMS power gating during $T_{ON}+T_{OFF}$ |
| $P_{Active}$ | Average power drawn from $V_{DD}$ during the "on-state" of FinFET or NEMS PG switch |
| $I_{Average}$ | Average current drawn from $V_{DD}$ during the "on-state" of FinFET or NEMS PG switch |
| $f_{Logic}$ | Switching frequency of the logic block |
| $f_{PG}$ | Frequency of the power gating (PG) signal |
| $\alpha$ | Activity factor of the circuit in the "on-state" of PG switch |
| $I_{Static}$ | Leakage current drawn from $V_{DD}$ during the "on-state" of FinFET or NEMS PG switch |
| $I_{Leakage}$ | Leakage current drawn from $V_{DD}$ during the "off-state" of FinFET PG switch |
| $C_{G,F(N)}$ | Gate capacitance of the FinFET or NEMS PG switch |
| $C_L$ | Total capacitance of the digital logic block |
| $V_{PG}$ | Actuation voltage applied to the NEMS switch |

## 3.4 Circuit Description

### 3.4.1 Power distribution network

The parasitic resistance and capacitance of the on-chip power distribution network along with the decoupling capacitor are modeled as shown in Fig. 3.4. The switching current waveform gets shaped due to the low pass filtering action of the wire, which makes a more realistic effect of the typical current waveforms in a chip. Values of R, C, and $C_D$ are determined based on the current drawn by the logic block.

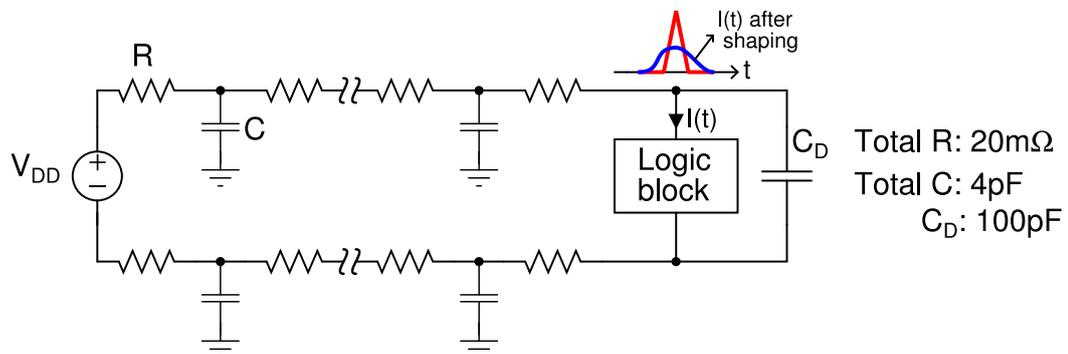

Fig. 3.4. Power distribution network and its effect on current waveform.

### 3.4.2 Logic device model

The logic block is implemented using bulk FinFET devices. The BSIM-CMG model [7] extracted from the TCAD device simulation [8] is calibrated to match the experimental results of the bulk FinFET device from [9]. The $I_D$-$V_{GS}$ plot of 20 nm gate length ($L_G$) FinFET device after calibration is shown in Fig. 3.5. This validated BSIM-CMG model is further used for the SPICE simulations. The important parameters of P-FinFET and the prospective scaled versions of the device with $L_G$=17 and 14 nm are summarized in Table 3.2. The scaling is done by keeping the $I_{OFF}$ value within the acceptable limits in comparison to the gate length (for $L_G$=10 nm, an $I_{OFF}$ of 100nA/µm is achieved in [10]).

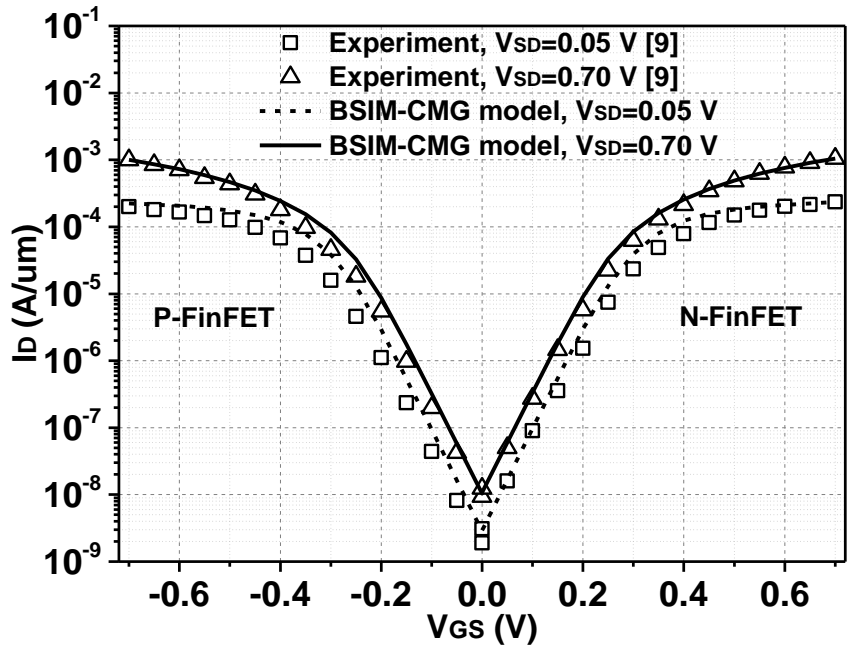

Fig. 3.5. $I_D$-$V_{GS}$ plot of the calibrated bulk FinFET device ($L_G$=20nm).

Table 3.2: Important parameters of P-FinFET device

| $L_G$ (nm) | $T_{FIN}$ (nm) | $H_{FIN}$ (nm) | EOT (nm) | $I_{ON}$ (mA/µm) | $I_{OFF}$ (nA/µm) | $I_{ON}/I_{OFF}$ |
|---|---|---|---|---|---|---|
| 20 | 8 | 42 | 0.85 | 1.00 | 10 | $1.00 \times 10^5$ |
| 17 | 7 | 42 | 0.80 | 1.10 | 37 | $2.97 \times 10^4$ |
| 14 | 5 | 35 | 0.75 | 1.23 | 64 | $1.92 \times 10^4$ |

### 3.4.3 Generic logic block

The energy dissipation for most of the digital logic gates follows that of a simple inverter. So as a generic representation, we have considered an inverter buffer chain of 10 stages, each with a fan-out of 4, as shown in Fig. 3.6. The $1^{st}$ stage (1×) represents a minimum sized inverter possible at a particular technology, and the $2^{nd}$ stage (4×) represents four minimum sized inverters in parallel. The frequency of the waveform generator in Fig. 3.6 is scaled so as to keep the dynamic power the same across all technologies at an activity factor of 1.

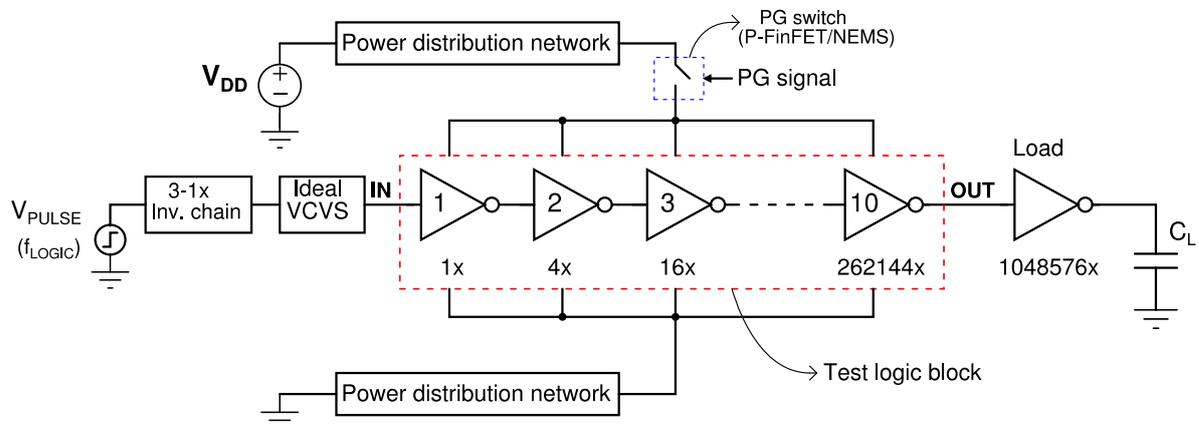

Fig. 3.6. Generic digital block considered for the analysis of power gating.

### 3.4.4 Power gating switch

In this chapter, we have considered the gating of $V_{DD}$ (using header switch). The size of the power gating (PG) switch (for both FinFET and NEMS relay cases) is chosen to satisfy the criteria of only *10% increment in the delay between the input and output terminal of the test logic block at the given clock frequency*. Based on the above criteria, the average voltage drop across the power gating switch is 2% of $V_{DD}$. For FinFET based power gating, the header P-channel FinFET is designed using the same process technology as that of the logic block to reduce the cost. The specifications of the logic block after power gating is shown in Table 3.3. The parameters of FinFET and NEMS PG switch for the same on-state performance are given in Table 3.4.

Table 3.3: Specifications of the logic block

| Technology | 20nm | 17nm | 14nm |
|---|---|---|---|
| $V_{DD}$ (V) | 0.7 | 0.7 | 0.7 |
| $f_{Logic}$ (GHz) | 2.00 | 2.75 | 3.50 |
| $I_{Average}$ at α =1 (mA) | 535 | 535 | 535 |
| $I_{Leakage}$ without PG switch (μA) | 340 | 1170 | 1710 |
| $I_{Leakage}$ with FinFET PG switch (μA) | 99 | 315 | 502 |
| $I_{Leakage}$ with NEMS PG switch (μA) | 0 | 0 | 0 |
| Logic block area (μm$^2$) | 3405 | 2965 | 2105 |

Table 3.4: Parameters of the power gating switch

| Technology | 20 nm | 17 nm | 14 nm |
|---|---|---|---|
| **P-FinFET PG Switch** | | | |
| Total W/L (mm/nm) | 10.05/20 | 11.43/17 | 10.24/14 |
| Gate Capacitance $C_{G,F}$ (pF) | 13.9 | 12.3 | 10.2 |
| On-Resistance (mΩ) | 22 | 20.6 | 23.3 |
| Total area (μm$^2$) | 351 | 355 | 278 |
| **NEMS PG Switch** | | | |
| No. of parallel NEMS switches | 4505 | 4840 | 4310 |
| Gate Capacitance $C_{G,N}$ (fF) | 225 | 242 | 215 |
| On-Resistance (mΩ) | 22.2 | 20.7 | 23.2 |
| Total area (μm$^2$) | 1802 | 1936 | 1724 |

## 3.5 Comparison of FinFET and NEMS Power Gating

The energy gain ($E_G$) obtained with NEMS power gating over the FinFET gating is plotted for various on-off time ratios in Fig. 3.7 (a). Since the on-state dynamic energy is kept constant across different technologies (by increasing clock frequency), so the increment in the $E_G$ with scaling is only due to the higher off-state leakage current. In reality, the clock frequency in a chip doesn't increase continuously with scaling (due to power concerns), so the on-state dynamic energy eventually falls and hence provides two-fold improvement in $E_G$. The on-off time ratio permissible for getting the same $E_G$ in turn increases with the scaling.

Increment in $f_{PG}$ by six orders doesn't affect $E_G$ for typical $T_{ON}/T_{OFF}$ values, indicating the negligible dynamic energy consumed by the PG switch as compared to logic block. At very high $T_{ON}/T_{OFF}$ values, $E_G$ tends to one, because the total energy consumption is dominated by the on-state dynamic energy of the logic block, which is the same for FinFET as well as NEMS power gating. Table 3.5 provides the percentage of energy savings obtained using the NEMS switch and also the $T_{ON}/T_{OFF}$ that is permissible if 10% energy saving is needed.

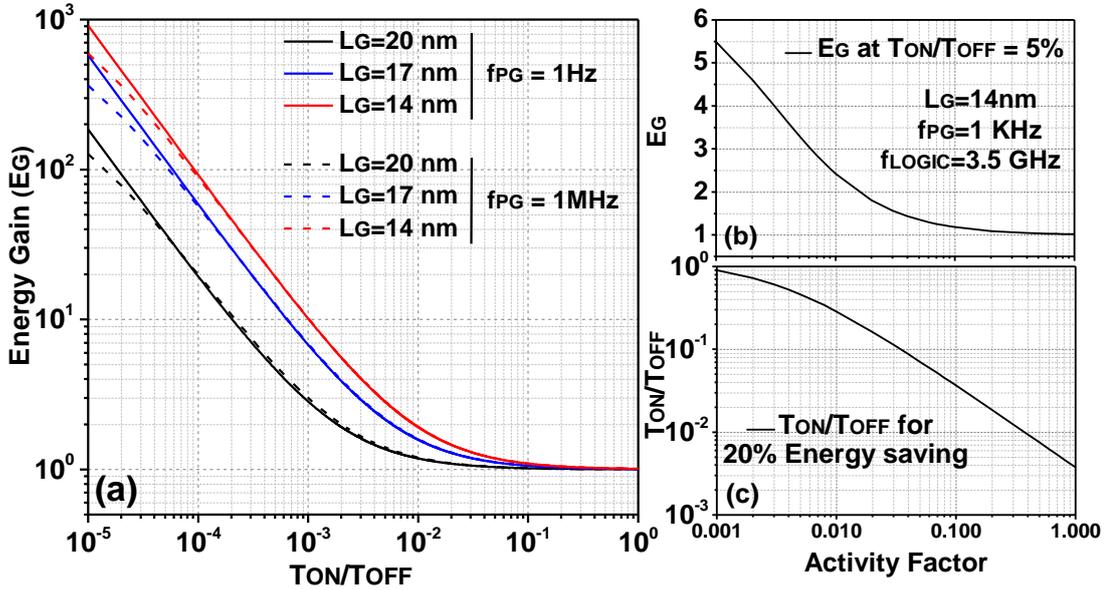

Fig. 3.7. (a) $E_G$ vs $T_{ON}/T_{OFF}$ at $\alpha = 0.1$. $f_{Logic}$ is scaled as given in Table 2.3, (b) $E_G$ vs $\alpha$ and (c) $T_{ON}/T_{OFF}$ vs $\alpha$ for $L_G = 14$ nm device.

Table 3.5: Inference from $E_G$ curve with $\alpha=0.1$

| Technology | 20 nm | 17 nm | 14 nm |
|---|---|---|---|
| Percentage of Net Energy Saving at $T_{ON}/T_{OFF}$=5% | 3.54% | 10.35% | 15.54% |
| Maximum "$T_{ON}/T_{OFF}$" for 10% Energy Saving | 1.7% | 5.2% | 8.4% |

The reduction in on-state energy consumption (either due to lower "$\alpha$" or "$f_{Logic}$") leads to higher $E_G$. As an example, the variation of $E_G$ with respect to $\alpha$ is plotted in Fig. 3.7 (b) and Fig. 3.7 (c) shows the on-off time ratio that is needed for 20% energy saving. Leakage currents considered in the aforementioned sections were at room temperature. The efficiency of the NEMS gating improves with the increase in the temperature, which is reflected in the $E_G$ curve shown in Fig. 3.8.

If the number of logic gates in the circuit increases, the possible increment in the $E_G$ due to the rise in the leakage current gets offset by the increase in the on-state dynamic energy (due to increased capacitance) of the logic block. The leakage current, as well as the capacitance, almost linearly varies with the number of gates. So previously obtained results doesn't vary significantly if the number of gates in the logic block changes. In Fig. 3.9, $E_G$ remains almost constant for a change in the number of logic gates by 1024 times (i.e., from 7 stages (~5.46K gates) to 12 stages (~5.6M gates)). Hence we can analyze the NEMS power gating efficiency for a circuit, irrespective of the number of logic gates present in them. While changing the number of stages, the PG switch size was varied according to the delay criteria given in section 3.4.4.

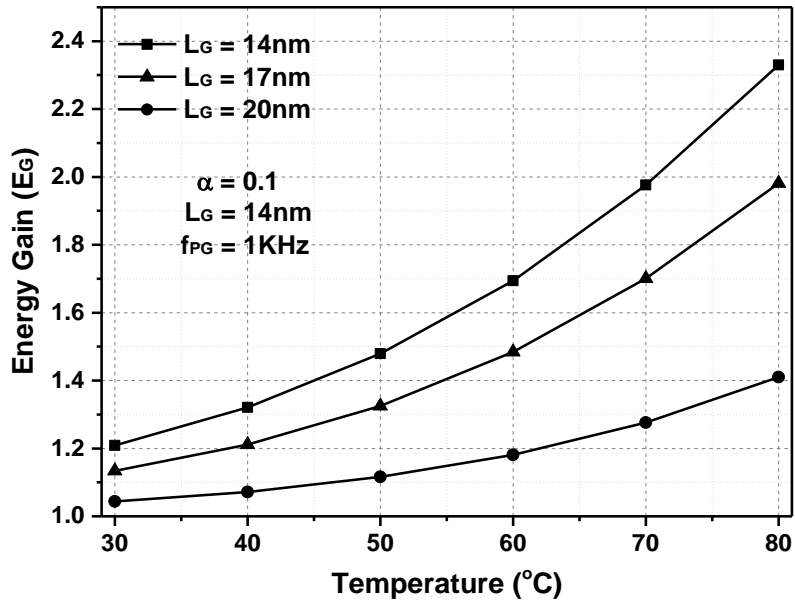
Fig. 3.8. Variation of $E_G$ with temperature.

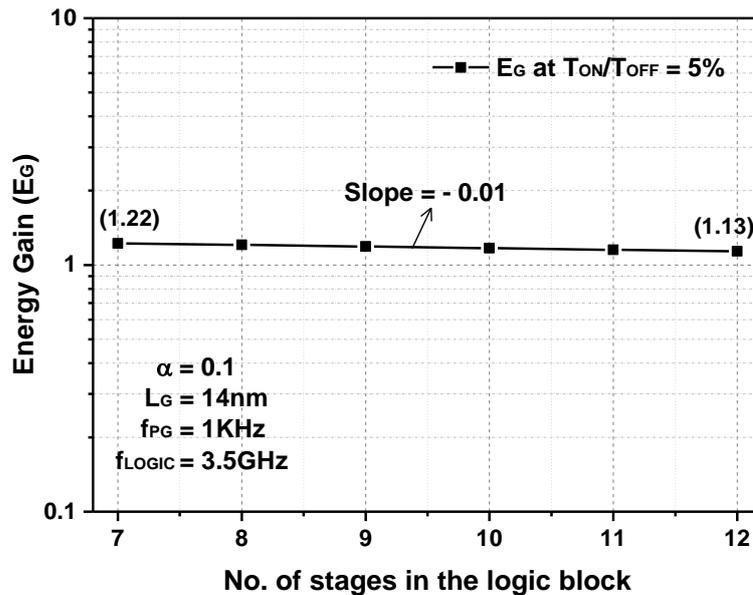
Fig. 3.9. Variation of $E_G$ with no. of logic stages (for $L_G$=14 nm).

As the complexity of the logic gate increases, many transistors are stacked between the $V_{DD}$ and $V_{SS}$ supply rails. The stacking of transistors increases the capacitive load for the previous block that drives it. The body-biasing effect in a stacked transistor is negligible here due to the fully depleted fins of the FinFET [11]. The effect of stacking is incorporated in the buffer chain by using a combination of 3-input static NAND and NOR gates configured as inverters. Due to stacking, the on-state energy consumption of logic increases, as well the leakage current (due to increased PG switch size to meet the delay criteria), though the former is more dominant, leading to a net decrement in the $E_G$ as shown in Fig. 3.10.

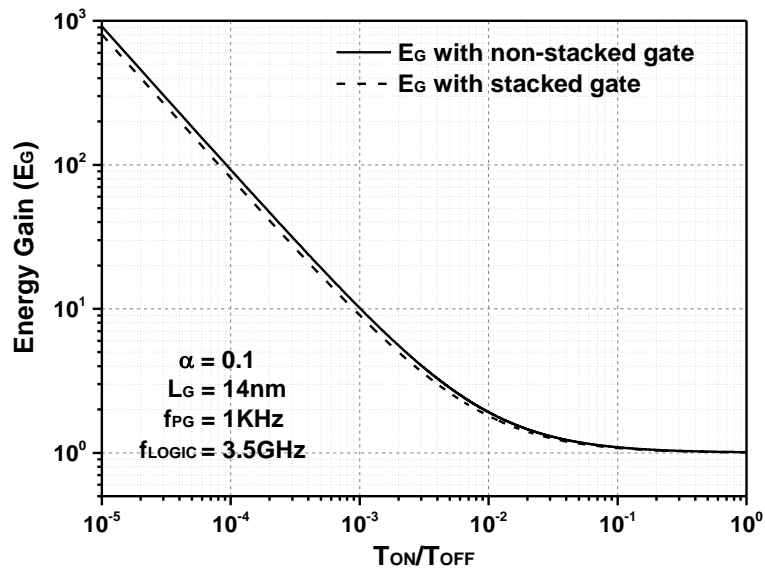

Fig. 3.10. $E_G$ after including stacking effect in the logic block (for $L_G$=14 nm).

## 3.6 Applicability of NEMS Power Gating in System on Chip (SoC)

In a System-on-chip (SoC) many functional units are integrated on the same chip and have specific operating conditions like activity factor, on-off time ratio, and clock frequency. As discussed in section 3.5, the insensitivity in the $E_G$ with respect to the logic block size can be exploited to study the NEMS power gating only with respect to the operating conditions of the units. For a case study, NEMS gating is evaluated to figure out its potential in reducing the overall energy consumption of an SoC at 14 nm gate length FinFETs. The value of clock frequency is chosen in a pessimistic approach by keeping in mind the futuristic scenario that may arise in a mobile SoC. The percentage of energy-saving obtained with NEMS gating in various functional units is summarized in Table 3.6. NEMS switch can be used for gating only in units where significant energy saving is achieved.

Table 3.6: Possible energy reduction using NEMS power gating in

SoC made with $L_G$ = 14nm FinFET

| Functional units in a SoC for mobile platform | Values used for | | | % of Net Energy saving using NEMS power gating |
|---|---|---|---|---|
| | α | $T_{ON}/T_{OFF}$ | $f_{Logic}$ (GHz) | |
| Application processor | 0.1 | 10% | 3.5 | 12.2% |
| Cache memory [+] | 0.05[+] | 10% | 3.5 | 23.4% |
| Clock distribution * | 1 | 10% | 3.5 | 1.6% |
| DSP/GPU | 0.1 | 10% | 1.0 | 29.5% |
| Baseband processor | 0.5 | 5% | 1.0 | 16.3% |

Temperature = 40°C. * Non-stacked gate. [+] Includes spatial activity also.

## 3.7 Integration of NEMS PG Switches in the ASIC Design

### 3.7.1 Process level integration

The development of NEMS switches has advanced significantly in recent years due to the tremendous benefits it provides with its ideal switch characteristics [13], [14]. Highly scaled NEMS switches with sub-2V pull-in voltages having a mechanical delay on the order of a few nano-seconds are recently demonstrated [15], [16]. The NEMS switches are fabricated using a different process flow as compared with the CMOS devices. Nevertheless, tremendous research has resulted in a closer integration of CMOS-NEMS devices [17]. In [18] - [20], NEMS switches are integrated into the CMOS back-end of line (BEOL) metal stacks using post-fabrication process steps involving the selective removal of the inter-layer dielectric material. In [21]-[23], the NEMS switches are integrated with the CMOS chip in a 3D-IC manner using through-silicon vias (TSV). Package level co-integration of the NEMS and CMOS chip is demonstrated in [17]. NEMS switches are more suitable for power gating at a coarse level due to the limitations of their physical size. As an example, the core-level power gating in a microprocessor implemented in [24] could be realized using the NEMS PG switches. However, CMOS PG switches are still needed to power gate the blocks at a finer level within the microprocessor core [25].

## 3.7.2 Design level integration

Apart from the physical integration of the NEMS PG switches, their inclusion during the ASIC design phase is also equally important. Before considering the inclusion of the NEMS PG switch in the ASIC design phase, let's review the design process for the inclusion of PMOS PG switches in the standard ASIC design flow [26]. Fig. 3.11 summarises the key steps in the ASIC design flow.

The RTL code describes the digital logic functionality of the chip. The description of the power management functions present in the chip is provided by the unified power format (UPF) script written using the tool command language (TCL). Typical power management description includes the designation of the power gated domains, power gating signals, isolation cells between power gated and always-on domains, the designation of the multi-$V_{DD}$ domains, level shifters between multi-$V_{DD}$ domains, and the allowed power states of the different modules. The RTL code, along with the UPF script, is verified for ensuring the correctness of various power states in the chip. After which, gate-level synthesis, place and route, and IR drop analysis in the power rails are carried out. The PMOS PG switches (present as standard cells in the technology library) are physically placed during the place and route step. The number of parallel standard cells that have to be placed is decided based on the on-resistance of the single unit and the current ratings of the logic block that is power gated. The above-described design flow may be reiterated multiple times to meet the power and timing specifications of the chip.

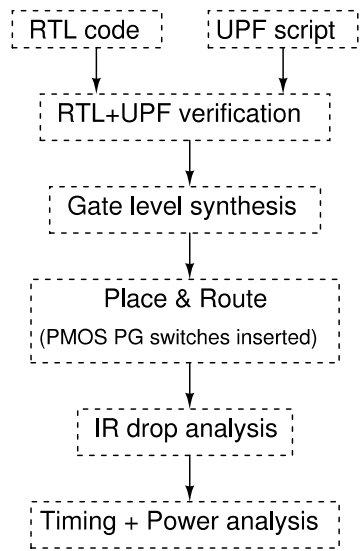

Fig. 3.11. Simplified representation of the key steps involved in the digital ASIC design.

The inclusion of the NEMS PG switch in the digital ASIC design flow is performed as described in Fig. 3.12. The design flow outlined in Fig. 3.12 is similar to the standard ASIC design flow described in Fig. 3.11. Instead of placing PMOS PG switches in parallel, only one PMOS PG standard cell is instantiated in the place and route step. The electrical parameters of the single PMOS PG standard cell are replaced with that of the parallel NEMS PG switches. The required number of parallel NEMS switches is based on the on-resistance of the single switch and the current ratings of the logic block that is power gated. In this way, the NEMS PG is included in the standard ASIC design flow without requiring any modifications.

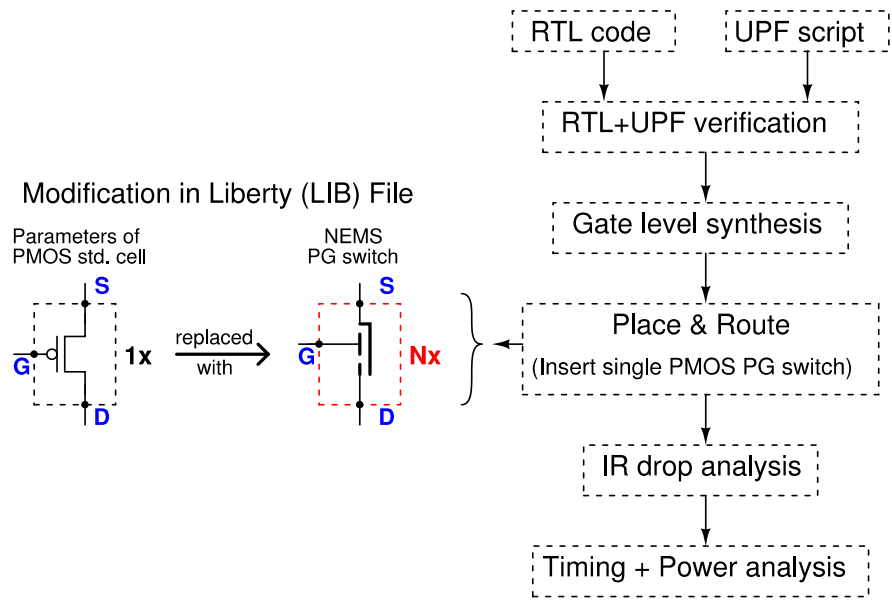

Fig. 3.12. Inclusion of NEMS power gating in the design phase of the digital ASIC design.

## 3.8 Conclusion

This chapter presents the conditions in which NEMS will be an efficient power gating switch when compared to FinFET in reducing static energy consumption. Using circuit-level simulations, the trends in energy-saving under various scenarios were shown. For a mobile SoC made using 14nm gate length FinFET, the effectiveness of NEMS gating was evaluated. The inclusion of NEMS PG switches during the digital ASIC design phase, and its process integration with CMOS ICs is addressed.

# 3.9 References

[1] N. S. Kim, T. Austin, D. Baauw, T. Mudge, K. Flautner, J. S. Hu, M. J. Irwin, M. Kandemir and V. Narayanan, "Leakage current: Moore's law meets static power," in *Computer*, Dec. 2003, vol.36, no.12, pp.68-75.

[2] Raychowdhury, J. Kim, D. Peroulis and K. Roy, "Integrated NEMS Switches for Leakage Control of Battery Operated Systems," in *Proceedings of IEEE Custom Integrated Circuits Conference (CICC),* Sep.2006, vol.457, no.460, pp. 10-13.

[3] M. B. Henry and L. Nazhandali, "NEMS-Based Functional Unit Power- Gating: Design, Analysis, and Optimization," *IEEE Transactions on Circuits and Systems-I: Regular Papers*, vol.60, no.2, pp.290-302, Feb. 2013.

[4] H. Fariborzi, M. Spencer, V. Karkare, Jaeseok Jeon, R. Nathanael, C. Wang, F. Chen, H. Kam, V. Pott, T. J. King Liu, E. Alon, V. Stojanovic and D. Markovic, "Analysis and demonstration of MEM-relay power gating," in *Proceedings of IEEE Custom Integrated Circuits Conference (CICC),* Sep. 2010, vol.1, no.4, pp.19-22.

[5] K. Akarvardar, D. Elata, R. Parsa, G. C. Wan, K. Yoo, J. Provine, P. Peumans, R. T. Howe, and H.-S. P. Wong, "Design considerations for complementary nanoelectromechanical logic gates," in *IEEE International Electron Devices Meeting (IEDM)*, Dec. 2007, pp. 299–302.

[6] H. Alrudainy, A. Mokhov and A. Yakovlev, "A scalable physical model for Nano-Electro-Mechanical relays," in *24th International Workshop on Power and Timing Modeling, Optimization and Simulation (PATMOS)*, Sep. 2014, pp.1-7.

[7] *BSIM-Common Multi-Gate Model (BSIM-CMG)*, 2016. [Online]. Available: http://bsim.berkeley.edu/models/bsimcmg/

[8] U. S. Kumar and V. R. Rao, "A Thermal-Aware Device Design Considerations for Nanoscale SOI and Bulk FinFETs," *IEEE Transactions on Electron Devices*, vol.63, no.1, pp.280-287, Jan. 2016.

[9] S. Natarajan *et al.*, "A 14 nm logic technology featuring 2[nd] generation FinFET, air-gapped interconnects, self-aligned double patterning and a 0.0588μm$^2$ SRAM cell size," in *IEEE International Electron Devices Meeting (IEDM)*, Dec. 2014, pp.3.7.1-3.7.3.

[10] P. Zheng, D. Connelly, F. Ding and T. J. K. Liu, "Inserted-oxide FinFET (iFinFET) design to extend CMOS scaling," *in IEEE International Symposium on VLSI-Technology, Systems and Applications (VLSI-TSA)*, Apr. 2015, pp. 1-2.

[11] C. Auth, "22-nm fully-depleted tri-gate CMOS transistors," in *Proceedings of IEEE Custom Integrated Circuits Conference (CICC)*, Sep. 2012, pp. 1–6.

[12] S. Sankar, U. S. Kumar, M. Goel, M. S. Baghini and V. R. Rao, "Considerations for Static Energy Reduction in Digital CMOS ICs Using NEMS Power Gating," in *IEEE Transactions on Electron Devices*, vol. 64, no. 3, pp. 1399-1403, March 2017.

[13] M. Spencer *et al.*, "Demonstration of Integrated Micro-Electro-Mechanical Relay Circuits for VLSI Applications," in *IEEE Journal of Solid-State Circuits*, vol. 46, no. 1, pp. 308-320, Jan. 2011.

[14] Z. A. Ye *et al.*, "Demonstration of 50-mV Digital Integrated Circuits with Microelectromechanical Relays," in *IEEE International Electron Devices Meeting (IEDM)*, Dec. 2018, pp. 4.1.1-4.1.4.

[15] S. Saha, M. S. Baghini, M. Goel and V. R. Rao, "Sub-50-mV Nanoelectromechanical Switch Without Body Bias," *IEEE Transactions on Electron Devices*, vol. 67, no. 9, pp. 3894-3897, Sept. 2020.

[16] S. Saha, A. Singh, M. S. Baghini, M. Goel and V. R. Rao, "Stand-by Power Reduction Using Experimentally Demonstrated Nano-Electromechanical Switch in CMOS Technologies," Accepted in *IEEE Transactions on Electron Devices*, DOI: 10.1109/TED.2020.3041434 (Early access).

[17] A. C. Fischer, F. Forsberg, M. Lapisa, S. J. Bleiker, G. Stemme, N. Roxhed and F. Niklaus, Integrating MEMS and ICs, *Nature Microsystems and Nanoengineering*, 1, 15005 (2015), DOI: https://doi.org/10.1038/micronano.2015.5.

[18] R. Gaddi, R. Van Kampen, A. Unamuno, V. Joshi, D. Lacey, M. Renault, C. Smith, R. Knipe, D. Yost, "MEMS technology integrated in the CMOS back end," *Elsevier Microelectronics Reliability*, vol. 50, no. 9-11, pp 1593-1598, 2010.

[19] S. T. Wipf, A. Göritz, C. Wipf, M. Wietstruck, A. Burak, E. Türkmen, Y. Gürbüz and M. Kaynak, "240 GHz RF-MEMS switch in a 0.13 μm SiGe BiCMOS Technology," in *IEEE Bipolar/BiCMOS Circuits and Technology Meeting (BCTM)*, Oct. 2017, pp. 54-57.

[20] U. Sikder, G. Usai, T. Yen, K. Horace-Herron, L. Hutin and T. K. Liu, "Back-End-of-Line Nano-Electro-Mechanical Switches for Reconfigurable Interconnects," in *IEEE Electron Device Letters*, vol. 41, no. 4, pp. 625-628, April 2020.

[21] T. Zhang and G. Sun, "Using NEM relay to improve 3DIC cost efficiency," *IEEE International 3D Systems Integration Conference (3DIC)*, 2012, pp. 1-4.

[22] H. S. Yang and M. S. Bakir, "3D integration of CMOS and MEMS using mechanically flexible interconnects (MFI) and through silicon vias (TSV)," *Proceedings 60th Electronic Components and Technology Conference (ECTC)*, 2010, pp. 822-828.

[23] T. Geßner *et al.*, "3D integration technologies for MEMS," *13th IEEE International Conference on Solid-State and Integrated Circuit Technology (ICSICT)*, 2016, pp. 334-337.

[24] S. Rusu *et al.*, "A 45 nm 8-Core Enterprise Xeon Processor," in *IEEE Journal of Solid-State Circuits*, vol. 45, no. 1, pp. 7-14, Jan. 2010.

[25] P. A. Meinerzhagen *et al.*, "An Energy-Efficient Graphics Processor in 14-nm Tri-Gate CMOS Featuring Integrated Voltage Regulators for Fine-Grain DVFS, Retentive Sleep, and $V_{MIN}$ Optimization," in *IEEE Journal of Solid-State Circuits*, vol. 54, no. 1, pp. 144-157, Jan. 2019.

[26] M. Keating, D. Flynn, R. Aitken, A. Gibbons, and K. Shi, "Low Power Methodology Manual for System-on-Chip Design", *Springer*, 2007.

# Chapter 4

# Discrete-Time Signal Amplification Using NEMS Devices

In this chapter, a novel technique of realizing Discrete-Time (D-T) signal amplification using Nano-Electro-Mechanical systems (NEMS) is presented. The amplifier uses mechanical switches instead of traditional solid-state devices and acts as an inherent sample and hold amplifier. The proposed NEMS D-T amplifier operates on a wide dynamic range of signals without consuming DC power. Moreover, the proposed amplifier does not suffer from the leakage current and the non-linearity associated with the sampling ohmic switch. As a proof of concept, the proposed NEMS D-T amplifier is demonstrated in circuit simulations using the calibrated Verilog-A models of the NEMS device. Minimum estimated area required for implementing the NEMS switches in the proposed NEMS D-T amplifier with differential outputs is 1250 µm$^2$. The non-idealities present in the proposed amplifier are highlighted, and possible ways to overcome them are discussed. Finally, the design considerations required for the NEMS D-T amplifier is described. Parts of this chapter have been published in [1].

## 4.1 Motivation for D-T Signal Amplification using NEMS devices

Consider a generic IoT system depicted in Fig. 4.1 (a). The sensor's output signal is generally very small in amplitude and has low bandwidth. Hence, it must be amplified and sampled before converting it into a digital signal using an analog-to-digital converter (ADC). Fig. 4.1 (a) depicts a case where the continuous-time (C-T) analog amplifier is followed by a dedicated sample and hold circuit [2]. In Fig. 4.1 (b), the sensor output signal is directly fed to a discrete-time (D-T) amplifier, which performs sample and hold function, as well as inherent amplification [3], [4]. Then the digital signal is further processed and transmitted. The "sample and hold amplification" function generally precedes the ADC and uses an operational- amplifier (op-amp) [3]-[5], as shown in Fig. 4.2.

In the sampling phase ($\phi_1$), the input "$V_{in}$" is sampled across capacitor $C_1$. In the hold phase ($\phi_2$), the charge "$C_1 V_{in}$" present in capacitor $C_1$ is transferred to the capacitor $C_2$ by the op-amp. Hence, if $C_1 > C_2$, one can obtain a voltage gain equal to $C_1/C_2$. Limitations of the op-amp (finite open-loop DC gain and bandwidth, voltage swing limit, non-linearity, and noise) greatly influence the performance of the D-T amplifier [5], and it is also the most power-hungry block of the circuit. Moreover, metal oxide semiconductor (MOS) switches used for the sampling and charge transfer suffers from the charge injection problem and non-linear on-resistance, which affects the overall performance [5]. The key idea in the switched capacitor amplifier is as follows: sample the input voltage across a large capacitor and dump the same charge on to a smaller capacitor. Hence, the input voltage gets amplified by a factor equal to the ratio of the capacitances.

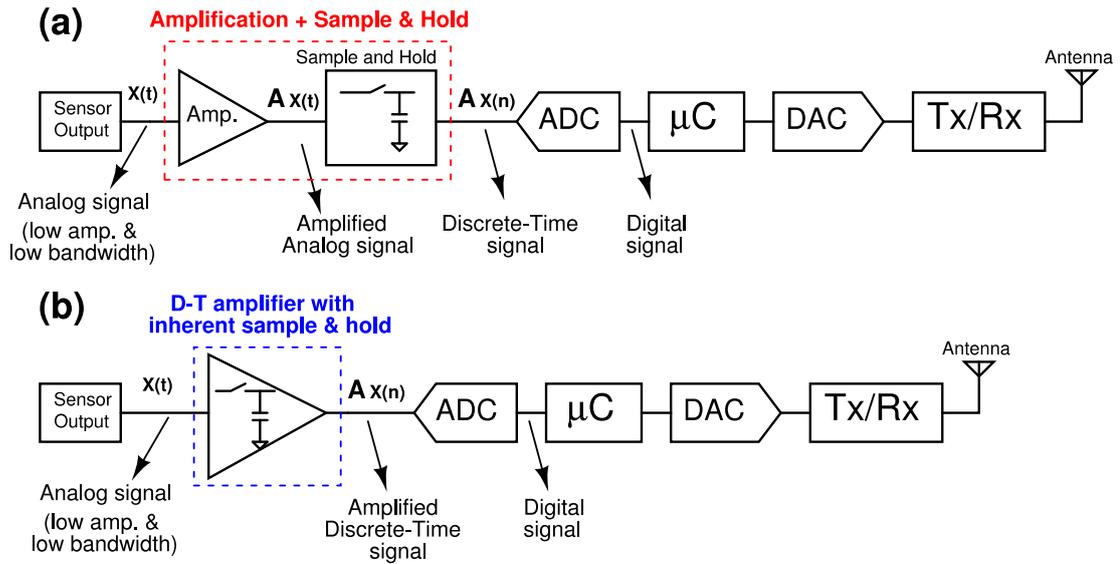

Fig. 4.1. Simplified picture of an IoT sensor system with (a) continuous-time (C-T) analog amplifier followed by a dedicated sample & hold amplifier, and (b) discrete-time (D-T) amplifier with inherent sample & hold.

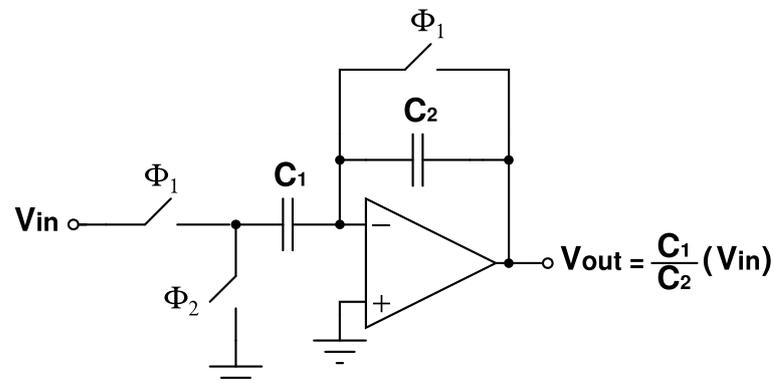

Fig. 4.2. Traditional swiched-capacitor D-T amplifier using op-amp.

In [6], metal-oxide semiconductor (MOS) capacitor is utilized for D-T signal amplification, as shown in Fig. 4.3 (a). The input is sampled at $\phi_1$ across the largest capacitance available in a MOS capacitor (i.e., oxide capacitance "$C_{OX}$" in strong inversion) as shown in Fig. 4.3 (b). In the hold phase ($\phi_2$), the gate terminal is left floating, and an external voltage source ($V_{PULL}$) is used to remove all the inversion layer charges from the channel, as shown in Fig. 4.3 (c). Thus gate-to-body capacitance becomes a series connection of $C_{OX}$ and $C_{DEP}$ (depletion capacitance), which becomes approximately equal to $C_{DEP}$ ($<C_{OX}$). Hence, the gate-to-body capacitance is reduced while keeping the gate charge constant, thus amplifying the input voltage at the gate by a factor of $C_{OX}/C_{DEP}$.

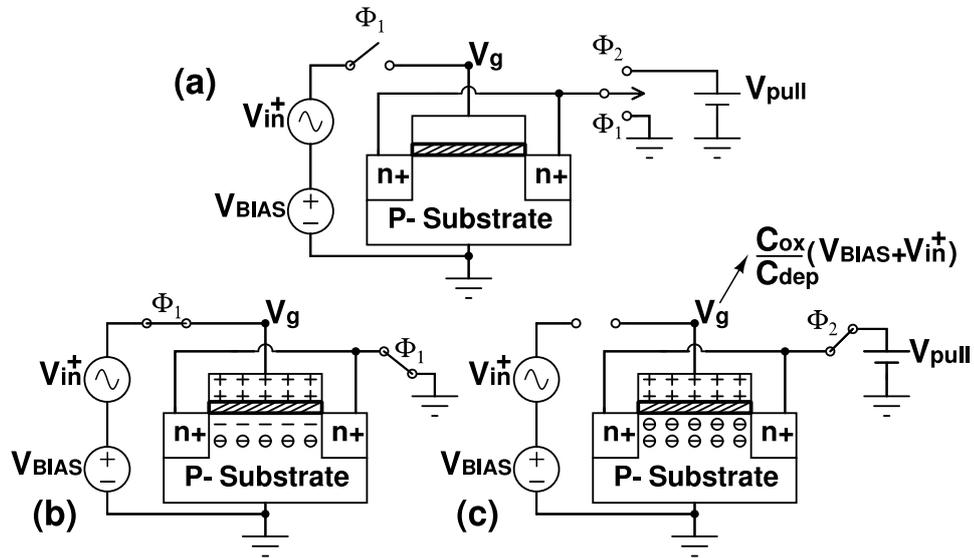

Fig. 4.3. (a) Discrete-Time (D-T) MOS parametric amplifier, (b) during the sample phase, and (c) during the hold phase.

The achievable gain using MOS capacitor-based D-T amplification is typically within the range of 5-10 [6]. For obtaining a determined gain, the MOS device needs to be biased, and the input signal is applied over the bias. In the hold phase, the DC bias also gets amplified, which causes the gate voltage to go to high values. This puts restrictions on the voltage swing limits and causes reliability issues. In a differential implementation, even for low gain (using low DC bias), the input swing is limited since $V_{BIAS} \pm V_{in}^+$ should be high enough to create strong inversion. Moreover, $V_{PULL}$ (Fig. 4.3 (c)) has to be high enough such that the amplified gate voltage should not cause the channel inversion during $\phi_2$.

The development of integrated circuits using mechanical switches has attracted significant interest in recent years due to its high ratio of resistances in the on and off-state, respectively. Hence, the NEMS switch helps in reducing the energy consumption of the system [7]-[9]. In [10], a MEMS resonator was used for harnessing the vibration energy. Continuous-time signal amplification utilizing the parametric effect of the MEMS capacitor is demonstrated in [11]. MEMS switch as an analog trans-conductance element is proposed in [12].

In this work, NEMS devices are used for performing D-T analog signal amplifier operation. Two types of NEMS devices are used: NEMS capacitive switch and NEMS ohmic switch. NEMS device, due to its movable mass, creates a capacitance that varies with the displacement. The variable capacitance of the NEMS capacitive switch is used for amplifying the signal. Hence, the proposed NEMS D-T amplifier does not consume any DC power. The on-resistance of the NEMS ohmic switch is highly linear, and the switch operation doesn't suffer from the charge injection problem. So, the NEMS ohmic switch is used for the sampling and charge transfer. The proposed NEMS D-T amplifier circuit is verified using circuit simulations deploying the calibrated Verilog-A model of the NEMS device.

## 4.2 Description of the NEMS Switch

### 4.2.1 NEMS Capacitive Switch

The structure of the NEMS capacitive switch [13], and its practical designed dimensions are shown in Fig. 4.4 (a). The beam clamped at the two ends forms the top plate of the capacitor. Between the top plate and the bottom plate, there is a bi-layer consisting of air and a dielectric layer ($\varepsilon_d$). If the electrostatic attractive force (due to applied voltage) between the two plates is higher than the spring restoring force of the beam, then the beam gets pulled-in, and the switch is closed, else it is open. For the on-condition, applied voltage $|V|>V_{PI,C}$ (pull-in voltage) and for the off-condition, $|V|<V_{PO,C}$ (pull-out voltage). Fig. 4.4 (b) depicts the off and on-state of the switch, respectively.

The NEMS capacitive switch described in Fig. 4.4 is designed in CoventorWare (FEM MEMS CAD tool) [14] for the purpose of accurate extraction of its relevant parameters, which are provided in Table 4.1. The Verilog-A model [15] of the NEMS capacitive switch is calibrated to match the results from CoventorWare and is then subsequently used for the circuit simulation. The C-V curve and transient characteristics of the NEMS capacitive switch using the calibrated Verilog-A model are plotted in Fig. 4.5 (a) and (b).

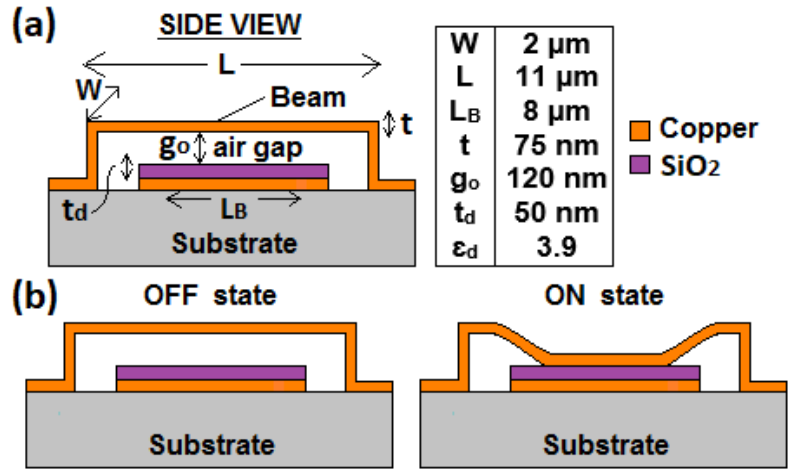

Fig. 4.4. (a) Schematic of NEMS capacitive switch, (b) Switch in off and on-state.

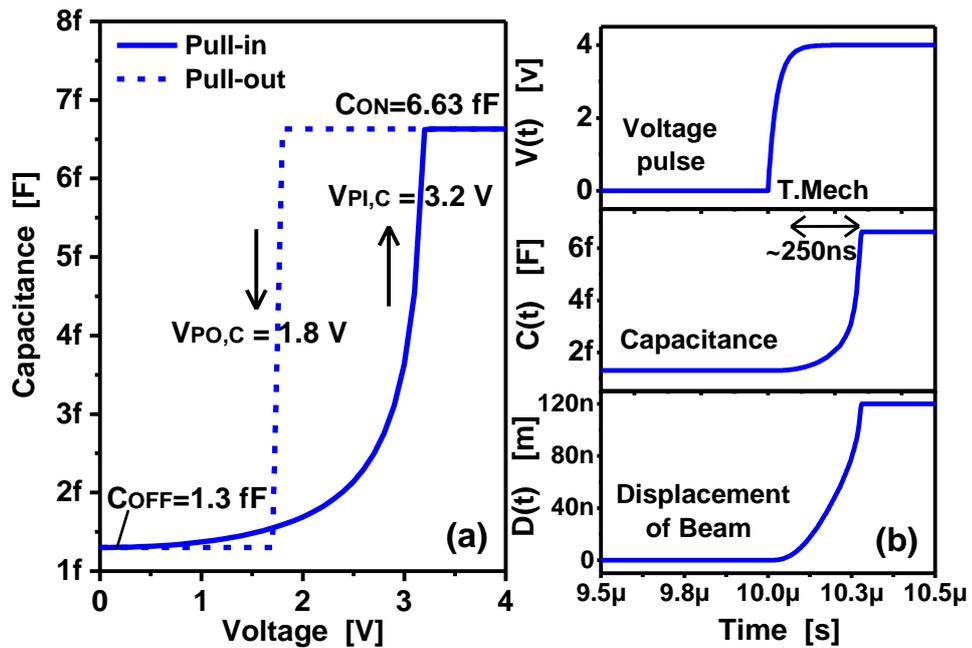

Fig. 4.5. Simulated characteristics of NEMS capacitive switch using the calibrated Verilog-A model (a) C-V curve, (b) Transient characteristics.

Table 4.1: Extracted parameters of the NEMS Capacitive switch

| Parameters | Value |
| --- | --- |
| Pull-in Voltage ($V_{PI,C}$) | 3.2 V |
| Pull-out Voltage ($V_{PO,C}$) | 1.8 V |
| on-state capacitance ($C_{ON}$) | 6.63 fF |
| off-state Capacitance ($C_{OFF}$) | 1.30 fF |
| Mechanical delay ($T_{Mech}$) | 250 ns |

### 4.2.2 NEMS Ohmic Switch

The side view and top view of the NEMS ohmic switch [13], [16], along with its practical dimensions, are shown in Fig. 4.6 (a) and (b), respectively. The cantilever beam fixed at one end forms the gate terminal, to which the floating channel is attached through the dielectric material. The electrostatic attractive force between the gate and the body sets the state of the switch. For the on-condition, $|V_{GB}|>V_{PI,R}$ (pull-in voltage) and for the off-condition, $|V_{GB}|<V_{PO,R}$ (pull-out voltage). Fig. 4.6 (c) depicts the switch in off and on-state, respectively. The relevant parameters of the switch extracted from CoventorWare are provided in Table 4.2. I-V characteristic curve and the transient characteristics of the NEMS ohmic switch using the calibrated Verilog-A model are plotted in Fig. 4.7 (a) and (b).

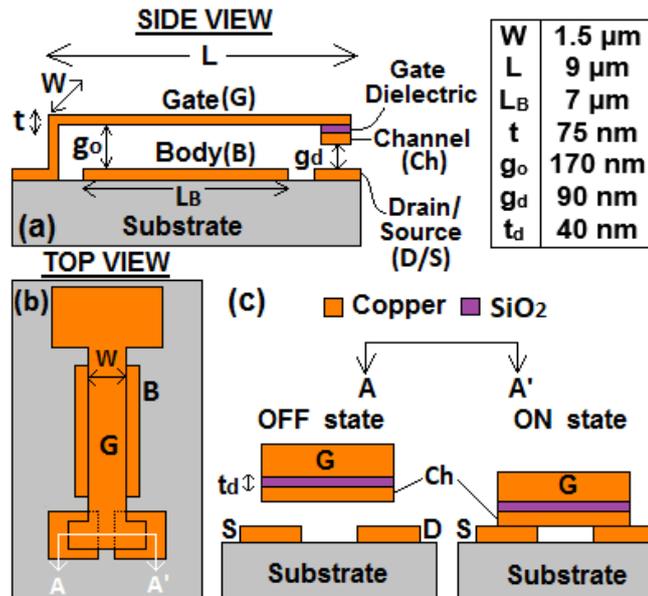

Fig. 4.6. (a) Side view of NEMS Ohmic switch, (b) Top view of the switch, (c) cross section of the switch along A-A' during off and on-state.

Table 4.2: Extracted parameters of the NEMS Ohmic switch

| Parameters | Value |
|---|---|
| Pull-in Voltage ($V_{PI,R}$) | 1.5 V |
| Pull-out Voltage ($V_{PO,R}$) | ~1.5 V |
| On-resistance ($R_{ON}$) | ~10 Ω |
| Mechanical delay ($T_{Mech}$) | 600 ns |

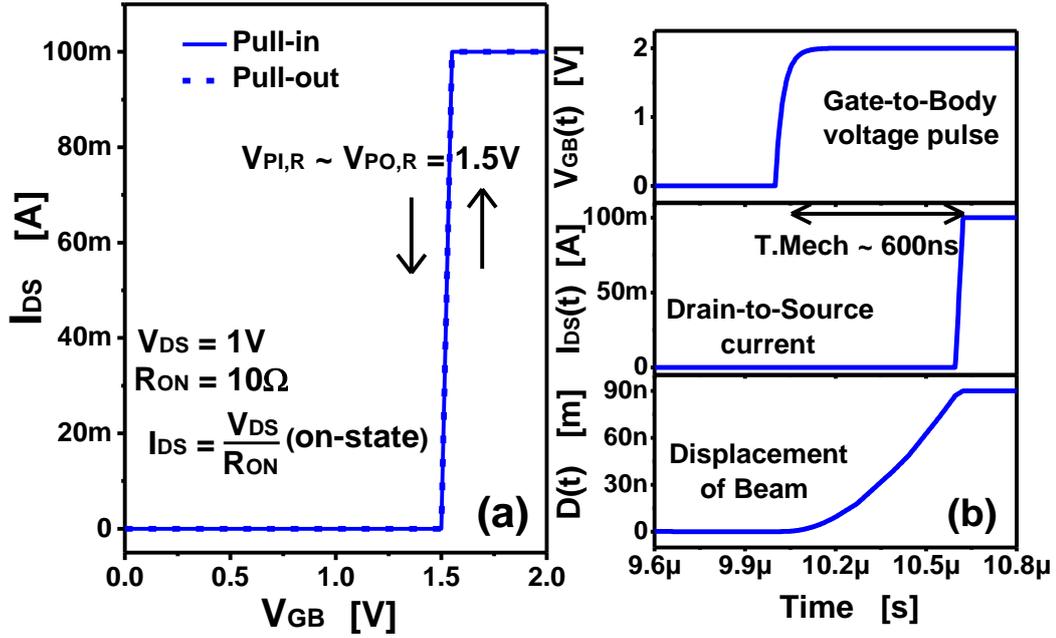

Fig. 4.7. Simulated characteristics of NEMS ohmic switch using the calibrated Verilog-A model (a) I-V curve, (b) Transient characteristics.

## 4.3 Principle of NEMS D-T Signal Amplification

Consider the schematic of the NEMS D-T amplifier shown in Fig. 4.8. "$V_{BIAS}$" and "$V_{in}$" correspond to the DC bias and the input signal of the amplifier, with $V_{in} < V_{BIAS}$. Let's assume two voltages, "$V_{in}+V_{BIAS}$" and "$V_{in}-V_{BIAS}$", are available. In the next section, the technique of generating them is described. The mechanism of amplification is as follows.

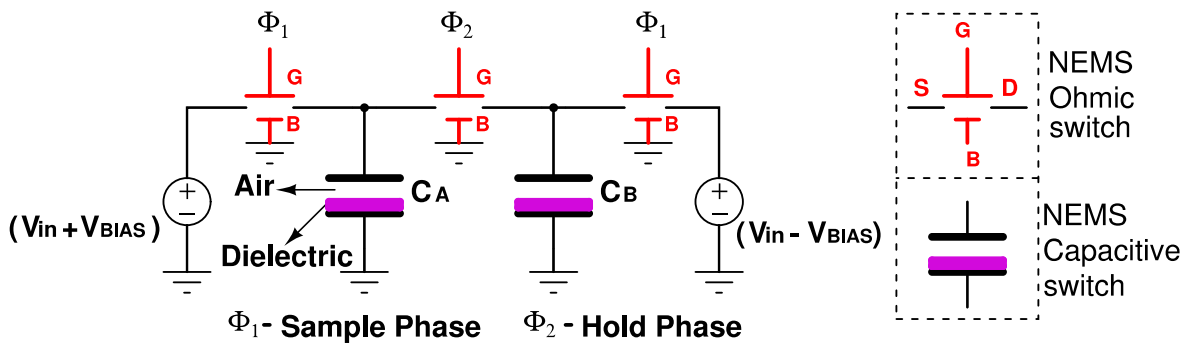

Fig. 4.8. Simplified circuit of the NEMS D-T amplifier.

**Step (i):** During the sample phase ($\phi_1$) the voltages "$V_{in}+V_{BIAS}$" and "$V_{in}-V_{BIAS}$" are sampled across the two NEMS capacitive switch, as shown in Fig. 4.9. If $|V_{in}\pm V_{BIAS}| > V_{PI,C}$, the beams get pulled-in due to the always attractive nature of the electrostatic forces. Hence $C_A$

= $C_B$ = $C_{ON}$, which is of relatively high value (Fig. 4.5 (a)). The voltages are sampled and stored across $C_A$ and $C_B$. The top plate of $C_B$ has negative charges, since $V_{in}$ < $V_{BIAS}$.

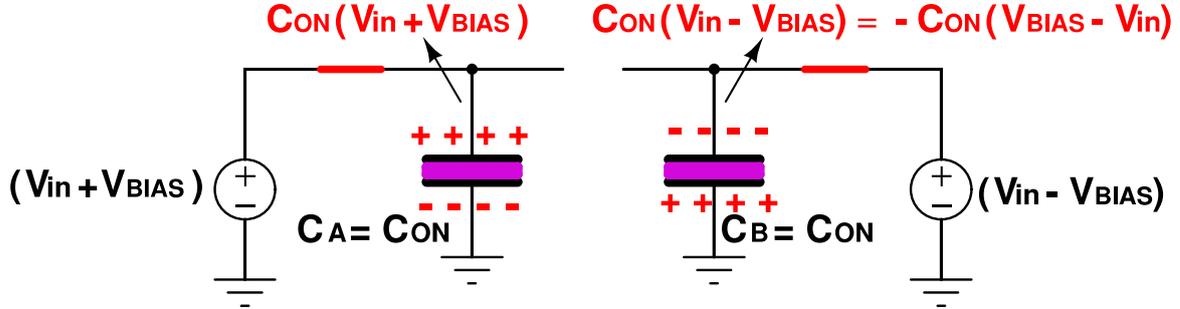

Fig. 4.9. NEMS D-T amplifier circuit during the sample phase.

**Step (ii):** In the hold phase ($\phi_2$), the moment at which the two capacitors are shorted is depicted in Fig. 4.10 (a). Total charge in the top plates of the capacitors is given by $Q_{TOTAL}$ = $C_{ON}$ ($V_{in}$+$V_{BIAS}$) + $C_{ON}$ ($V_{in}$-$V_{BIAS}$). There will be transfer of charge between $C_A$ (left side) and $C_B$ (right side) due to the potential difference. Therefore, the charges due to $V_{BIAS}$ get cancelled.

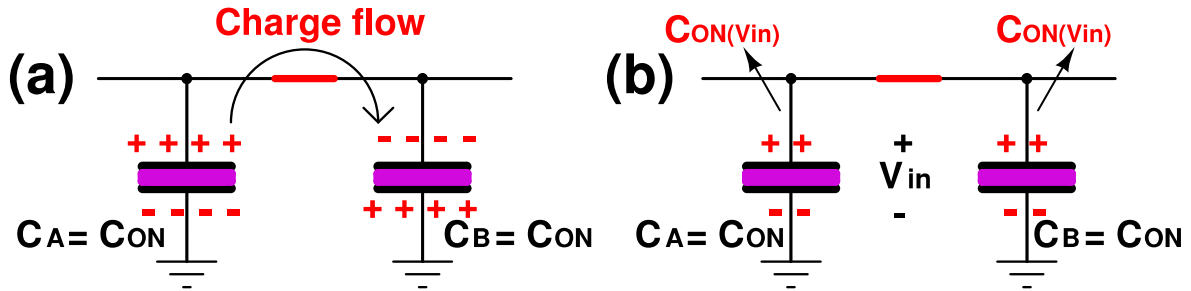

Fig. 4.10. NEMS D-T amplifier circuit during (a) the beginning of hold phase, and (b) the middle of hold phase.

**Step (iii):** The total charge in the top plates is now given by $Q_{TOTAL}$ = 2 × $C_{ON}$ ($V_{in}$) since the two capacitors are in parallel (Fig. 4.10 (b)).

**Step (iv):** For obtaining amplification, $V_{in}$ × ($C_{ON}$/$C_{OFF}$) must be less than the pull-out voltage of the NEMS capacitive switch ($V_{PO,C}$). Since $V_{in}$ × ($C_{ON}$/$C_{OFF}$) < $V_{PO,C}$ the beams get released due to the fact that the spring restoring force is now stronger than the electrostatic force between the plates. As a result, the value of capacitors $C_A$ and $C_B$ is reduced to $C_{OFF}$ as indicated in Fig. 4.11. For the charge conservation to hold, $Q_{TOTAL}$ = 2 × $C_{OFF}$ ($V_{out}$) = 2 × $C_{ON}$ ($V_{in}$). Thus, $V_{in}$ is amplified and the theoretical gain ($A_V$) of the NEMS D-T Amplifier is given by the equation (4.1).

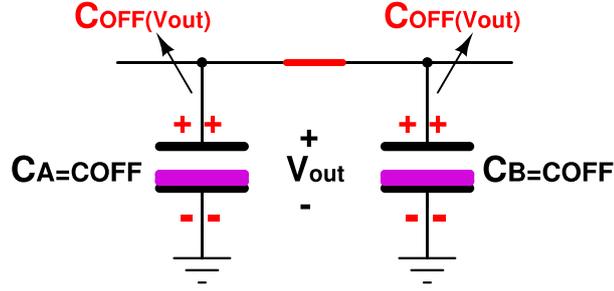

Fig. 4.11. NEMS D-T amplifier circuit in the final stage of amplification during the hold phase.

$$A_V = \frac{V_{out}}{V_{in}} = \frac{C_{ON}}{C_{OFF}} \quad (4.1)$$

## 4.4 Implementation of NEMS D-T Amplifier

Implementation of the NEMS D-T amplifier in the Cadence Spectre circuit simulator is given in Fig. 4.12. The NEMS capacitive and ohmic switches are implemented using the calibrated Verilog-A models as discussed in Section 4.2. To obtain a voltage equal to $V_{in}^+ \pm V_{BIAS}$, two large capacitors (normal parallel plate devices) are used to sample the voltage $V_{BIAS}$ (with appropriate polarity) during the hold phase (forming two floating DC sources) and add them in series with the input signal during the sample phase. The clock signals ($\phi_1$, $\phi_2$) are entirely non-overlapping with each other. The clock frequency ($f_{CLK}$), DC bias ($V_{BIAS}$) and $C_{large}$ are chosen to be 100 kHz, 4 V and 30 pF, respectively, throughout the work.

As an example, assume $V_{in}^+$ to be a DC signal equal to 0.2 V (Fig. 4.12). Fig. 4.13 shows the corresponding transient waveforms of the D-T amplifier. In the sample phase, "0.2V + 4V" is sampled across $C_A$ and "0.2V - 4V" is sampled across $C_B$. In the hold phase, the voltages $V_A^+$ and $V_B^+$ are equal to 1 V as shown in the Fig. 4.13, thus providing a gain equal to 5. The capacitance of the NEMS capacitive switch and the displacement of the beam during the sample and hold phase are clearly depicted in Fig. 4.13.

The bias voltage ($V_{BIAS}$) of the amplifier doesn't appear at the nodes $V_A^+$ and $V_B^+$ at the end of hold phase. Hence, only the amplified signal will appear as an input to the subsequent stages (For ex: CMOS buffer) and thus proper DC voltage isolation is achievable. For a sinusoidal input signal, the sampled and amplified differential output waveform are plotted in

Fig. 4.14. The differential outputs were generated by taking the difference between the outputs of the two separate amplifiers having inputs as $V_{in}^+$ and $V_{in}^-$, respectively.

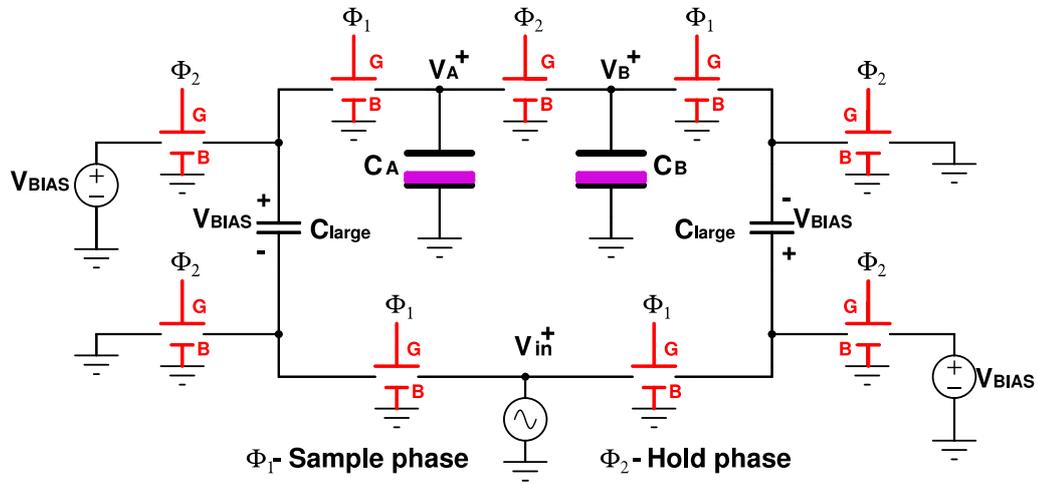

Fig. 4.12. Circuit implementation of the NEMS D-T amplifier.

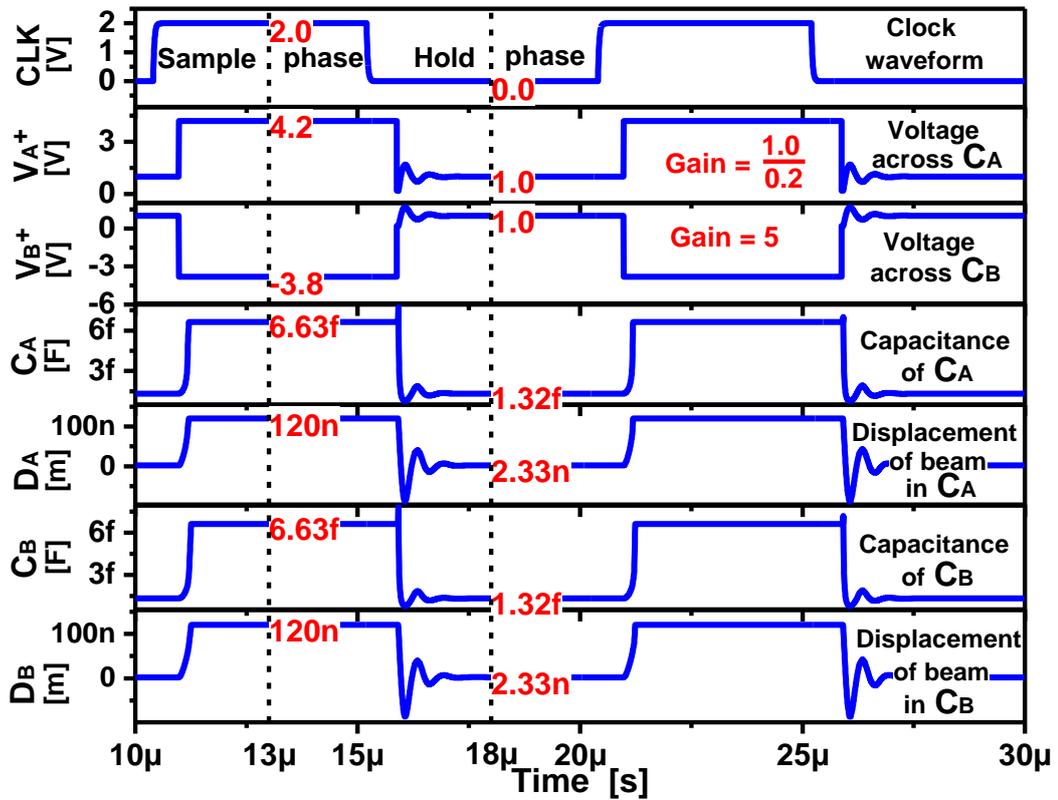

Fig. 4.13. Transient simulation waveforms of the NEMS D-T amplifier using Cadence Spectre for $V_{in}^+ = 0.2$ V, $V_{BIAS} = 4$ V and $f_{CLK} = 100$ kHz.

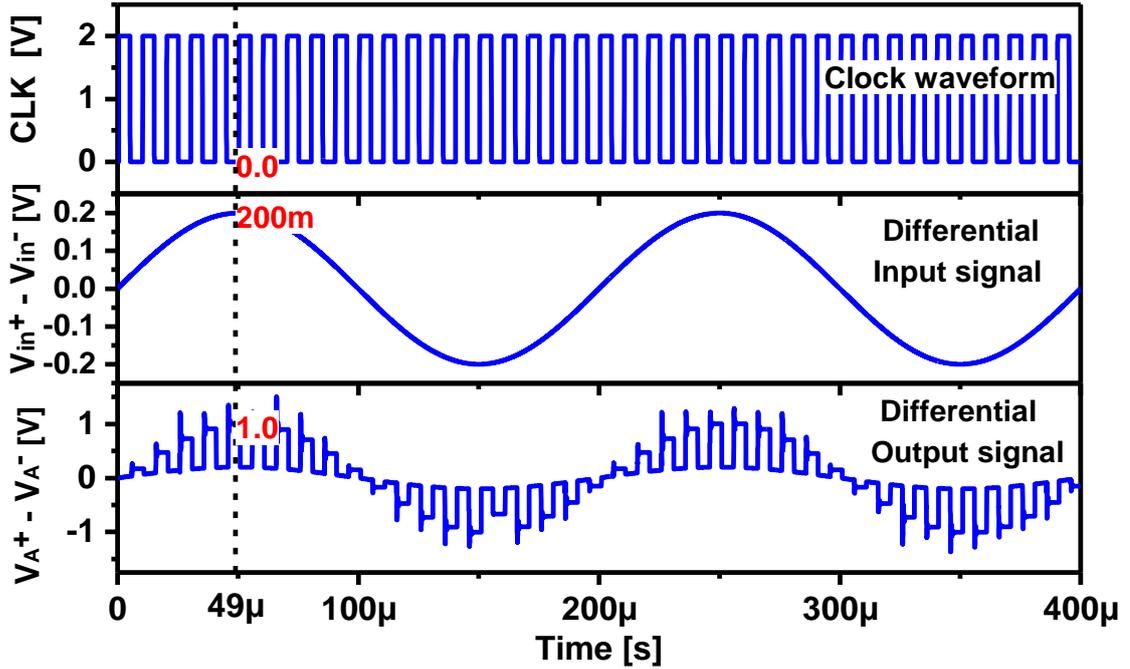

Fig. 4.14. Sampled and amplified sinusoidal input ($f_{IN}$ = 5 KHz) using differential NEMS D-T amplifier.

## 4.5 Non-Idealities in the NEMS D-T Amplifier

### 4.5.1 Non-linearity

Two of the major non-idealities present in the NEMS D-T amplifier are identified and discussed below. The first one being the variation of gain with input signal amplitude. This is because, the amplified signal ($V_{out}$) will cause the beam of the NEMS capacitive switch to displace and change the value of capacitance during the end of hold phase. From Fig. 4.5 (a), we observe that the capacitance of the NEMS capacitive switch varies with its terminal voltage in a weak fashion until the pull-in point is reached. Hence, the capacitance of the NEMS capacitive switch during the hold phase (at the end of the amplification), i.e. $C_{OFF}$, is a function of the output voltage. The plot of gain (*denoted in red color*) versus input signal amplitude shown in Fig. 4.17 indicates the reduction in gain by 4.7% for an increase in the input amplitude from 1 mV to 325 mV.

### 4.5.2 Parasitic capacitances of the NEMS ohmic switch

The other non-ideality is the presence of parasitic capacitances in the NEMS ohmic switch. The parasitic capacitance model of the NEMS ohmic switch is described in Fig. 4.15 (a). The parasitic capacitances load the capacitance $C_A$ and $C_B$, and share their charge. To reduce the clock feedthrough due to the gate-to-source and gate-to-drain capacitances of the ohmic switch [5], the clock signal can be applied to the body terminal instead of gate terminal as indicated in Fig. 4.15 (b).

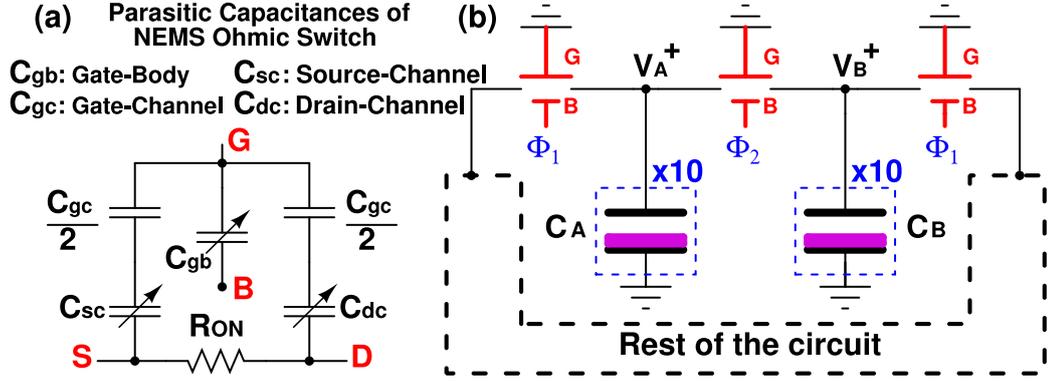

Fig. 4.15. (a) Parasitic capacitances in NEMS ohmic switch (b) Modified NEMS D-T amplifier.

Furthermore, using multiple NEMS capacitive switches in parallel as shown in Fig. 4.15 (b), will lower the gain reduction. The loaded gain for the modified NEMS D-T amplifier with 10 parallel switches (Fig. 4.15 (b)) is now given by equation (4.2).

$$A_{V,loaded} = \frac{(10 \times C_{ON}) + C_{GS(D),ON} + C_{GS(D),OFF}}{(10 \times C_{OFF}) + C_{GS(D),ON} + C_{GS(D),OFF}} \qquad (4.2)$$

In equation (4.2), $C_{GS(D),ON}$ and $C_{GS(D),OFF}$ represent the parasitic capacitance between gate to source/drain terminal in the NEMS ohmic switch during the on and off-states, respectively. They load the capacitances $C_A$ and $C_B$ during both the phases. The values of $C_{GS(D),ON}$ and $C_{GS(D),OFF}$ are 1 fF and 0.13 fF, respectively. Hence, the loaded gain for the modified NEMS D-T amplifier is equal to 4.77. Fig. 4.16 shows the waveforms of the modified NEMS D-T amplifier for a 0.2 V input DC signal, confirming the reduction in loaded gain. The NEMS D-T amplifier could also be co-designed by including the parasitic capacitances of the ohmic switch for obtaining gain.

The variation of gain (*denoted in blue color*) with the input amplitude for the modified NEMS D-T amplifier (Fig. 4.15 (b)) is shown in Fig. 4.17. The gain reduces by 5.4% at an input amplitude of 325mV. If the differential gain is considered, then the effect of non-linearity in the gain is considerably reduced due to the cancellation of even order non-linear terms. This is evident in the plot of gain (*denoted in black color*) in Fig. 4.17 for the differential implementation of the modified NEMS D-T amplifier, where the gain reduces only by 2% (at 325mV input). The non-linearity of the proposed NEMS D-T amplifier is influenced by the C-V characteristics of the NEMS capacitive switch, which in turn is dependent on the physical parameters of the switch considered. The effect of physical parameters of the NEMS capacitive switch on the amplifier non-linearity is not much explored as a part of this thesis. Moreover, the evaluation of the proposed NEMS D-T amplifier for the use in applications like ADC, sensor front-end signal chains etc., is beyond the scope of this Chapter.

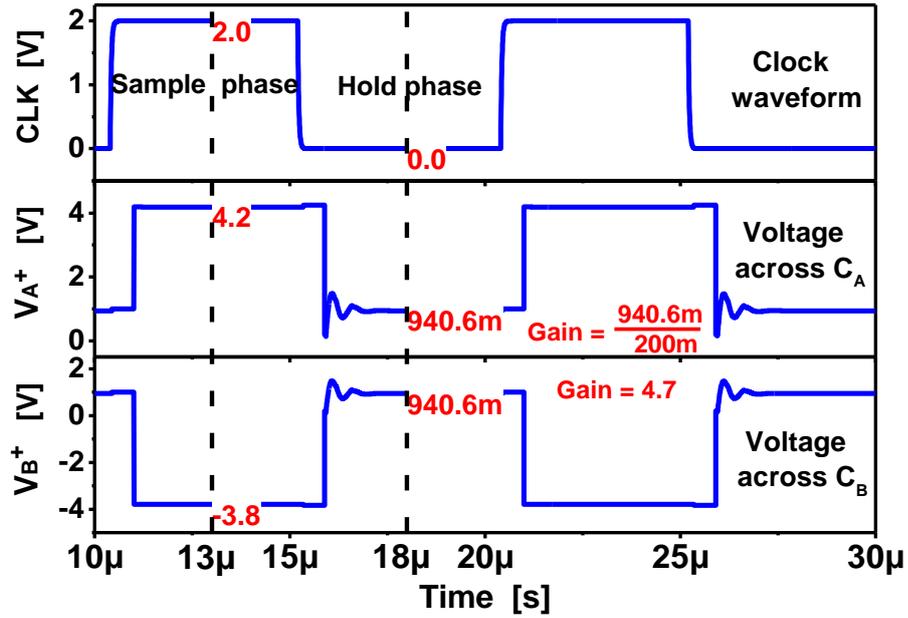

Fig. 4.16. Transient waveforms depicting the loaded gain in the modified NEMS D-T amplifier ($V_{in}^+ = 0.2$ V, $V_{BIAS} = 4$ V).

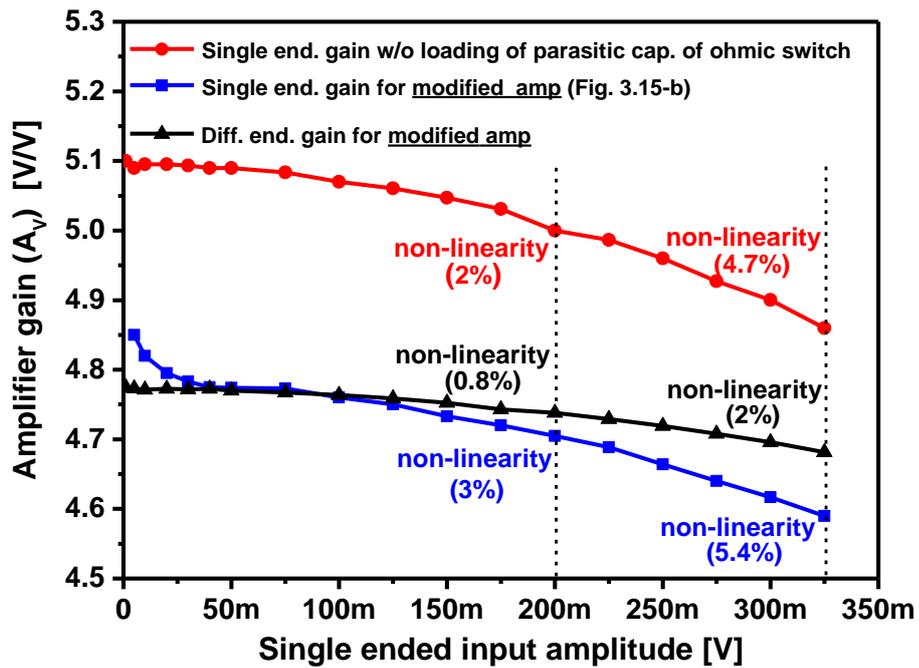

Fig. 4.17. Variation of gain with respect to the input signal amplitude.

## 4.6 Design Considerations for NEMS D-T Amplifier

### 4.6.1 Voltage gain

The gain of the NEMS D-T amplifier depends on the value of $C_{ON}$ and $C_{OFF}$ of the NEMS capacitive switch and is given by equation (4.3). The theoretical voltage gain ($A_V$) is defined by the equation (4.4).

$$C_{ON} = \frac{\varepsilon_o \varepsilon_d A}{t_d} \; ; \; C_{OFF} = \frac{\varepsilon_o A}{(g_o + \frac{t_d}{\varepsilon_d})} \tag{3.3}$$

$$A_V = \frac{C_{ON}}{C_{OFF}} = 1 + \frac{g_o \varepsilon_d}{t_d} \tag{4.4}$$

The switch parameters $g_o$, $t_d$ and $\varepsilon_d$ (Fig. 4.4) determine the gain. In reality, the presence of parasitic fringe capacitances and curvature of the beam causes the gain to reduce from the theoretical value. This effect is presented by including parameter $\gamma$ ($\gamma < 1$) in the description for the practical gain as given in equation (4.5). For the NEMS capacitive switch considered in this work, the theoretical gain possible is 10.36, but the practical gain is 5.1 (=6.63 fF/1.3 fF). The maximum achievable voltage gain is limited by the fabrication process of the NEMS switch.

$$A_{V,practical} = \left(1 + \frac{g_o \varepsilon_d}{t_d}\right) \times \gamma \tag{4.5}$$

### 4.6.2 Voltage range

The input and output voltage swings needed in the amplifier determine the condition on the pull-in ($V_{PI,C}$) and the pull-out voltage ($V_{PO,C}$) of the NEMS capacitive switch. The equations (4.6) and (4.7) describe the conditions, respectively.

$$V_{BIAS} \geq V_{PI,C} + V_{in,max} \tag{4.6}$$

$$V_{out,max} < V_{PO,C} \tag{4.7}$$

The condition given by equation (4.6) ensures that both the NEMS capacitive switches are pulled-in during the sampling phase. The condition given by equation (4.7) ensures that the NEMS capacitive switches are pulled-out during the hold phase. The expression for $V_{PI,C}$ and $V_{PO,C}$ for the NEMS capacitive switch is given by the equation (4.8) [17].

$$V_{PI,C} = \sqrt{\frac{8\, K_{eff}\left(g_o + \frac{t_d}{\varepsilon_d}\right)^3}{27\varepsilon_o A}};\ V_{PO,C} = \sqrt{\frac{2\, K_{eff}\, g_o t_d^2}{\varepsilon_d^2\, \varepsilon_o A}} \tag{4.8}$$

In equation (4.8), $K_{eff}$ is the effective spring constant of the beam and "A" is the overlap area which is given by $W \times L_B$ (Fig. 4.4 (a)). Based on the equations (4.4) and (4.8), the expressions for $V_{PI,C}$ and $V_{PO,C}$ are given by the equations (4.9) and (4.10), respectively.

$$V_{PI,C} = \alpha_1 \sqrt{\left(\frac{A_V t_d}{\varepsilon_d}\right)^3} \sqrt{\frac{K_{eff}}{A}} \tag{4.9}$$

$$V_{PO,C} = \alpha_2 \sqrt{(A_V - 1)\left(\frac{t_d}{\varepsilon_d}\right)^3} \sqrt{\frac{K_{eff}}{A}} \tag{4.10}$$

In equation (4.9) and (4.10), $\alpha_1$ and $\alpha_2$ is given by $\sqrt{\frac{8}{27\varepsilon_o}}$ and $\sqrt{\frac{2}{\varepsilon_o}}$, respectively. For the NEMS capacitive switch considered in this work, the term $\frac{K_{eff}}{A}$ is given by equation (4.11), where "E" is the Young's Modulus of the beam material. The term $\frac{K_{eff}}{A}$ is dependent on the structure of the switch and can be controlled independently without affecting the gain. Based on the maximum output voltage swing needed, $\frac{K_{eff}}{A}$ can used to set the $V_{PO,C}$. The terms $V_{PI,C}$ and $V_{PO,C}$ are inter-related with each other. This is quantified by the equation (4.12), which provides the ratio of $V_{PO,C}/V_{PI,C}$ as a function of the amplifier gain. Hence, when the term $\frac{K_{eff}}{A}$ is used to set $V_{PO,C}$, $V_{PI,C}$ is also fixed.

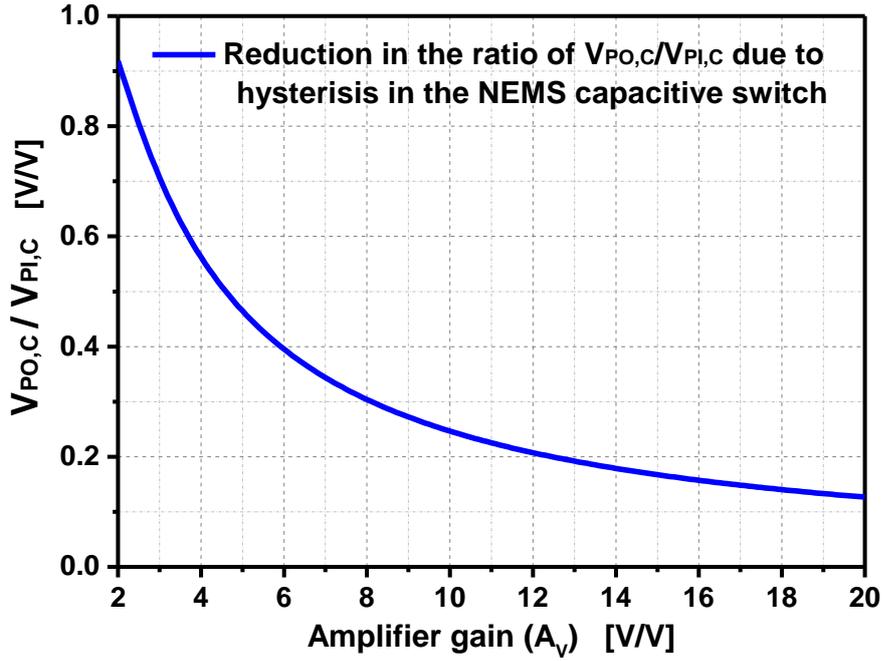

Fig. 4.18. Ratio of $V_{PO,C}/V_{PI,C}$ versus the amplifier gain.

$$\frac{K_{eff}}{A} \approx \frac{32\,E}{L_B}\left(\frac{t^3}{L^3}\right) \quad (4.11)$$

$$\frac{V_{PO,C}}{V_{PI,C}} = \left(\frac{\alpha_2}{\alpha_1}\right)\sqrt{\frac{A_V-1}{A_V^3}} \quad (4.12)$$

As shown in Fig. 4.18, for the higher voltage gain, the ratio of $V_{PO,C}/V_{PI,C}$ falls due to the increment in the hysteresis window of the NEMS capacitive switch. Hence, it is desirable to have a capacitive switch with low hysteresis so as to increase the usable range of the amplifier. Finally the bias voltage ($V_{BIAS}$) required for the amplifier is chosen to satisfy the condition required for the pull-in, based on maximum $V_{in}$ as given by the equation (4.6). The NEMS switches in this work are chosen to keep the structure simple enough. The design procedure discussed in this section can be extended to different structures of the NEMS switch as well.

### 4.6.3 Noise analysis

Consider the simplified representation of the noise model in the NEMS D-T amplifier shown in Fig. 4.19. During the sample phase, the integrated noise on the capacitor $C_{ON}$ is denoted as $V_{n,s}$ as shown in Fig. 4.19 (a). The value of the integrated noise power $V_{n,s}$ during the sample phase is given by equation (4.13).

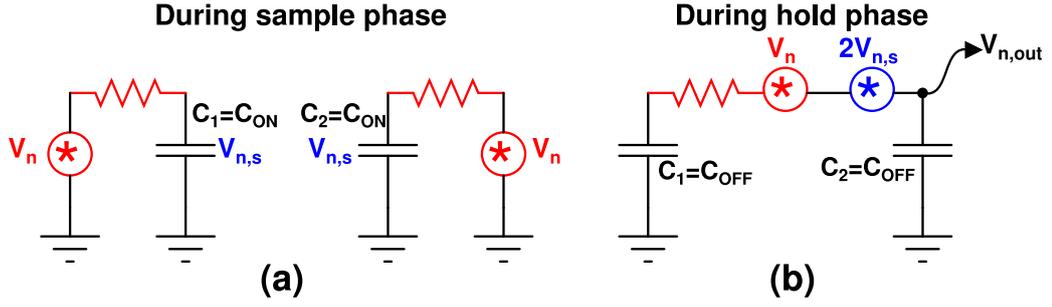

Fig. 4.19. Simplified noise model of the NEMS D-T amplifier during (a) sample phase, and (b) hold phase. "Vn" denotes the thermal noise due to the on-resistance of the NEMS ohmic switch.

$$V_{n,s} = \frac{kT}{C_{ON}} \qquad (4.13)$$

In the hold phase, the simplified noise model in the NEMS D-T amplifier is shown in Fig. 4.19 (b). The term "$2V_{n,s}$" appears due to the un-correlated nature of the noise in the two capacitors during the sample phase. The total integrated output noise power during the hold phase is denoted by the term "$V_{n,out}$" in Fig. 4.19 (b). The expression for the term $V_{n,out}$ is approximated as given by equation (4.14).

$$V_{n,out} \approx \left(\frac{2kT}{C_{OFF}} + \frac{2kT}{C_{ON}}\right) = \frac{2kT}{C_{ON}}(1 + A_V) \qquad (4.14)$$

where the term "$A_V$" is the voltage gain of the NEMS D-T amplifier, $C_{ON}$ is the total on-state capacitance of the NEMS capacitive switch. For a differential implementation of the NEMS D-T amplifier, the total integrated output noise at the end of the hold phase is given by equation (4.15).

$$V_{n,out}(\text{diff}) = \frac{4kT}{C_{ON}}(1 + A_V) \qquad (4.15)$$

Based on (4.15), the minimum value of the total $C_{ON}$ required in the NEMS D-T amplifier can be calculated to meet the given noise specification. Based on the total $C_{ON}$ required, equivalent number of parallel NEMS capacitive switches can be used appropriately.

### 4.6.4 Scaling and process integration

If all the geometrical parameters of the NEMS capacitive switch in the proposed D-T amplifier are scaled by a factor "ŋ", then the voltage gain remains unchanged and the switch parameters $V_{PI,C}$, $V_{PO,C}$, $C_{ON}$, $C_{OFF}$ and $T_{Mech}$ are linearly scaled by "ŋ" as given in equation (4.16).

$$f'_n = \eta \times f_n \tag{4.16}$$

In equation (4.16), terms $f_n$ and $f'_n$ represent the parameters before and after scaling. The scaling of the NEMS ohmic switch follows a similar trend. The downscaling of the vertical dimensions (beam & airgap) of the switch pose fabrication challenges. However, the tremendous progress in the NEMS fabrication has resulted in low voltages, low footprint [18], [19] along with considerable improvement in the reliability of scaled switches [20], [21].

NEMS switches of the proposed D-T amplifier can be integrated with the CMOS die using one of the three methods as follows. Currently NEMS devices are integrated with CMOS circuits at PCB [25] or die level [26]. At PCB level [25], the integration of NEMS ICs and CMOS ICs is relatively inexpensive. However, it consumes more board area. At the die level [26], the integration of NEMS & CMOS dies within the same package increases the cost compared to the PCB integration. However, the die level integration reduces the PCB footprint considerably.

Tremendous efforts have been invested in integrating NEMS devices along with CMOS devices in the back-end of line (BEOL) using post-processing steps [27] or by using 3D-ICs [22], [23]. This will reduce the area footprint significantly, however at an increased overall fabrication cost.

Considering the three approaches mentioned here, integration of NEMS and CMOS devices at the die level appears to provide an optimum solution in reducing the additional cost and the area overhead.

### 4.6.4 Power dissipation

The power dissipation in the NEMS DT-amplifier is entirely dynamic in nature. For a differential implementation of the modified NEMS D-T amplifier, the power dissipated in the amplifier ($P_{Amp}$) and the power dissipated in driving the switches ($P_{sw}$) are given by equation (4.17) and (4.18), respectively.

$$P_{amp} = 2 \times (20C_{ON} + 4C_{GS(D),ON} + 4C_{GS(D),OFF}) V_{BIAS}^2 f_{CLK} \tag{4.17}$$

$$P_{sw} = 2 \times (6C_{PAR,ON} + 3C_{GB,ON}) V_{GB}^2 f_{CLK} \tag{4.18}$$

In equation (4.18), $C_{PAR,ON}$ $(=C_{GS,ON} + C_{GD,ON} + C_{GB,ON})$ is the total parasitic capacitance looking into the gate terminal of the NEMS ohmic switch in the ON state. Some amount of power ($P_{clk}$) is dissipated in the non-overlap clock generator circuit. Hence the total power dissipation is given by $P_{Amp} + P_{sw} + P_{clk}$. For the differential implementation of the modified NEMS D-T amplifier, the total power consumption is 0.6 µW.

### 4.6.5 Performance comparison

Table 4.3 compares the performance of NEMS D-T amplifier with the op-amp and MOSCAP based D-T amplifier. As we see from the table, the proposed NEMS D-T amplifier handles a maximum differential input signal of 0.65V and consumes only 0.6 µW of power for a sampling frequency of 100 kHz. Hence, the proposed NEMS D-T amplifier is well suited for the energy constrained IoT applications, where reducing the static bias currents in the analog front-end system is of utmost importance.

Table 4.3: Performance comparison of the NEMS D-T amplifier

|  | **Traditional D-T Amplifier (designed & simulated based on [5])** | | **MOSCAP D-T Amplifier (designed & simulated based on [6])** | **Dynamic amplifier (designed & simulated based on [24])** | **This work** |
|---|---|---|---|---|---|
| **Description/Topology** | 2-stage miller op-amp | Folded cascode op-amp | NMOS MOSCAP based D-T amplifier | NMOS Dynamic source follower based D-T amplifier | Modified NEMS D-T amp. |
| **Process technology** | 180 nm CMOS | 180 nm CMOS | 180 nm CMOS | 180 nm CMOS | NEMS |
| **Gain** | 5 | 5 | ~6.8 | ~4.7 | ~4.7 |
| **Input & output type** | Differential input & differential output | | | | |
| **DC voltage required [V]** | $V_{DD}$ = 1.8 | $V_{DD}$ = 3.3 | $V_{BIAS}$ = 0.6, $V_{PULL}$ = 3.3 | $V_{DD}$ = 1.8 | $V_{BIAS}$ = 4 |
| **Amp. of clock signal [V]** | 1.8 | 3.3 | 3.3 | 1.8 | 2 |
| **Clock frequency [kHz]** | 100 | 100 | 100 | 100 | 100 |
| **Amplification of bias voltage** | No | No | Yes | Yes | No |
| **Non-linearity* in diff. gain at 0.1V diff. input [%]** | 0.012 | 0.004 | 0.097 | 0.36 | 0.105 |
| **Max. diff. input in [V] for 2% non-linearity* in diff. gain** | 0.33 | 0.53 | 0.46 | 0.17 | 0.65 |
| **THD [dB] at $f_{IN}$~10kHz & 0.5Vp-p diff. out** | -100.1 | -113.5 | -102.3 | -65.9 | -73.9 |
| **$P_{amp}$ [µW]** | 48.6 | 62.2 | 0.86 | 8.41 | 0.44 |
| **$P_{clk}+P_{sw}$ [µW]** | 0.22 | 0.59 | 0.43 | 0.39 | 0.16 |
| **Total power [µW]** | 48.82 | 62.79 | 1.29 | 8.8 | 0.60 |
| **% of power reduction w.r.t D-T amp using FC op-amp** | 22.24 | - | 97.94 | 85.98 | 99.04 |

* defined as the change in gain for DC input.

## 4.7 Conclusion

In this chapter a technique of realizing the D-T signal amplification using NEMS devices is presented, and subsequently verified the same using the calibrated Verilog-A models. The presented NEMS D-T amplifier opens up a new avenue for utilizing NEMS switches in analog applications, which further motivates the development of low voltage and low footprint devices. In IoT sensor nodes, output signal of the sensor is typically of low amplitude and bandwidth. This paper shows the proposed NEMS D-T amplifier can receive weak sensor output signals as low as 1mV and acts as a front end sample and hold amplifier preceding the ADC. The proposed NEMS D-T amplifier consumes only dynamic power, thereby reducing the quiescent current consumption in the analog front-end block. As a result, the overall power consumption is reduced and the battery life of IoT sensor nodes is enhanced.

## 4.8 References

[1]   S. Sankar, M. Goel, M. S. Baghini and V. R. Rao, "A Novel Method of Discrete-Time Signal Amplification Using NEMS Devices," in *IEEE Transactions on Electron Devices*, vol. 65, no. 11, pp. 5111-5117, Nov. 2018.

[2]   X. Zou, X. Xu, L. Yao and Y. Lian, "A 1-V 450-nW Fully Integrated Programmable Biomedical Sensor Interface Chip," in *IEEE Journal of Solid-State Circuits*, vol. 44, no. 4, pp. 1067-1077, April 2009.

[3]   Y. Huang *et al.*, "A Self-Powered CMOS Reconfigurable Multi-Sensor SoC for Biomedical Applications," in *IEEE Journal of Solid-State Circuits*, vol. 49, no. 4, pp. 851-866, April 2014.

[4]   P. Kimtee, D. M. Das and M. S. Baghini, "A mismatch insensitive reconfigurable discrete time biosignal conditioning circuit in 180 nm MM CMOS technology," in *20th IEEE International Symposium on VLSI Design and Test (VDAT)*, 2016, pp. 1-2.

[5]   B. Razavi, "Design of Analog CMOS Integrated circuits", McGraw-Hill, 2002.

[6]   S. Ranganathan and Y. Tsividis, "Discrete-time parametric amplification based on a three-terminal MOS varactor: Analysis and experimental results," *IEEE Journal of Solid-State Circuits*, vol. 38, no. 12, pp. 2087–2093, Dec. 2003.

[7]   M. Spencer, *et al.*, "Demonstration of integrated micro-electro-mechanical relay circuits for VLSI applications," *IEEE Journal of Solid-State Circuits.*, vol. 46, no. 1, pp. 308–320, Jan. 2011.

[8]   D. Lee, W. S. Lee, C. Chen, F. Fallah, J. Provine, S. Chong, J. Watkins, R. T. Howe, H. S. P. Wong and S. Mitra, "Combinational logic design using six-terminal NEM relays," *IEEE Transactions on Computer-Aided Design Integrated Circuits Systems*, vol. 32, no. 5, pp. 653–666, May 2013.

[9]   R. Venkatasubramanian, S. K. Manohar, and P. T. Balsara, "NEM relay based sequential logic circuits for low-power design," *IEEE Transactions on Nanotechnology*, vol. 12, no. 3, pp. 386–398, May 2013.

[10]  S. Meninger, J. O. Mur-Miranda, R. Amirtharajah, A. P. Chandrakasan, and J. H. Lang, "Vibration-to-electric energy conversion," *IEEE Transactions on Very Large Scale Integration Systems*, vol. 9, no. 1, pp. 64–76, Feb. 2001.

[11]  J. P. Raskin, A. R. Brown, B. T. Yakub, and G. M. Rebeiz, "A novel parametric-effect MEMS amplifier," *IEEE Journal of Microelectromechanical Systems*, vol. 9, no. 6, pp. 528–537, Dec. 2000.

[12]  K. Akarvardar and H.-S. P. Wong, "Analog nanoelectromechanical relay with tunable transconductance," *IEEE Electron Device Letters*, vol. 30, no. 11, pp. 1143–1145, Nov. 2009.

[13]  G. M. Rebeiz and J. B. Muldavin, "RF MEMS switches and switch circuits," in *IEEE Microwave Magazine*, vol. 2, no. 4, pp. 59-71, Dec. 2001.

[14]  MEMS FEM CAD tool [Online]. Available: www.coventor.com/mems-solutions/products/coventorware/

[15]  K. Van Caekenberghe, "Modeling RF MEMS Devices," *IEEE Microwave Magazine*, vol. 13, no. 1, pp. 83-110, 2012.

[16]  F. Chen, H. Kam, D. Markovic, T. K. Liu, V. Stojanovic and E. Alon, "Integrated circuit design with NEM relays," *IEEE/ACM International Conference on Computer-Aided Design*, 2008, pp. 750-757.

[17]  G. M. Rebeiz, "RF MEMS: Theory, Design, and Technology". New York: Wiley, 2003.

[18] C. Qian, A. Peschot, B. Osoba, Z. A. Ye, and T.-J. K. Liu, "Sub-100 mV computing with electro-mechanical relays," *IEEE Transactions on Electron Devices.*, vol. 64, no. 3, pp. 1323–1329, Mar. 2017.

[19] J. O. Lee, Y. H. Song, M. W. Kim, M. H. Kang, J. S. Oh, H. H. Yang and J. B. Yoon, "A sub-1-volt nanoelectromechanical switching device," *Nature Nanotechnology*, vol. 8, pp. 36–40, Jan. 2013.

[20] M. Ramezani, S. Severi, A. Moussa, H. Osman, H. A. C. Tilmans and K. D. Meyer, "Contact reliability improvement of a poly-SiGe based nano-relay with titanium nitride coating," in *Proceedings of 18th IEEE Transducers*, 2015, pp. 576-579.

[21] Y. Chen, R. Nathanael, J. S. Jeon, J. Yaung, L. Hutin, and T.-J. King Liu, "Characterization of contact resistance stability in MEM relays with tungsten electrodes," *IEEE Journal of Microelectromechanical Systems*, vol. 21, no. 3, pp. 511–513, Jun. 2012.

[22] W. Y. Choi and Y. J. Kim, "Three-Dimensional Integration of Complementary Metal-Oxide-Semiconductor-Nanoelectromechanical Hybrid Reconfigurable Circuits," in *IEEE Electron Device Letters*, vol. 36, no. 9, pp. 887-889, Sept. 2015.

[23] A. C. Fischer, F. Forsberg, M. Lapisa, S. J. Bleiker, G. Stemme, N. Roxhed and F. Niklaus, Integrating MEMS and ICs, *Nature Microsystems and Nanoengineering*, 1, 15005, 2015.

[24] J. Hu, N. Dolev and B. Murmann, "A 9.4-bit, 50-MS/s, 1.44-mW Pipelined ADC Using Dynamic Source Follower Residue Amplification," *IEEE Journal of Solid-State Circuits*, vol. 44, no. 4, pp. 1057-1066, April 2009.

[25] S. Sankar, M. Goel, P. -H. Chen, V. R. Rao, and M. S. Baghini, "Switched-Capacitor-Assisted Power Gating for Ultra-Low Standby Power in CMOS Digital ICs," *IEEE Trans. Circuits Syst. I, Reg. Papers*, vol. 67, no. 12, pp. 4281-4294, Dec. 2020.

[26] E. Carty, P. Fitzgerald, and P. McDaid, "The Fundamentals of Analog Devices' Revolutionary MEMS Switch Technology", *Analog Devices Technical Article*.

[27] M. Ramezani, *et al.*, "Submicron three-terminal SiGe-based electromechanical ohmic relay," in *IEEE 27th International Conference on Micro Electro Mechanical Systems (MEMS)*, San Francisco, CA, 2014, pp. 1095-1098.

# Chapter 5

# Efficient Inductive Rectifier for Piezo Energy Harvesting

Inductor-based rectifiers have been developed to overcome the limitations of the traditional diode-based rectifiers for piezo energy harvesting (PEH). This chapter proposes a new efficient method of extracting power from the piezoelectric transducer using the inductive rectifier. In the proposed method, the internal capacitance of the transducer is initially pre-charged, and the generated charges in response to the mechanical vibration are accumulated on the capacitor. When the accumulated voltage reaches a maximum allowed value ($V_{MAX}$), the total energy stored in the capacitor is transferred to the output, and the process is repeated. The proposed method ensures that the voltage swing across the transducer is always maximum for extracting the highest possible output power for given CMOS technology. The operational steps are self-timed, and the negative voltage swing across the transducer is avoided. The proposed inductive rectifier is realized using the buck-boost power stage of the DC energy harvesting system for enabling efficient multi-source harvesting. Test chip is fabricated in the 180 nm CMOS technology, having a $V_{MAX}$ of 3.3 V. Total silicon area occupied by the proposed inductive rectifier circuit in the 180nm CMOS technology is 1 mm$^2$. The proposed inductive rectifier extracts power in a single stage, even at a lower rectified output voltage $\leq$ 1 V. For a piezo open-circuit voltage of 1 V, the proposed rectifier extracts 3.68× more power than the maximum output power of a full-bridge rectifier with ideal diodes. This is despite the voltage loss occurring during accumulation in the power stage implementation. Parts of this chapter have been published in [23].

## 5.1 Introduction

Energy constrained applications such as wireless sensor nodes, wearable electronic devices, medical implants, and Internet-of-Things (IoT) devices need to be energy-autonomous for avoiding replacement or frequent charging of batteries. Reducing the inherent energy consumption is one approach in prolonging the battery life [1]. However, it may not be sufficient. Energy harvesting from ambient sources (solar, thermal, vibrations, and RF) in conjunction with the battery can enable energy-autonomous systems. However, the amount of harvested energy depends on the ambient conditions and is not always constant. Hence, an efficient multi-source energy harvesting system along with the provision to store the excess energy and reuse it later is required [2].

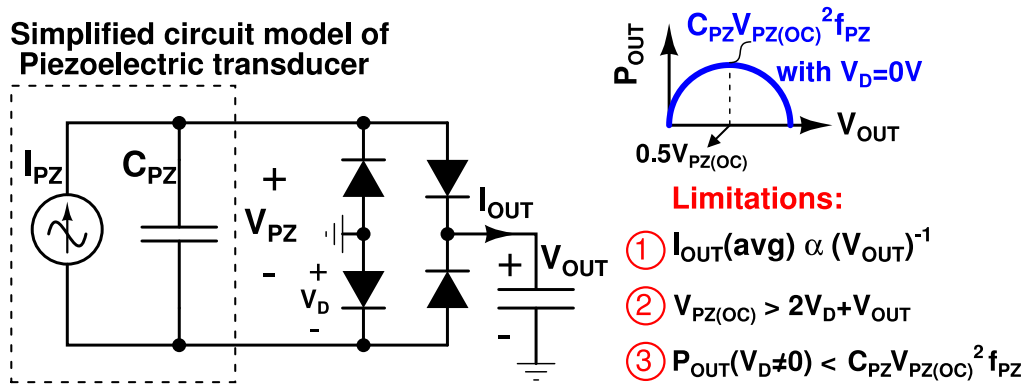

Fig. 5.1. Limitations of the conventional diode-based rectifier [7].

Several multi-source energy harvesting systems [3]-[6] have been proposed. In [3]-[6], vibrational energy was inefficiently extracted from the piezoelectric transducer using diode-based rectifiers, as shown in Fig. 5.1. Rectifiers using inductors [7]-[20] overcome the limitations of the diode-based rectifier by avoiding the power loss due to the discharge of the internal piezo capacitor ($C_{PZ}$) during the zero crossings of the AC piezo current ($I_{PZ}$).

Several methods of extracting power from the piezoelectric transducer using inductor-based rectifiers have been proposed. They are, namely the bias-flip rectifier or the synchronous switch harvesting on inductor (SSHI) [7]-[11], sense-and-set rectifier [12], synchronous electrical charge extraction (SECE) rectifier [13]-[15], and the inductive rectifier [16]-[20].

For improved energy efficiency, circuits are typically operated at supply voltages lower than 1 V. Hence, to harvest weaker vibrations and power the circuits, the rectifier must have good power extraction capability even at lower rectified output voltages ($V_{OUT} \leq 1V$).

Bias-flip rectifiers [7]-[11] and the sense-and-set rectifier [12] have good power extraction capability at higher rectified output voltages. However, their power extraction capability reduces significantly at lower rectified output voltages. SECE rectifiers [13]-[15] are more suitable for harvesting power from mechanical shocks rather than periodic vibrations. Inductive rectifiers [16]-[20] have good power extraction capability independent of the rectified output voltage.

When the voltage swing across the piezoelectric transducer ($V_{PZ}$) is higher, more electrical power can be extracted from it [17]. The works from [7]-[19] do not always ensure that $V_{PZ}$ is maximum for extracting the highest possible output power for a given CMOS technology. Moreover, they require a dedicated inductive power stage for PEH. This makes multi-source energy harvesting inefficient, since two power stages are required for harvesting the AC and DC sources. Inductive rectifier should also adapt to the variations in the transducer capacitance or inductor values over time. Since, changes in the resonance time-period reduces the output power significantly [10].

The work from [20] allows for higher voltage swing across the piezoelectric transducer and extracts higher output power. However, the open-circuit voltage ($V_{PZ,OC}$) of the piezoelectric transducer must be known prior hand. This is not always possible, since $V_{PZ,OC}$ depends on the strength of vibration. In case of aperiodic vibrations, $V_{PZ,OC}$ varies with time. In other words, the work in [20] extracts higher power only at a known input condition, or else the extracted power drastically reduces.

To overcome the drawbacks of the prior works, the new contributions in the proposed inductive rectifier are as follows,

1) An efficient method of extracting power from the piezoelectric transducer without any prior knowledge of its open-circuit voltage ($V_{PZ,OC}$) is proposed. The key idea is to pre-charge and accumulate $V_{PZ}$ till $V_{MAX}$, and transfer the total energy to output. It results in the highest output power extraction for given CMOS technology.
2) The operational steps are self-timed using zero-current detection (ZCD) and zero-voltage detection (ZVD) techniques. This avoids the need for manual tuning of the timing generator for different piezoelectric transducer and inductor values or variations in their values.
3) The negative voltage swing in the bias-flip step is avoided, which simplifies the gate driver circuit.
4) The proposed inductive rectifier is realized using the buck-boost power stage of the existing solar and thermal energy harvesting system [2]. This enables the harvesting of AC and DC sources together using the same power stage.

## 5.2 Prior Approaches in Piezo Energy Harvesting

### 5.2.1 Drawbacks of the diode based rectifier

Consider the full-bridge rectifier (FBR) circuit shown in Fig. 5.1. The output current ($I_{OUT}$) is inversely proportional to the output voltage ($V_{OUT}$) [7]. This causes the maximum extracted average output power ($P_{OUT(FBR)}$) to be limited to the value given by (5.1) for an ideal diode case (i.e., $V_D = 0$ V) [7].

$$P_{OUT-FBR} = C_{PZ} V_{PZ(OC)}^2 f_{PZ} \tag{5.1}$$

where $C_{PZ}$, $V_{PZ(OC)}$, and $f_{PZ}$ are the capacitance, open-circuit voltage and resonant frequency of the piezo transducer, respectively. With the real diode, the extracted output power will be lesser than the value given by (5.1). Moreover, when real diodes are used, the value of $V_{PZ(OC)}$ has to be greater than $2V_D + V_{OUT}$ for the diodes to be turned on. This sets a limitation on the minimum vibrational energy that could be harvested.

### 5.2.2 Principle of operation of the inductive rectifier

*A) Harvesting every half cycle of vibration*

Consider Fig. 5.2, which depicts the key steps involved in the inductive rectifier harvesting energy every half cycle [16]. The generated AC current ($I_{PZ}$) in response to the mechanical vibration is integrated into the internal capacitance $C_{PZ}$ of the piezo transducer. During the zero crossing of $I_{PZ}$, an inductor is connected across the piezo transducer, as shown in Fig. 5.2. The value of L is chosen such that the $LC_{PZ}$-tank resonance frequency $f_{LC}$ ($=1/T_{LC}$) is larger (more than 100×) compared to the vibration frequency $f_{PZ}$ ($=1/T_{PZ}$), as given by (5.2).

$$f_{LC} = \left(\frac{1}{2\pi\sqrt{LC_{PZ}}}\right) \gg f_{PZ} \tag{5.2}$$

As shown in Fig. 5.2, the energy stored in $C_{PZ}$ charges the inductor. When the inductor current reaches the peak value, the voltage across $C_{PZ}$ goes to zero. The inductor current is then dumped to the output, as depicted in Fig. 5.2. The total time it takes for the transfer of energy from the capacitor to the output (i.e., $0.25T_{LC} + t_{HAR}$) is negligible compared to the time period of vibration ($T_{PZ}$). This process continues again for the negative half cycle of $I_{PZ}$. Hence, the energy from the piezo transducer is transferred to the output twice in every period of vibration.

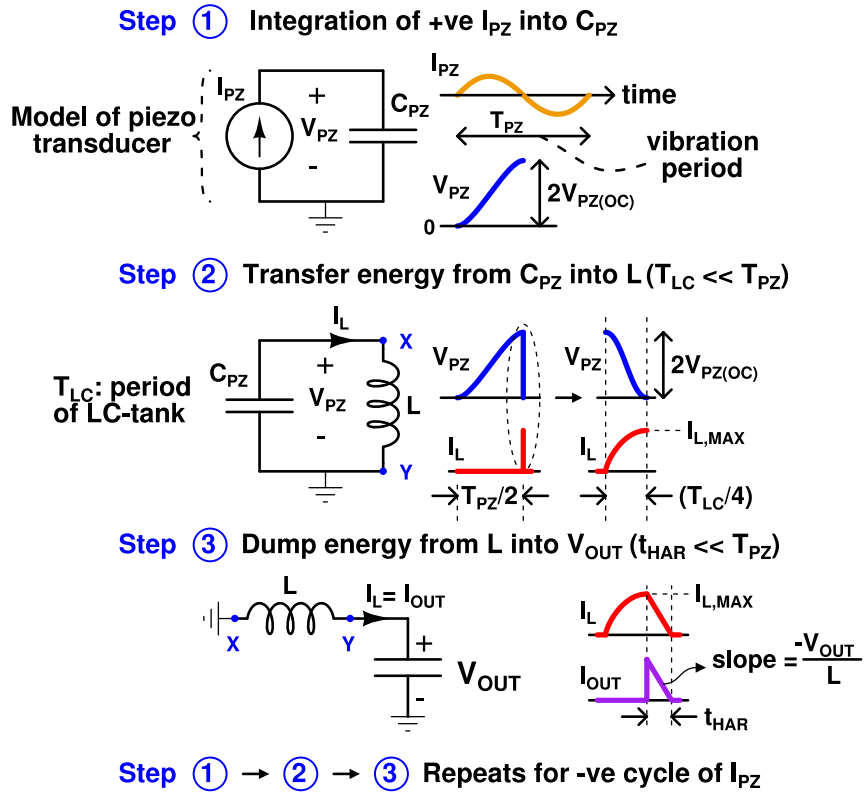

Fig. 5.2. Various steps in the inductive rectifier harvesting energy every half cycle of the mechanical vibration.

*B) Harvesting every one cycle of vibration*

In Fig. 5.3 (a), inductive rectification is performed to harvest energy every one cycle of vibration. After the integration of positive $I_{PZ}$, the voltage across $C_{PZ}$ is flipped during the bias-flip step [7], as depicted in Fig. 5.3 (b). The inductor current rises to the peak value and then starts charging $C_{PZ}$ in the opposite direction. For the ideal case (loss is neglected), the voltage $V_{PZ}$ is flipped to an equivalent negative value ($-V_{PZ}$), as shown in Fig. 5.3 (b). In the negative half cycle, $I_{PZ}$ reverses its direction, and the integration process continues. The total energy stored in $C_{PZ}$ is dumped to the output at the end of one cycle, as indicated in Fig. 5.3 (a).

## 5.2.3 Extracting more power by maximizing $V_{PZ}$

The comparison of $V_{PZ}$ and $I_{OUT}$ waveforms for two cases of harvesting discussed in Section 5.2.2 are shown in Fig. 5.4. The extracted average output power for each case is provided in Table 5.1. Table 5.1 shows that increasing the maximum voltage swing across the piezo transducer results in the higher extracted average output power. The physical origin for the increased output power is as follows. When the voltage swing across the piezo transducer increases, more amount of mechanical energy is converted into equivalent electrical energy [17], [20]. The voltage swing across the piezo is typically limited by the maximum voltage ($V_{MAX}$) that could be handled by the CMOS transistors present in the switch network.

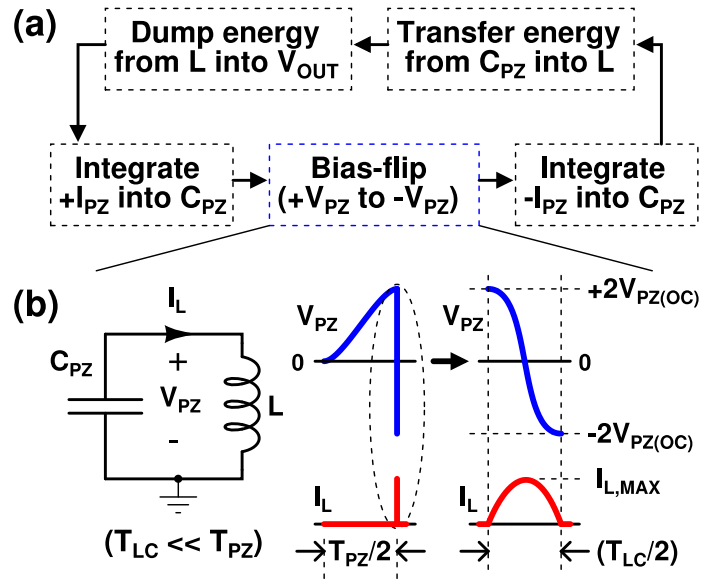

Fig. 5.3. (a) Various steps in the inductive rectifier harvesting energy every one cycle of the mechanical vibration, and (b) expanded view of the bias-flip step.

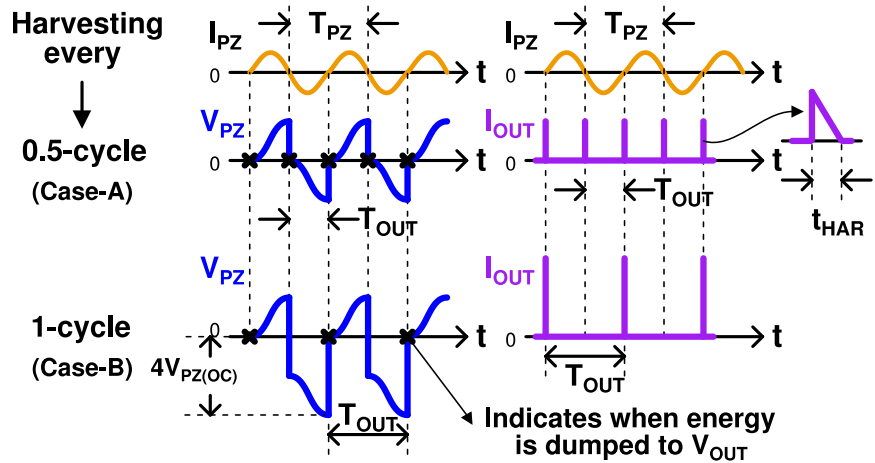

Fig. 5.4. Comparison of $V_{PZ}$ and $I_{OUT}$ waveforms for the two cases of harvesting.

Table 5.1: Comparison of $P_{OUT}$ for different voltage swings of $V_{PZ}$

| Case | Max. $V_{PZ}$ | $E_{OUT}$ | $(1/T_{OUT})$ | $P_{OUT}=(E_{OUT}/T_{OUT})$ |
|---|---|---|---|---|
| A | $2V_{PZ,OC}$ | $\frac{C_{PZ}}{2}(2V_{PZ,OC})^2$ | $2f_{PZ}$ | $4C_{PZ}V_{PZ,OC}^2 f_{PZ}$ |
| B | $4V_{PZ,OC}$ | $\frac{C_{PZ}}{2}(4V_{PZ,OC})^2$ | $f_{PZ}$ | $8C_{PZ}V_{PZ,OC}^2 f_{PZ}$ |

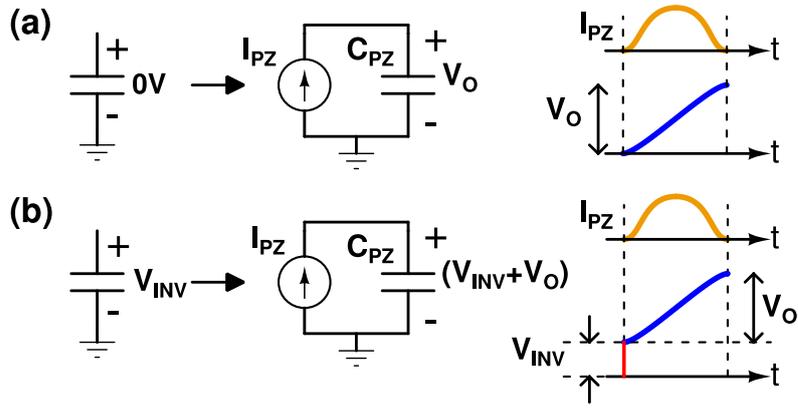

Fig. 5.5. Extracting more energy using investments [17]. (a) $I_{PZ}$ flowing into an uncharged $C_{PZ}$, (b) $I_{PZ}$ charging $C_{PZ}$ which has initial voltage $V_{INV}$.

### 5.2.4 Extracting more power by investing energy into C_PZ

The work in [17] proposed to use energy investment in $C_{PZ}$ to increase the voltage swing across the piezo transducer. Consider Fig. 5.5 (a), where $I_{PZ}$ flows into an uncharged capacitor $C_{PZ}$ to increase its voltage to $V_O$. In Fig. 5.5 (b), $I_{PZ}$ flows into the capacitor $C_{PZ}$ which is already charged to a voltage $V_{INV}$. The final voltage across $C_{PZ}$ is equal to $V_{INV} + V_O$. The net energy extracted from $I_{PZ}$ ($E_{IPZ}$) in Fig. 5.5 (b) is given by (5.3).

$$E_{IPZ} = \frac{C_{PZ}}{2} \times \{(V_{INV} + V_O)^2 - (V_{INV})^2\}$$

$$= \frac{1}{2} C_{PZ} V_O^2 + C_{PZ} V_{INV} V_O \quad (5.3)$$

As observed from (5.3), $I_{PZ}$ delivers more energy ($C_{PZ} V_{INV} V_O$) while increasing the voltage of a charged capacitor compared to the case with $V_{INV} = 0V$. The physical origin for this extra energy is the same as explained in Section 5.2.3.

### 5.2.5 Prior implementations of the inductive rectifier

In [17], the energy was invested in increasing the potential across $C_{PZ}$ by a value $V_{INV}$ during the bias-flip step as shown in Fig. 5.6 (a). In this case, energy was invested from the output battery. Since the capacitor $C_{PZ}$ is already charged to a value $2V_{PZ(OC)}$ (in the previous half-cycle), increasing its voltage further by a value $V_{INV}$ requires more amount of investment energy. This is due to the quadratic dependence of energy on the capacitor voltage. The net extracted average output power ($P_{OUT-INV}$) using the investment technique [17] is given by (5.4).

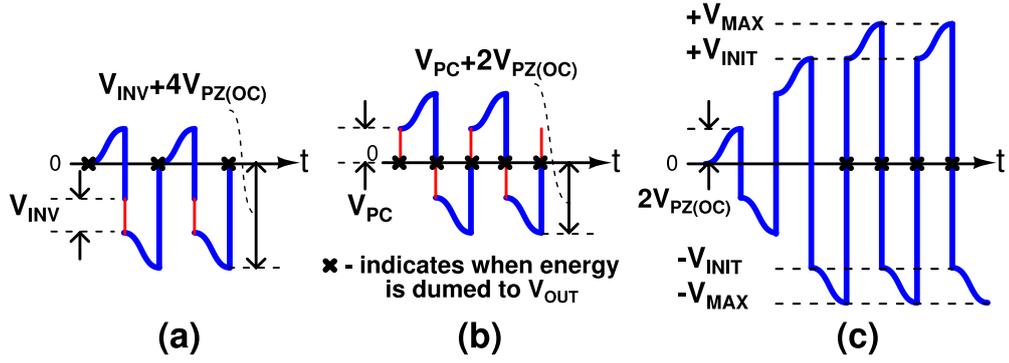

Fig. 5.6. Prior approaches in inductive rectifier. (a) Investment method [17], (b) pre-charging method [18]-[19], and (c) double pile-up resonance method [20].

$$P_{OUT-INV} = 2C_{PZ}V_{PZ(OC)}(V_{INV} + 4V_{PZ(OC)})f_{PZ} \tag{5.4}$$

The work in [18] and [19] overcomes the earlier drawback by investing the energy initially when $C_{PZ}$ is uncharged. As seen from Fig. 5.6 (b), $C_{PZ}$ is charged to $V_{PC}$ at the beginning of every cycle. Later the integration of $I_{PZ}$ continues for half-cycle, after which the total energy is transferred to the output. The extracted average output power ($P_{OUT-PC}$) using the pre-charge technique [18]-[19] is given by (5.5).

$$P_{OUT-PC} = 4C_{PZ}V_{PZ(OC)}(V_{PC} + V_{PZ(OC)})f_{PZ} \tag{5.5}$$

As discussed in Section 5.2.3, the voltage swing across the piezoelectric transducer ($V_{PZ}$) should be $V_{MAX}$ for extracting maximum possible output power for given CMOS technology. However, the works in [17]-[19] does not always ensure that the maximum value of $V_{PZ}$ equals $V_{MAX}$.

In [20], the voltage swing across the piezo transducer is initially accumulated and is limited between $V_{MAX}$ and $V_{INIT}$ during the steady-state, as shown in Fig. 5.6 (c). The bias-flip is restricted by flipping the voltage from $\pm V_{MAX}$ to $\mp V_{INIT}$. Since the voltage is not flipped entirely in magnitude, the remaining inductor current in the bias-flip step (Fig. 5.3 (b)) is dumped to the output. The extracted average output power ($P_{OUT-DPR}$) using the double pile-up resonance method [20] is given by (5.6).

$$P_{OUT-DPR} = 4C_{PZ}V_{PZ(OC)}(V_{INIT} + V_{PZ(OC)})f_{PZ} \tag{5.6}$$

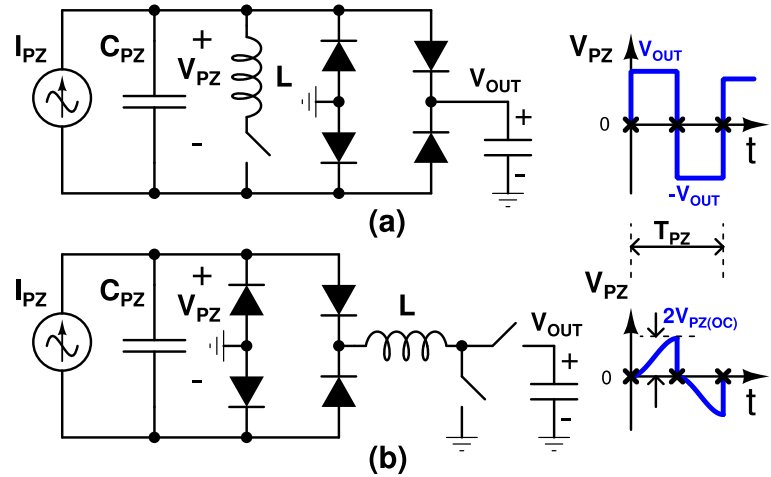

Fig. 5.7. Simplified representation of (a) bias-flip (or) SSHI rectifier [7]-[11], and (b) SECE rectifier [13]-[15].

where $V_{INIT}$ is a voltage carefully chosen as $V_{MAX} - 2V_{PZ(OC)}$ in [20] to allow for maximum voltage swing across the piezo transducer. As discussed in Section 5.1, the value of $V_{PZ(OC)}$ depends on the strength of the vibration and is not always a known parameter. For aperiodic vibrations, $V_{PZ(OC)}$ varies with time. The maximum voltage swing of $V_{PZ}$ in [20] could be written as $V_{INIT} + 2V_{PZ(OC)}$. If $V_{INIT}$ is lesser than $V_{MAX} - 2V_{PZ(OC)}$, the voltage swing across the piezo transducer reduces below $V_{MAX}$ and results in lower output power.

## 5.2.6 Other rectifier implementations utilizing inductor

Apart from the truly inductive rectifiers discussed in Section 5.2.5, other rectifier architectures utilizing inductors and diodes are also used for PEH. The bias-flip rectifier or the synchronous switch harvesting on inductor (SSHI) topology is shown in Fig. 5.7 (a) [7]-[11]. The extracted average output power ($P_{OUT-BF}$) using the bias-flip rectifier [7] is given by (5.7).

$$P_{OUT-BF} = 4C_{PZ}V_{PZ(OC)}V_{OUT}f_{PZ} \qquad (5.7)$$

As observed from (5.7), the output power of the bias-flip rectifier is low at the lower rectified output voltage ($V_{OUT}$). This is because $V_{PZ}$ is limited to $V_{OUT}$ as observed from Fig. 5.7 (a). To overcome this trade-off, bias-flip rectifiers are operated at higher rectified $V_{OUT}$ followed by a large step-down ratio DC-DC converter for powering the low voltage circuits [7]-[9], [11]. This causes additional power loss due to the multi-stage power extraction. Sense and set rectifier [12] is similar to the bias-flip rectifier, wherein it maintains the voltage swing across the piezo transducer to be equal to the required maximum power point voltage. Nevertheless, it suffers from the same limitations as that of the bias-flip rectifier in terms of the extracted $P_{OUT}$. The synchronous electrical charge extraction (SECE) rectifier [13]-[15] is shown in Fig. 5.7 (b). The extracted average output power using the SECE rectifier [14] is given by (5.8). The output power of the SECE rectifier is only 4× higher than full-bridge rectifier discussed in Section 5.2.1. The usage of diodes in the bias-flip and the SECE rectifier topology limits the output power extraction capability, as compared with the truly inductive rectifiers discussed in Section 5.2.5.

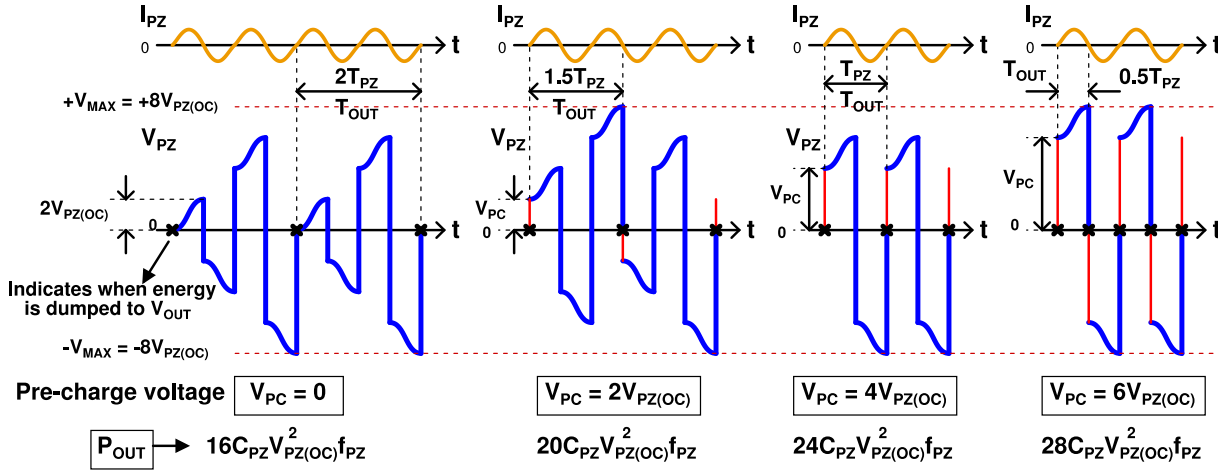

Fig. 5.8. Concept of the proposed inductive rectifier using pre-charge and accumulation of $V_{PZ}$ till $V_{MAX}$. $V_{PZ}$ waveforms for multiple pre-charge voltages ($V_{PC}$) are shown, where $P_{OUT}$ increases linearly with $V_{PC}$.

$$P_{OUT-SECE} = 4C_{PZ}V_{PZ(OC)}^2 f_{PZ} \tag{5.8}$$

## 5.3 Proposed Inductive Rectifier

In the proposed inductive rectifier, the drawbacks in the works from [17]-[20] are solved to extract more output power under all the conditions. The operation of the proposed inductive rectifier is as follows,

1) The capacitance of the piezo-transducer ($C_{PZ}$) is pre-charged to a voltage $V_{PC}$ at the beginning.

2) $I_{PZ}$ is then integrated into $C_{PZ}$. During the zero-crossings of $I_{PZ}$, the bias-flip process flips the voltage polarity of $V_{PZ}$ and helps in the accumulation of voltage. This process continues till $V_{PZ}$ reaches $V_{MAX}$.

3) Once $V_{PZ}$ reaches $V_{MAX}$, the total energy stored in $C_{PZ}$ is transferred to the output. The

process repeats once again from step-1.

Fig. 5.8 illustrates the $V_{PZ}$ waveforms in the proposed technique for different pre-charge voltages ($V_{PC}$). By pre-charging at the beginning, energy is invested in $C_{PZ}$ to raise its voltage from zero to a value $V_{PC}$. This reduces the amount of investment energy ($E_{INV}$) needed for increasing the voltage across $C_{PZ}$, as compared to the work in [17]. Moreover, the voltage $V_{PZ}$ is accumulated till the maximum possible value of $V_{MAX}$ irrespective of the pre-charge voltage ($V_{PC}$) or the open-circuit voltage ($V_{PZ(OC)}$), as indicated in Fig. 5.8. This ensures that the voltage swing across the transducer always reaches $V_{MAX}$. Hence, the problem of reduced voltage swing across the piezo transducer under non-optimal conditions is avoided as experienced in [17]-[20]. With the given CMOS process technology (i.e., given $V_{MAX}$), the proposed rectifier always extracts the highest possible output power.

The extracted average output power ($P_{OUT}$) in the proposed inductive rectifier is defined by (5.9). The average time period ($T_{OUT}$) in which the harvested energy is transferred to the output is provided by (5.10). In other words, $T_{OUT}$ is the time taken to accumulate the voltage across the piezo from $V_{PC}$ to $V_{MAX}$. Integrating every half cycle of $I_{PZ}$ into $C_{PZ}$ causes $V_{PZ}$ to increase by $2V_{PZ(OC)}$, and accumulating it over one vibration period ($T_{PZ}$) results in the term $4V_{PZ(OC)}$ appearing in the denominator of (5.10). From (5.9) and (5.10), the final expression for $P_{OUT}$ is given by (5.11).

$$P_{OUT} = \frac{(E_{OUT}-E_{INV})}{T_{OUT}} = \frac{C_{PZ}}{2} \times \frac{(V_{MAX}^2-V_{PC}^2)}{T_{OUT}} \tag{5.9}$$

$$T_{OUT} = \frac{(V_{MAX}-V_{PC})}{4V_{PZ(OC)}} \times T_{PZ} \tag{5.10}$$

$$P_{OUT} = 2C_{PZ}V_{PZ(OC)}(V_{MAX} + V_{PC})f_{PZ} \tag{5.11}$$

To compare the power extraction capabilities of various rectifier architectures, the work in [17] proposed a Figure of Merit (FoM) by normalizing the extracted average output power of the proposed circuit with the power obtained using an ideal FBR. The theoretical FoM for the proposed inductive rectifier is given by (5.12). As seen from (5.12), using a CMOS process technology that can handle higher voltages (i.e., higher $V_{MAX}$) improves the power extraction capability. The theoretical FoM for the proposed method and various other rectifier architectures are plotted in Fig. 5.9 (a) and (b) for two different $V_{MAX}$ and $V_{PZ(OC)}$. As observed in Fig. 5.9, the proposed method always guarantees higher $P_{OUT}$ compared with the other techniques. This is because the voltage swing across the piezo transducer (i.e. $V_{PZ}$) always reaches $V_{MAX}$ irrespective of the value of open-circuit voltage ($V_{PZ(OC)}$) or the pre-charge voltage ($V_{PC}$).

$$\text{FoM}_{(\text{Theoretical})} = \frac{2(V_{MAX}+V_{PC})}{V_{PZ(OC)}} \tag{5.12}$$

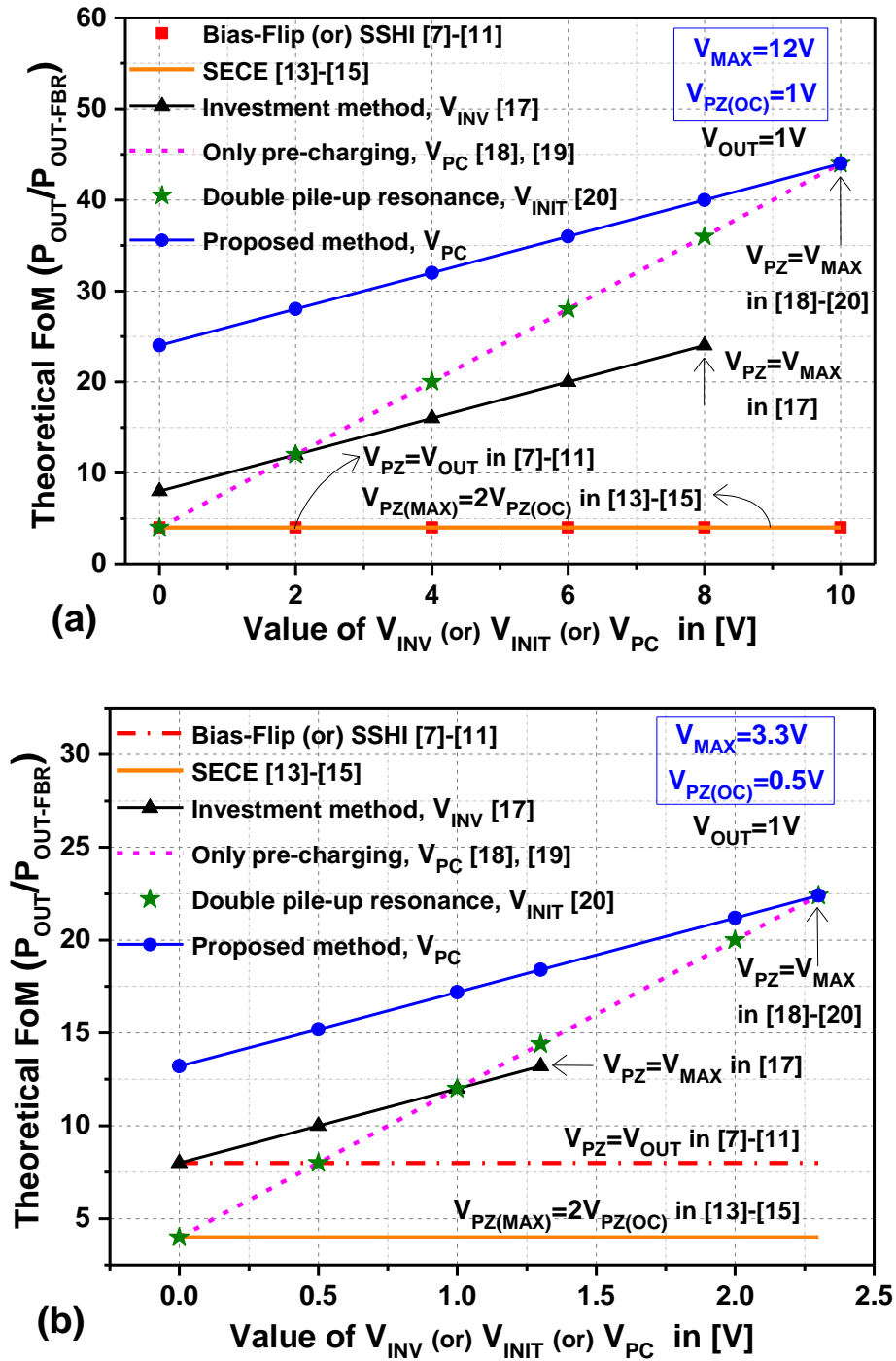

Fig. 5.9. Comparison of theoretical FoM for various inductor based rectifiers at $V_{OUT}$ of 1V. (a) $V_{MAX}$ = 12V, $V_{PZ(OC)}$ = 1V, and (b) $V_{MAX}$ = 3.3V, $V_{PZ(OC)}$ = 0.5V.

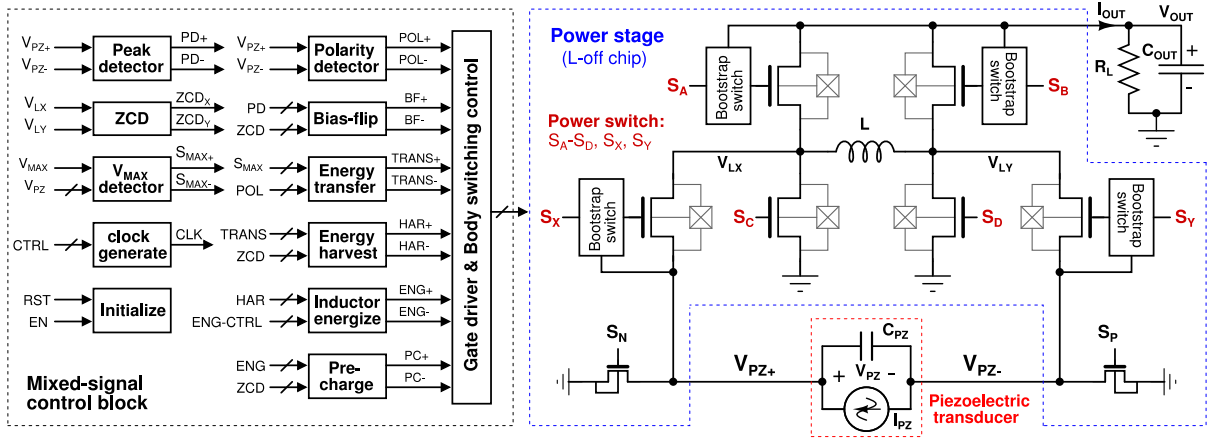

Fig. 5.10. System level block diagram of the proposed inductive rectifier including the power stage. $S_P$ and $S_N$ are used for defining 0V reference for the transducer.

## 5.4 Circuit Implementation

Fig. 5.10 shows the system level block diagram of the proposed inductive rectifier. For enabling an efficient multi-source harvesting interface, the proposed inductive rectifier utilizes the buck-boost power stage of the existing solar and thermal energy harvesting system [2], as discussed in Section 5.1. The inductor is required only during the exchange of energy with the capacitor and the output. Based on (5.2), the inductor utilization time for the PEH is typically one percent of the vibration time period ($T_{PZ}$). Hence, the power stage can be utilized for solar and thermal energy harvesting most of the time. However, in this work, only the PEH system is implemented.

### 5.4.1 Power stage configuration

Fig. 5.11 (a) shows the various possible switch configurations associated with the proposed inductive rectifier. The configurations depicted in Fig. 5.11 (b)-(g) occurs during the positive half cycle of $I_{PZ}$. When $V_{PZ}$ is more than 0V, switch $S_P$ is ON ($S_N$ is OFF), and $I_{PZ}$ gets integrated into $C_{PZ}$, as shown in Fig. 5.11 (b). The power stage configuration during the bias-flip step (BF+) is shown in Fig. 5.11 (c).

At the start of the bias-flip step BF+, $V_{PZ}$ (=$V_{PZ+}$) is more than 0V, and the switch $S_P$ is ON ($S_N$ is OFF). The inductor gets energized from $C_{PZ}$ until $V_{PZ}$ (or $V_{PZ+}$) reduces to 0V. During the instant $V_{PZ}$ reduces to 0V, the switch $S_P$ is turned OFF, and the switch $S_N$ is turned ON. Since the inductor current continues to flow in the same direction, $C_{PZ}$ now gets charged to an opposite polarity as indicated in Fig. 5.11 (c). $V_{PZ-}$ node voltage now becomes positive, while the node $V_{PZ+}$ remains at 0V. Hence, the node voltages ($V_{PZ+}$ and $V_{PZ-}$) *remain either positive or zero during the entire bias-flip period*. This avoids the need for a complicated negative gate driver (NGD) circuit as used in the previous works [20], [17], and [7]. The integration of $I_{PZ}$ into $C_{PZ}$ continues in the negative half cycle.

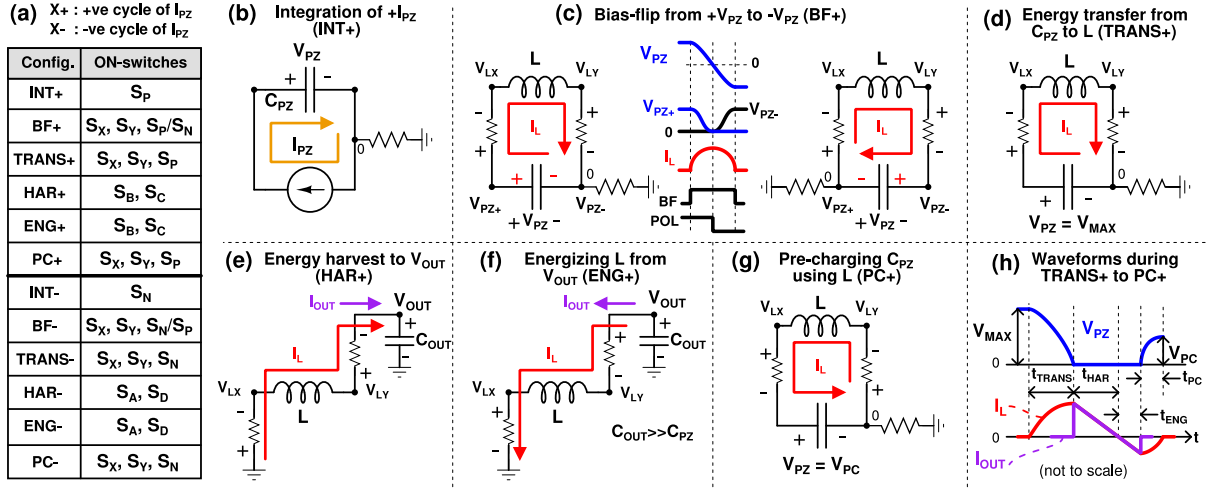

Fig. 5.11. (a) Possible switch configurations in the power stage, (b) integration of positive $I_{PZ}$ into $C_{PZ}$ (INT+), (c) positive bias-flip step (BF+), (d) energy transfer from $C_{PZ}$ to L (TRANS+), (e) energy transfer to $V_{OUT}$ (HAR+), (f) energizing inductor using $V_{OUT}$ (ENG+), (g) pre-charging $C_{PZ}$ using L, (h) waveforms of $V_{PZ}$ and $I_L$ during TRANS+ to PC+ step. Power stage configurations shown in (b)-(g) occur during the positive half cycle of $I_{PZ}$.

When the voltage $V_{PZ}$ reaches $V_{MAX}$ during the integration of $I_{PZ}$, the total energy stored in $C_{PZ}$ is transferred to the inductor, as shown in Fig. 5.11 (d). The inductor energy is then dumped to the output, as depicted in Fig. 5.11 (e). The inductor is then energized for a time period $t_{ENG}$ from $V_{OUT}$ as shown in Fig. 5.11 (f). The energized inductor is now used to pre-charge $C_{PZ}$ to a value $V_{PC}$ as shown in Fig. 5.11 (g). The corresponding voltage and current waveforms are depicted in Fig. 5.11 (h). The power stage configurations occurring during the negative half-cycle of $I_{PZ}$ follows a similar manner and is not shown here.

### 5.4.2 Control circuit

The voltages at nodes $V_{PZ+}$ and $V_{PZ-}$ either remain positive or at 0V, as shown in Fig. 5.12 (a). Hence, for detecting the polarity of $V_{PZ}$ (=$V_{PZ+} - V_{PZ-}$), a PMOS-input hysteretic comparator is used, as shown in Fig. 5.12 (a). The control signals for switch $S_P$ and $S_N$ is generated based on the polarity of $V_{PZ}$ and the operational state of the rectifier, as shown in Fig. 5.12 (b). The peak-detector circuit [20] used to detect the zero-crossing of $I_{PZ}$ (i.e., the start of bias-flip step) is provided in Fig. 5.12 (c).

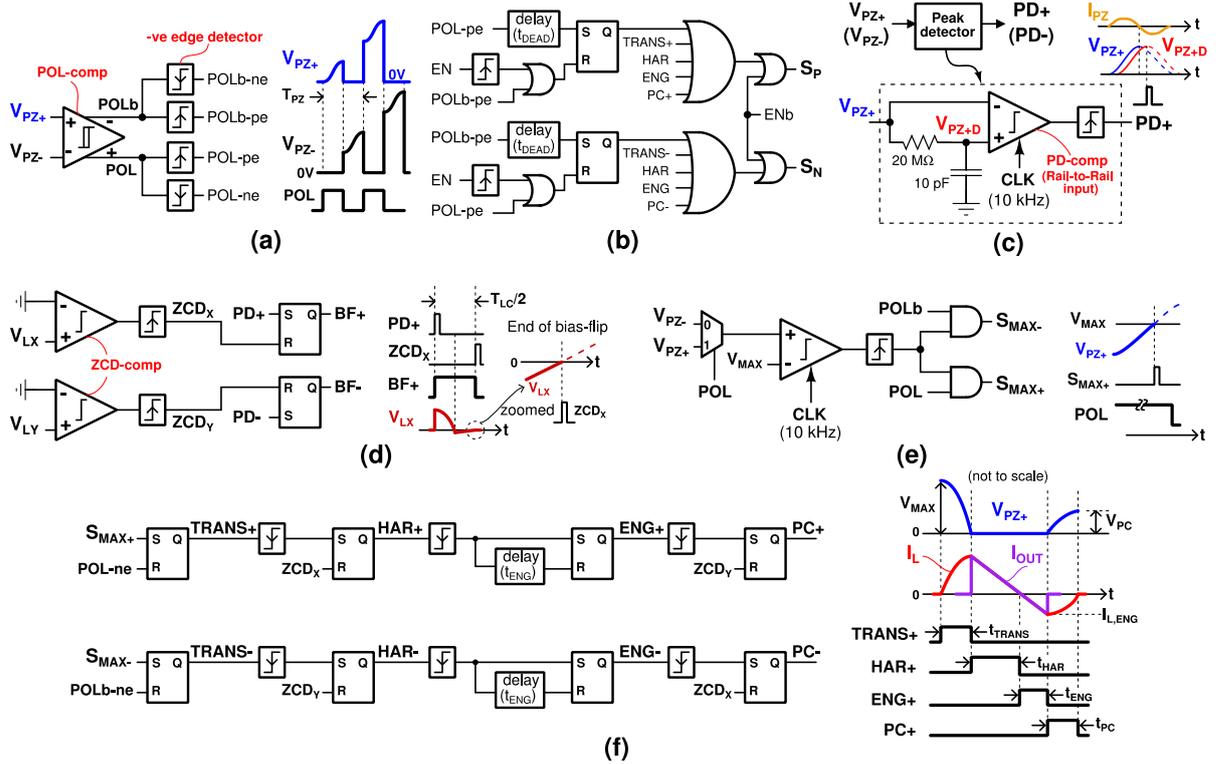

Fig. 5.12. Simplified representation of the control circuit. (a) polarity detector, (b) signal generation for $S_P$ and $S_N$, (c) peak detector [20], (d) bias-flip signal generation, (e) $V_{MAX}$ detection, and (f) generation of energy transfer (TRANS), energy harvest (HAR), inductor energise (ENG) and pre-charge (PC) signals.

The generation of the bias-flip signal is depicted in Fig. 5.12 (d). The peak-detector signal (PD+ or PD-) initiates the start of the bias-flip step. As shown in Fig. 5.11 (c), the bias-flip step ends when the inductor current goes to zero. Hence, the voltage polarity at the node $V_{LX}$ (or $V_{LY}$) is monitored to prevent the zero crossings of the inductor current during the bias-flip step BF+ (or BF-), as shown in Fig. 5.12 (d). Thus, the bias-flip duration in the proposed rectifier is self-timed, and does not require any on-chip timing reference or external adjustment as used in the prior implementations [7]-[10], [17], [20]. Modified common-gate input comparator from [2] is used for zero-crossing detection (ZCD-comp) of the inductor current, as shown in Fig. 5.13.

An NMOS-input strong-arm latch comparator is used to detect the time instant when the voltage $V_{PZ}$ reaches $V_{MAX}$ as shown in Fig. 5.12 (e). This indicates the beginning of the energy transfer phase, which continues till $V_{PZ}$ reduces to zero (indicated by the polarity signal), as depicted in Fig. 5.12 (f). Then the energy harvest phase begins and continues till the inductor current is entirely drained to the output. Fig. 5.11 (e) indicates the inductor current direction and the polarity of the node $V_{LX}$ during the harvest phase (HAR+). Hence, the polarity of the node voltage $V_{LX}$ (or $V_{LY}$) is monitored to prevent the zero-crossings of the inductor current during the harvest phase HAR+ (or HAR-), as shown in Fig. 5.12 (f).

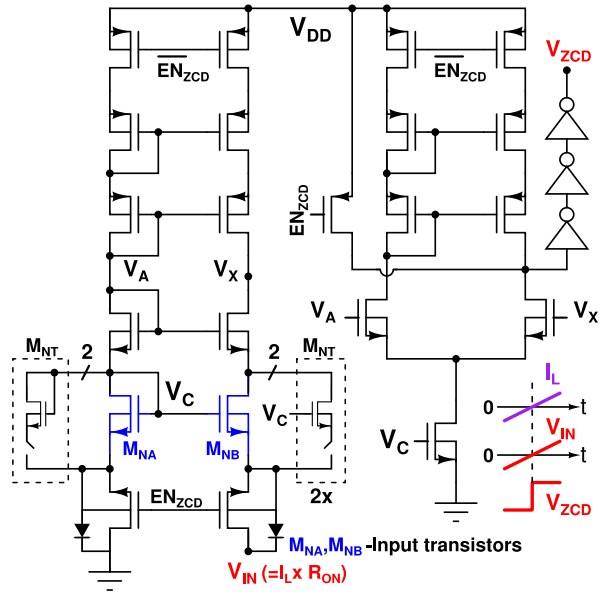

Fig. 5.13. Modified zero-crossing detector comparator (ZCD-comp) from [2].

The inductor energize time ($t_{ENG}$) is controlled as shown in Fig. 5.12 (f). The value of the inductor current ($I_{L,ENG}$) at the end of the energize phase is given by (5.13). The energized inductor then pre-charges $C_{PZ}$ to a voltage $V_{PC}$. The value of $V_{PC}$ is obtained by equating the energy stored in the inductor before pre-charge to the energy acquired by $C_{PZ}$ after pre-charge, as shown in (5.14). From (5.13) and (5.14), the expression for $V_{PC}$ is given by (5.15). The pre-charge step ends by monitoring the time instant the inductor current goes to zero, as shown in Fig. 5.12 (f). The sequence of steps in the proposed inductive rectifier is summarized by the flow chart given in Fig. 5.14.

$$I_{L,ENG} = \frac{V_{OUT}}{L} \times t_{ENG} \qquad (5.13)$$

$$\tfrac{1}{2} \times L \times I_{L,ENG}^2 = \tfrac{1}{2} \times C_{PZ} \times V_{PC}^2 \qquad (5.14)$$

$$V_{PC} = V_{OUT} \times \frac{t_{ENG}}{\sqrt{LC_{PZ}}} \qquad (5.15)$$

### 5.4.3 Gate and body driver circuit

The NMOS power switches $S_A$, $S_B$, $S_X$, and $S_Y$ shown in Fig. 5.10 are capacitively bootstrapped to keep the on-resistance independent of its source node voltage. The bootstrapped switch from [22] is modified to handle voltages as high as $2V_{DD}$ without any voltage stress, as shown in Fig. 5.15 (a). For the NMOS power switches $S_C$, $S_D$, $S_X$, and $S_Y$ shown in Fig. 5.10, the body terminal is connected to the node with the lowest potential using the arrangement depicted in Fig. 5.15 (a) [2]. However, the power switches $S_A$ and $S_B$ shown in Fig. 5.10, interface to the external output load. Hence, the body terminal for the NMOS power switches $S_A$ and $S_B$ are connected to the node with the lowest potential using the arrangement shown in Fig. 5.15 (b). The appropriate digital control signals driving the body-bias switches shown in Fig. 5.15 are generated based on the operational state of the inductive rectifier. All the switches in the power stage shown in Fig. 5.10 and Fig. 5.15 are implemented using thick-oxide devices which could handle a maximum of 3.3 V.

Fig. 5.14. Flow chart summarizing various steps in the proposed inductive rectifier implemented in this work. Alternate method of implementing pre-charge step is suggested in Section 5.6.

Fig. 5.15. (a) Modified bootstrapped power switch from [22] handling $2V_{DD}$ without voltage stress ($V_{DD}$=3V), and (b) body control for switches $S_A$ and $S_B$.

### 5.4.4 Post-layout simulation results

The piezo transducer used in this work is Mide PPA-1021. The characterized value of $C_{PZ}$, $f_{PZ}$ and $R_{PZ}$ for the piezo transducer in the clamp-0 location are 19 nF, 146 Hz, and 2 MΩ, respectively. The above parameters are used to model the piezo transducer in the SPICE simulations. The inductance value used in this work is 47 µH. The simulated $V_{PZ}$ waveforms for a $V_{PZ(OC)}$ of 1 V and an $V_{MAX}$ of 3.3 V in the 180 nm CMOS technology is shown in Fig. 5.16. As observed from Fig. 5.16, pre-charging the piezo transducer reduces the time it takes to accumulate $V_{PZ}$ till $V_{MAX}$.

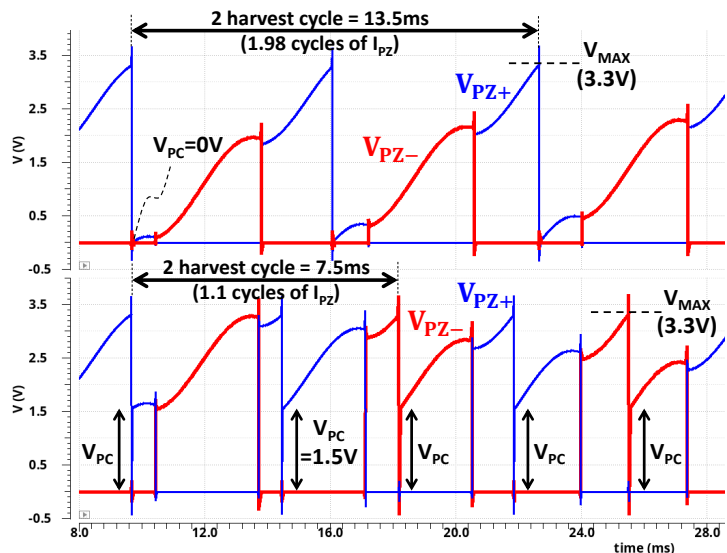

Fig. 5.16. Simulated $V_{PZ}$ waveforms in the proposed inductive rectifier with and without pre-charge. $V_{PZ(OC)} = 1$V and $V_{MAX} = 3.3$ V.

The post-layout simulation of the output power extraction capability for the proposed inductive rectifier in the worst-case process corner (SS) is shown in Fig. 5.17. As observed from Fig. 5.17 (b), the proposed rectifier can extract a maximum output power of 18.41 µW for an $V_{PZ(OC)}$ of 1 V. Whereas, the ideal FBR can only extract 2.77 µW of output power. This corresponds to a post-layout simulated FoM of 6.64× in the worst-case process corner (SS).

However, equation (5.12) predicts a theoretical FoM of 9.6× for the same input condition. This difference is due to the loss in $R_{PZ}$ and power stage conduction loss which were not included in (5.12). Fig. 5.18 (a) shows the effect of power stage conduction loss on the extracted $P_{OUT}$. As observed from Fig. 5.18 (a), using a larger inductor reduces the power stage conduction loss (due to reduced peak current) [7], [17]-[21]. However, it demands for a highly accurate zero-current detector circuit with sophisticated offset cancellation techniques. As a trade-off, 47 µH inductor is used for measurements in this work. Fig. 5.18 (b) shows the simulated result of the extracted $P_{OUT}$ for typical range of $R_{PZ}$ values. From Fig. 5.18 (b) it could be observed that, the power loss in $R_{PZ}$ reduces the extracted $P_{OUT}$.

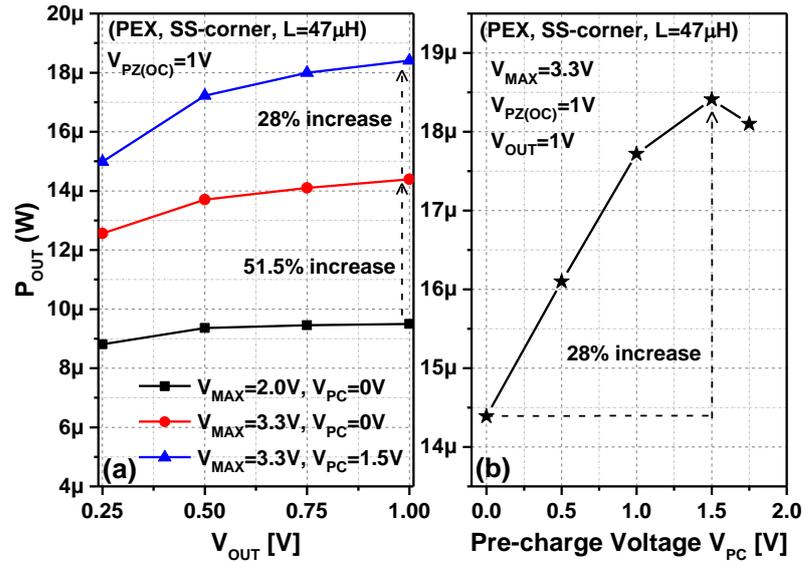

Fig. 5.17. Post-layout simulated $P_{OUT}$ of the proposed inductive rectifier with respect to (a) output voltage, and (b) pre-charge voltage. $V_{PZ(OC)} = 1V$.

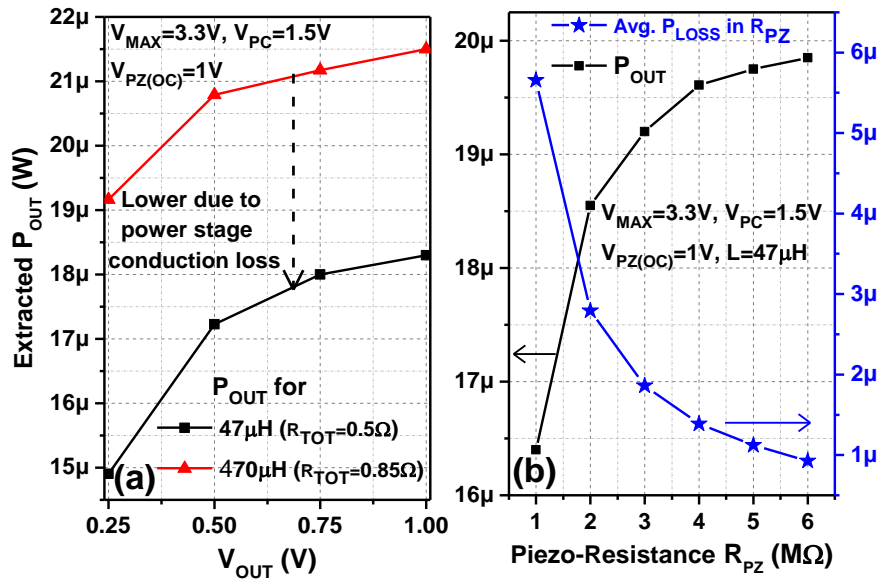

Fig. 5.18. Simulated effect of (a) power stage conduction loss on $P_{OUT}$, and (b) $R_{PZ}$ on $P_{OUT}$. $R_{TOT}$ includes the DCR of inductor and other parasitic resistances.

## 5.5 Chip Measurement Results

The proposed inductive rectifier circuit is implemented in the 180 nm mixed-mode CMOS technology. The chip micrograph and the measurement setup are shown in Fig. 5.19 (a) and (b), respectively. All the IO pins of the chip are ESD protected.

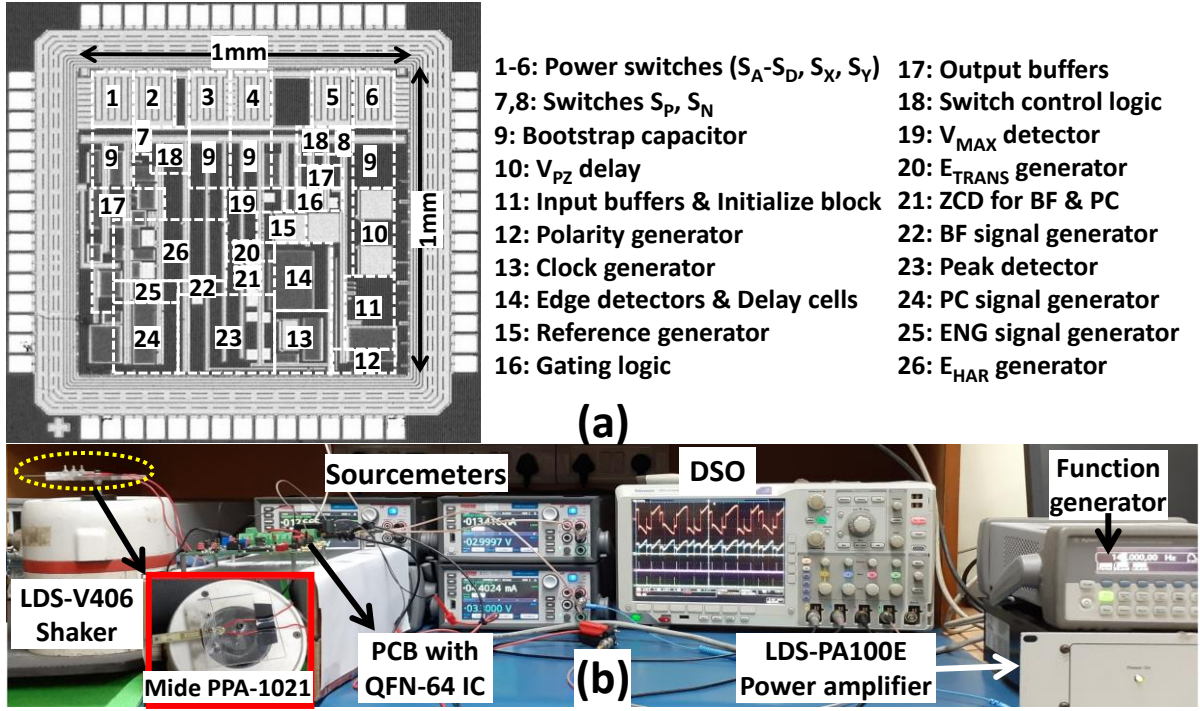

Fig. 5.19. (a) Chip micrograph, and (b) measurement setup.

### 5.5.1 $V_{PZ}$ and $I_L$ waveforms

The inductor current and the output current was measured using the current probe (Tektronix-TCP305A) as shown in Fig. 5.20. The measured $V_{PZ}$ waveforms without pre-charge and with pre-charge are shown in Fig. 5.21 (a) and (b), respectively. In Fig. 5.21 (a), $V_{PZ}$ is accumulated till $V_{MAX}$ of 3.3 V without any pre-charging. Whereas $V_{PZ}$ is pre-charged to 1.5 V, and is then accumulated till $V_{MAX}$ of 3.3 V, as shown in Fig. 5.21 (b). The measurement of $V_{PZ}$ and the inductor current during the bias-flip step (BF-) is shown in Fig. 5.21 (c). As explained in Section 5.4.1, the proposed inductive rectifier avoids the negative voltage swings across the piezo transducer during the bias-flip step. This is clearly evident from Fig. 5.21 (c), where the voltages $V_{PZ+}$ or $V_{PZ-}$ remain either at a positive value or at zero.

From Fig. 5.21, it is observed that there exists a voltage loss after every bias-flip step, which is discussed later in Section 5.5.3. Once the accumulated voltage across the piezo reaches $V_{MAX}$, the total energy stored in the piezo capacitor is transferred to the inductor. Hence, the voltage across the piezo falls to zero, and the inductor current rises to its peak value during the energy transfer step (TRANS-), as shown in Fig. 5.22 (a). The entire inductor current is then dumped to the output during the energy harvest phase (HAR-), as shown in Fig. 5.22 (b). The output node is connected to load resistance in parallel with the capacitor, as shown in Fig. 5.10. In the steady state, $V_{OUT}$ remains constant.

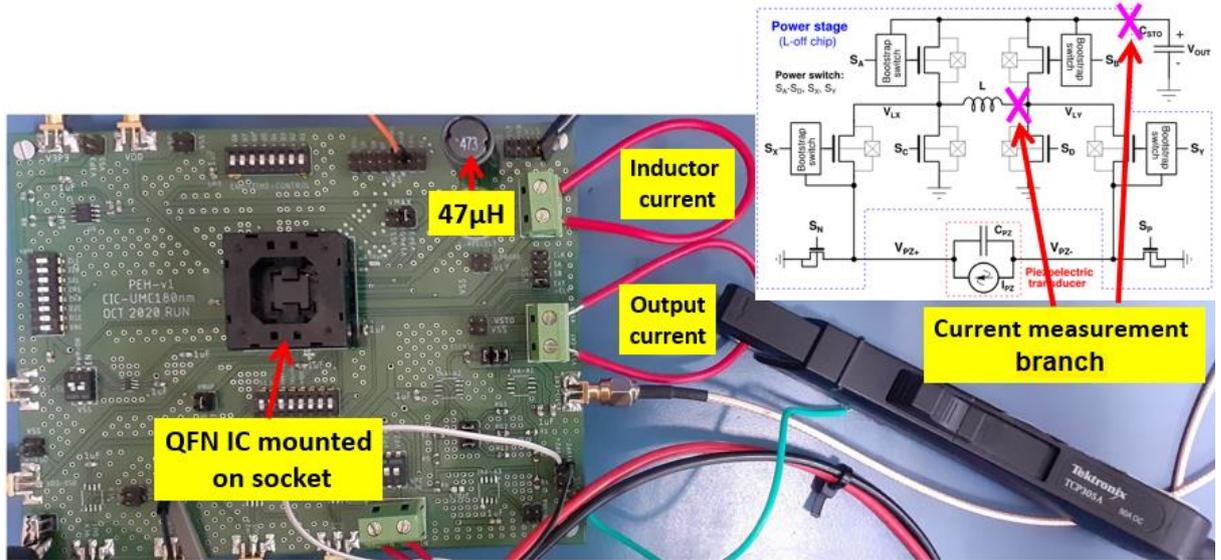

Fig. 5.20. Inductor current and output current measurement using the current probe. The current measurement branch is also highlighted in the inlet figure.

The inductor is energized from $V_{OUT}$ for a time-period $t_{ENG}$ as indicated in Fig. 5.22 (c). The inductor current rises to a value given by (5.13) during the inductor energize step (ENG-), as shown in Fig. 5.22 (c). The energized inductor is then used to pre-charge the piezo capacitance to a value $V_{PC}$ as provided by (5.15). Fig. 5.22 (d) shows the pre-charge step (PC-) where the piezo node voltage $V_{PZ-}$ rises to a value of 1.5 V.

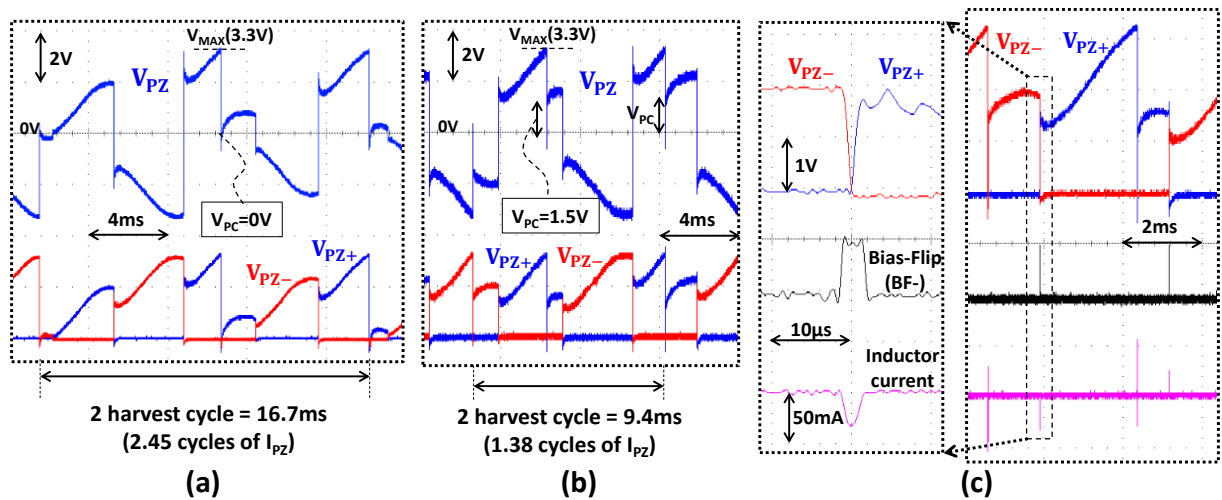

Fig. 5.21. Measurement of (a) $V_{PZ}$ waveform without precharge, (b) $V_{PZ}$ waveform with pre-charge, and (c) zoomed waveforms during the bias-flip (BF-) step. $V_{PZ(OC)} = 1$ V, $V_{MAX} = 3.3$ V, L = 47 µH, $R_L$ = 25 kΩ, and $C_{OUT}$ = 20 µF.

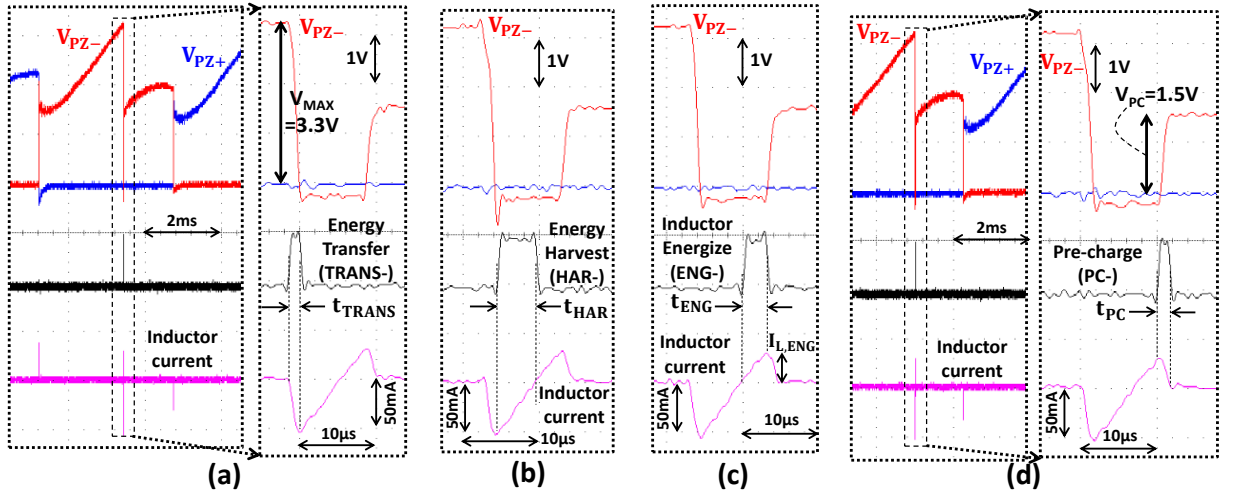

Fig. 5.22. Measurement of the inductor current waveform during (a) energy transfer, (b) energy harvest, (c) inductor energize, and (d) pre-charge step. $V_{PZ(OC)} = 1$ V, $V_{MAX} = 3.3$ V, L = 47 µH, $R_L = 25$ kΩ, and $C_{OUT} = 20$ µF.

### 5.5.2 $P_{OUT}$ extraction performance

As indicated in Section 5.5.1, the output voltage across the external load resistance $R_L$ in parallel with the output capacitor $C_{OUT}$ is constant in the steady-state. The extracted output power ($P_{OUT}$) is given by (5.16).

$$P_{OUT} = \frac{V_{OUT}^2}{R_L} \tag{5.16}$$

where $V_{OUT}$ is the steady-state voltage across $R_L$.

Measurement of $V_{OUT}$ with $R_L = 100$ kΩ and $C_{OUT} = 20$ µF is given in Fig. 5.23 for three different cases. In Fig. 5.23 (a), the voltage across the piezo is accumulated without any pre-charging till 1.65 V. Rectified $V_{OUT}$ in the steady-state is 730 mV, which corresponds to a $P_{OUT}$ of 5.33 µW. In Fig. 5.23 (b), the voltage across the piezo is accumulated without any pre-charging till a $V_{MAX}$ of 3.3 V. Rectified $V_{OUT}$ in the steady-state is 921 mV, corresponding to a $P_{OUT}$ of 8.48 µW.

For the case without pre-charging, increasing $V_{MAX}$ by a factor of 2× increases $P_{OUT}$ by the same factor theoretically based on (5.11). However, from the measurement of $V_{OUT}$ shown in Fig. 5.23 (a) and (b), $P_{OUT}$ increases by a factor of 1.6× when the value of $V_{MAX}$ is increased from 1.65 to 3.3 V.

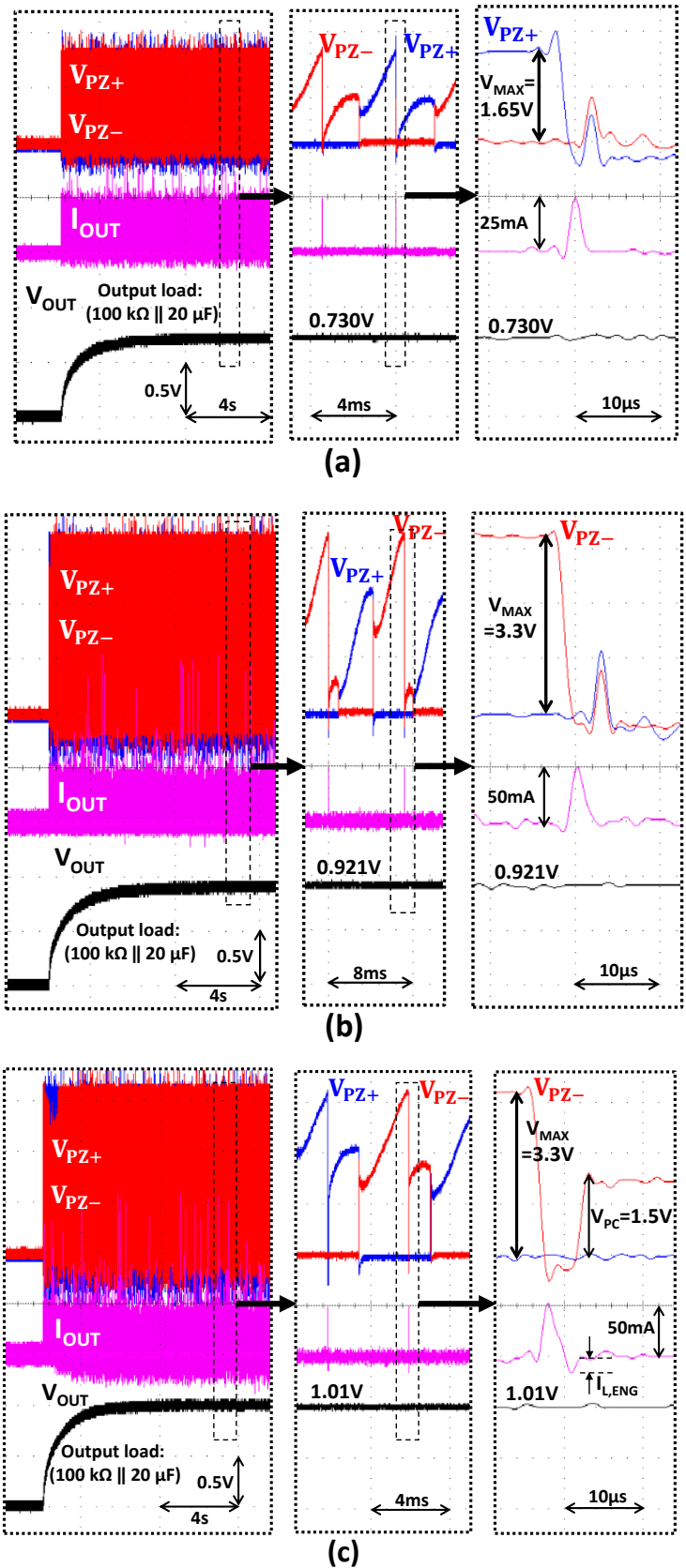

Fig. 5.23. Measurement of $V_{OUT}$ with (a) only accumulation till 1.65 V, (b) only accumulation till $V_{MAX}$=3.3 V, and (c) pre-charge to 1.5 V and accumulate till $V_{MAX}$=3.3 V. $R_L$=100 kΩ, $C_{OUT}$=20 µF, $V_{PZ(OC)}$=1 V for (a)-(c).

In Fig. 5.23 (c), the piezo transducer is pre-charged to 1.5 V and $V_{PZ}$ is then accumulated till 3.3 V. Rectified $V_{OUT}$ in the steady-state is 1.01 V, corresponding to a $P_{OUT}$ of 10.2 µW. By pre-charging to 1.5 V, the theoretical model from (5.11) predicts an increase in $P_{OUT}$ by 45% compared to the case without pre-charging. From the measurements shown in Fig. 5.23 (b) and (c), pre-charging to 1.5 V increases $P_{OUT}$ by 20% compared to the case without pre-charging.

### 5.5.3 Voltage loss occurring in accumulation

As discussed in Section 5.4.4, the theoretical expression for $P_{OUT}$ given by (5.11) does not include the conduction loss and the loss due to the parasitic resistance ($R_{PZ}$) of the piezo transducer. In addition, it could be observed from Fig. 5.21 (a) and (b), that voltage loss occurs after every bias-flip step. Fig. 5.24 clearly depicts the voltage loss for the case of accumulating $V_{PZ}$ from 0 V till the $V_{MAX}$ of 3.3 V. After every bias-flip, the piezo node voltage reduces momentarily (approximately by 0.8 V) before continuing with the subsequent integration of $I_{PZ}$. This results in more time to accumulate $V_{PZ}$ till $V_{MAX}$, thereby reducing the extracted average output power. As shown in Fig. 5.24, the time to accumulate till $V_{MAX}$ increases by approximately 56% in the measurements compared with the theoretical model. The voltage loss occurs due to the delayed turn-off of the bootstrapped power switches $S_X$ and $S_Y$ shown in Fig. 5.10. This delayed turn-off causes the capacitor $C_{PZ}$ to momentarily discharge through the inductor after every bias-flip step (since $T_{LC} \ll T_{PZ}$ from (5.2)).

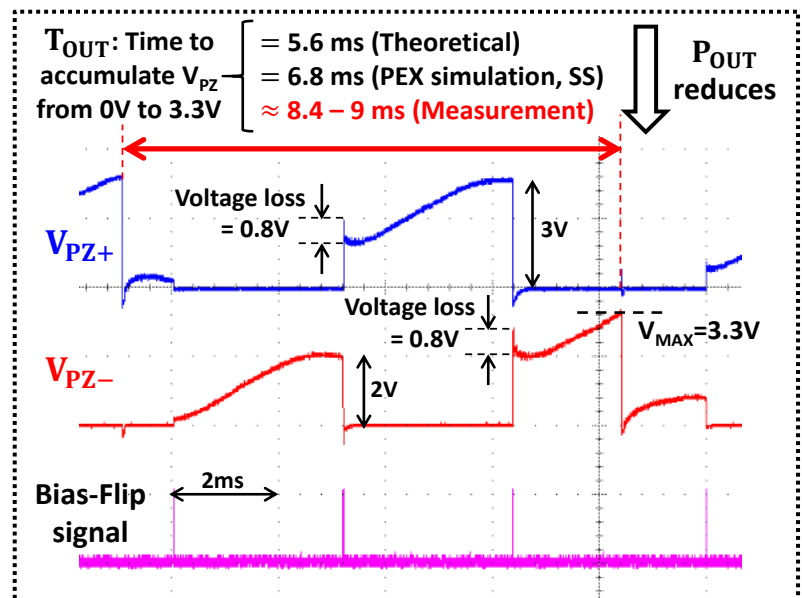

Fig. 5.24. Voltage loss resulting in increased time to accumulate till $V_{MAX}$. $V_{PZ(OC)} = 1$ V, $V_{PC} = 0$ V and $V_{MAX} = 3.3$ V.

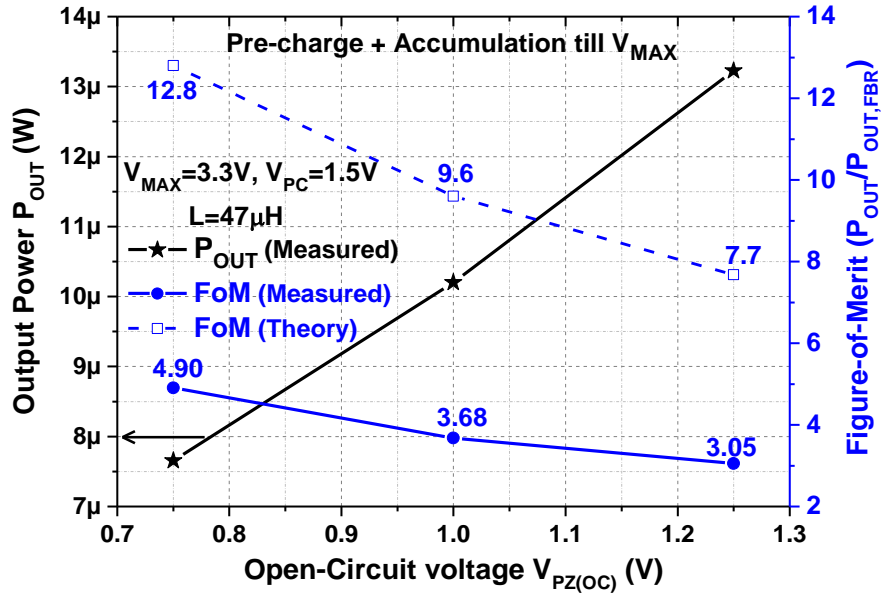

Fig. 5.25. Measured $P_{OUT}$ and FoM for different piezo open-circuit voltages (i.e. diff. vibration strengths). $V_{MAX} = 3.3V$, $L = 47\ \mu H$, $R_L = 100\ k\Omega$, and $C_{OUT} = 20\ \mu F$.

The voltage loss results in reduced voltage-flip efficiency of ~ 69% after the bias-flip step. The extracted $P_{OUT}$ and the measured FoM are plotted with respect to $V_{PZ(OC)}$ in Fig. 5.25. Loss in the parasitic resistance ($R_{PZ}$) of the piezo transducer, power stage conduction loss (as discussed in Section 5.4.4), and the voltage loss occurring in the test chip reduce the extracted output power from the theoretical value predicted by (5.12). This results in the difference between the measured and theoretical FoM plotted in Fig. 5.25.

### 5.5.4 Comparison with other inductor based rectifiers

Table 5.2 shows the comparison of the proposed inductive rectifier with the other inductor-based rectifiers for PEH. As observed from Table 5.2, only the proposed rectifier always achieves a maximum voltage swing of $V_{MAX}$ across the piezoelectric transducer for extracting the highest possible output power for given CMOS technology (Fig. 5.9). The bias-flip period is self-timed, avoiding the manual tuning of the timing generator for different piezoelectric transducer and inductor values or variations in their values. Moreover, the negative voltage swing across the piezoelectric transducer is avoided, simplifying the gate driver for the power stage.

The proposed inductive rectifier is realized using the buck-boost power stage from [2], thereby enabling multi-source harvesting of DC and AC sources together. For an $V_{PZ(OC)}$ of 1 V, the proposed inductive rectifier extracts 3.68× (=10.2 μW/2.77 μW) more power than the maximum output power of a full-bridge rectifier with ideal diodes. This is despite the voltage loss during accumulation as described in Section 5.5.3.

The proposed rectifier extracts output power in a single stage, even at lower rectified $V_{OUT} \leq$ 1 V. The power extraction is also relatively independent of $V_{OUT}$ as observed from the measured FoM at $V_{OUT}$ of 1 V and 0.5 V. Whereas for bias-flip and the sense-and-set rectifier, the FoM reduces by the same factor as that of rectified $V_{OUT}$. The usage of PMOS power switches at the output load side prohibits the inductive rectifiers from [17] and [20] to harvest at a lower rectified $V_{OUT}$.

Table 5.2: Comparison with other on-chip inductive rectifies for piezo energy harvesting

| References | JSSC 2014 [17] | JSSC 2016 [8] | JSSC 2018 [13] | JSSC 2018 [20] | SSCL 2019 [9] | JSSC 2019 [10] | JSSC 2019 [12] | ISSCC 2020 [14] | TCAS-I 2021 [11] | This work 2021 |
|---|---|---|---|---|---|---|---|---|---|---|
| Process technology | 350 nm CMOS | 350nm CMOS | 40nm HV CMOS | 180nm BCD | 130nm CMOS | 130nm CMOS | 180nm CMOS | 600nm CMOS | 180nm CMOS | 180nm CMOS |
| $C_{PZ}$ (nF) | 15 | 20.8 | 43 | 20 | 20 | 14, 22 | 8 | 24 | 2 | 19 |
| $f_{PZ}$ (Hz) | 143 | 229.6 | 75.4 | 140 | 143 | 441, 432 | 53 | 56 | 415 | 146 |
| Inductor (µH) | 330 | 3400 | 2200 | 680 | 3300 | 47, 33 | 1000 | 220 | 100, 68 | 47 |
| $V_{PZ(OC)}$ | 2.6 V | 1.25 V | N.A | 0.6 V | 1.6 V | 2 V | 2.13 V | N.A | 1.02 V | 1 V |
| Max. amplitude of $V_{PZ}$ | 6 V | 5 V | 6 V | 3.5 V | 3.3 V | 3.3 V | 2 V | 12 V | 4.2 V | 3.3 V |
| $P_{CTRL}/P_{OUT}$ (µW/µW) | 0.63/46 | N.A/36.8 | N.A/82.6 | 0.96/15 | 5.18/30.53 | 2/100 | 0.23/0.87 | 0.8/275 | N.A/4.7 | 0.75/10.2 |
| Piezo energy extraction method | Inductive rectifier | Bias-flip rectifier | SECE | Inductive rectifier | Bias-flip rectifier | Bias-flip rectifier | Sense & set rectifier | P-SECE | Bias-flip rectifier | Inductive rectifier |
| Always max. $V_{PZ}$ Swing = $V_{MAX}$ ? | No | No | No | No | No | No | No | No | No | Yes |
| Value of $V_{PZ(OC)}$ to be known ? | No | No | No | Yes | No | No | No | No | No | No |
| Bias-flip time control | External adjust | External adjust | – | External adjust | External adjust | External adjust | – | – | Self-timed | Self-timed |
| Avoidance of NGD[&] (or) neg. voltage converter | No | No | No | No | No | No | Yes | No | No | Yes |
| Multi-source EH possible | No | No | No | Yes | No | No | No | No | No | Yes |
| Single stage $P_{OUT}$ extraction | Yes | No | No | Yes | No | Yes | Yes | No | No | Yes |
| Independent load voltage | | Yes | Yes | | Yes | No | Yes | Yes | Yes | |
| FoM* independent of rectified volt. | | No | Yes | | No | No | No | Yes | No | |
| Measured FoM* (at rectified voltage) | 3.2, 3.6 × (2.7-3.7V) | 4.93 × (2.25V) | 3.14 × (2V) | 14.52 × (4V) | 4.17 × (3.3V) | 4.48, 4.28 × (3.5V) | 5.12 × (1V) | 3.28 × (2-5V) | 5.44, 4.56 × (2.5-3V) | 3.68× (1V) 3.4× (0.5V) |

[&] NGD: Negative Gate Driver,
*FoM (periodic at resonance): (Measured $P_{OUT}$) / ($P_{OUT}$ of FBR with ideal diode = $C_{PZ}V_{PZ(OC)}^2 f_{PZ}$),

## 5.6 Conclusion

In this chapter, an efficient method of extracting power from the piezoelectric transducer is proposed. The proposed method ensures that the voltage swing across the transducer is always maximum for extracting the highest possible output power for given CMOS technology. The proposed inductive rectifier is entirely self-timed and avoids negative voltage swings across the transducer. Moreover, it allows for efficient multi-source energy harvesting of DC and AC sources together using the same power stage. For an $V_{PZ(OC)}$ of 1V, the proposed rectifier extracts 3.68× more power than the full-bridge rectifier, even at a lower rectified $V_{OUT}$ of 1V. Despite the voltage loss during accumulation, the proposed method of power extraction from the piezoelectric transducer is experimentally demonstrated. The conduction loss in the inductive rectifier could be reduced by using a larger inductor along with a sophisticated offset-cancelled ZCD comparator. A fraction of the inductor current after the energy transfer step could be used for pre-charging $C_{PZ}$, before dumping the remaining current to the output. In this way, a separate inductor energizing step could be avoided.

## 5.7 References

[1] S. Sankar, M. Goel, P. -H. Chen, V. R. Rao, and M. S. Baghini, "Switched-Capacitor-Assisted Power Gating for Ultra-Low Standby Power in CMOS Digital ICs," *IEEE Trans. Circuits Syst. I, Reg. Papers*, vol. 67, no. 12, pp. 4281-4294, Dec. 2020.

[2] P.-H. Chen, H.-C. Cheng, and C.-L. Lo, "A Single-Inductor Triple-Source Quad-Mode Energy-Harvesting Interface With Automatic Source Selection and Reversely Polarized Energy Recycling," *IEEE J. Solid-State Circuits*, vol. 54, no. 10, pp. 2671–2679, Oct. 2019.

[3] S. Bandyopadhyay and A. P. Chandrakasan, "Platform Architecture for Solar, Thermal, and Vibration Energy Combining With MPPT and Single Inductor," *IEEE J. Solid-State Circuits*, vol. 47, no. 9, pp. 2199–2215, Sep. 2012.

[4] G. Chowdary, A. Singh, and S. Chatterjee, "An 18 nA, 87% Efficient Solar, Vibration and RF Energy-Harvesting Power Management System With a Single Shared Inductor," *IEEE J. Solid-State Circuits*, vol. 51, no. 10, pp. 2501–2513, Oct. 2016.

[5] G. Saini and M. Shojaei Baghini, "A Generic Power Management Circuit for Energy Harvesters With Shared Components Between the MPPT and Regulator," *IEEE Trans. Very Large Scale Integr. (VLSI) Syst.*, vol. 27, no. 3, pp. 535-548, March 2019.

[6] A. Devaraj, M. Megahed, Y. Liu, A. Ramachandran, and T. Anand, "A Switched Capacitor Multiple Input Single Output Energy Harvester (Solar + Piezo) Achieving 74.6% Efficiency With Simultaneous MPPT," *IEEE Trans. Circuits Syst. I, Reg. Papers*, vol. 66, no. 12, pp. 4876–4887, Dec. 2019.

[7] Y. K. Ramadass and A. P. Chandrakasan, "An Efficient Piezoelectric Energy Harvesting Interface Circuit Using a Bias-Flip Rectifier and Shared Inductor," *IEEE J. Solid-State Circuits*, vol. 45, no. 1, pp. 189–204, Jan. 2010.

[8] D. A. Sanchez, J. Leicht, F. Hagedorn, E. Jodka, E. Fazel, and Y. Manoli, "A Parallel-SSHI Rectifier for Piezoelectric Energy Harvesting of Periodic and Shock Excitations," *IEEE J. Solid-State Circuits*, vol. 51, no. 12, pp. 2867–2879, Dec. 2016.

[9] S. Li, A. Roy, and B. H. Calhoun, "A Piezoelectric Energy-Harvesting System With Parallel-SSHI Rectifier and Integrated Maximum-Power-Point Tracking," *IEEE Solid-State Circuits Lett.*, vol. 2, no. 12, pp. 301–304, Dec. 2019.

[10] S. Javvaji, V. Singhal, V. Menezes, R. Chauhan, and S. Pavan, "Analysis and Design of a Multi-Step Bias-Flip Rectifier for Piezoelectric Energy Harvesting," *IEEE J. Solid-State Circuits*, vol. 54, no. 9, pp. 2590–2600, Sep. 2019.

[11] B. Ciftci, S. Chamanian, A. Koyuncuoglu, A. Muhtaroglu, and H. Külah, "A Low-Profile Autonomous Interface Circuit for Piezoelectric Micro-Power Generators," *IEEE Trans. Circuits Syst. I, Reg. Papers*, vol. 68, no. 4, pp. 1458–1471, 2021.

[12] Y. Peng, K. D. Choo, S. Oh, I. Lee, T. Jang, Y. Kim, J. Lim, D. Blaauw, and D. Sylvester, "An Efficient Piezoelectric Energy Harvesting Interface Circuit Using a Sense-and-Set Rectifier," *IEEE J. Solid-State Circuits*, vol. 54, no. 12, pp. 3348–3361, Dec. 2019.

[13] A. Morel, A. Quelen, P. Gasnier, R. Grézaud, S. Monfray, A. Badel, and G. Pillonnet, "A Shock-Optimized SECE Integrated Circuit," *IEEE J. Solid-State Circuits*, vol. 53, no. 12, pp. 3420–3433, Dec. 2018.

[14] A. Morel, A. Quelen, C. A. Berlitz, D. Gibus, P. Gasnier, A. Badel, and G. Pillonnet, "Self-Tunable Phase-Shifted SECE Piezoelectric Energy-Harvesting IC with a 30nW MPPT Achieving 446% Energy-Bandwidth Improvement and 94% Efficiency," in *IEEE Int. Solid-State Circuits Conf. (ISSCC) Dig. Tech. Papers*, Feb. 2020, pp. 488–490.

[15] G. Shi, Y. Xia, X. Wang, L. Qian, Y. Ye, and Q. Li, "An Efficient Self-Powered Piezoelectric Energy Harvesting CMOS Interface Circuit Based on Synchronous Charge Extraction Technique," *IEEE Trans. Circuits Syst. I, Reg. Papers*, vol. 65, no. 2, pp. 804–817, Feb. 2018.

[16] D. Kwon and G. A. Rincon-Mora, "A Single-Inductor AC-DC Piezoelectric Energy-Harvester/Battery-Charger IC Converting ±(0.35 to 1.2V) to (2.7 to 4.5V)," in *IEEE Int. Solid-State Circuits Conf. Dig. Tech. Papers*, Feb. 2010, vol. 53, pp. 494–495.

[17] D. Kwon and G. A. Rincón-Mora, "A Single-Inductor 0.35 µm CMOS Energy-Investing Piezoelectric Harvester," *IEEE J. Solid-State Circuits*, vol. 49, no. 10, pp. 2277–2291, Oct. 2014.

[18] S. Yang and G. A. Rincón-Mora, "Energy-Harvesting Piezoelectric-Powered CMOS Series Switched-Inductor Bridge," *IEEE Trans. Power Electron.*, vol. 34, no. 7, pp. 6489–6497, Jul. 2019.

[19] S. Yang and G. A. Rincón-Mora, "Efficient Power Transfers in Piezoelectric Energy-Harvesting Switched-Inductor Chargers," *IEEE Trans. Circuits Syst. II, Exp. Brief*, vol. 68, no. 4, pp. 1248-1252, April 2021.

[20] K. Yoon, S. Hong, and G. Cho, "Double Pile-Up Resonance Energy Harvesting Circuit for Piezoelectric and Thermoelectric Materials," in *IEEE J. Solid-State Circuits*, vol. 53, no. 4, pp. 1049-1060, April 2018.

[21] S. Yang and G. A. Rincón-Mora, "Piezoelectric CMOS Charger: Highest Output-Power Design," in *Proc. 21st Int. Symp. Qual. Electron. Design (ISQED)*, 2020, pp. 292-297.

[22] B. Razavi, "The Bootstrapped Switch [A Circuit for All Seasons]," *IEEE Solid-State Circuits Mag.*, vol. 7, no. 3, pp. 12–15, Sep. 2015.

[23] S. Sankar, P. -H. Chen and M. S. Baghini, "An Efficient Inductive Rectifier Based Piezo-Energy Harvesting Using Recursive Pre-Charge and Accumulation Operation," *in IEEE Journal of Solid-State Circuits (Early Access)*, doi: 10.1109/JSSC.2022.3153590.

# Chapter 6

# Conclusion

This chapter summarises the work presented in this thesis. Four different techniques of energy-saving are presented for energy-constrained CMOS circuits and systems. The proposed techniques help in improving the energy autonomy or extending the battery life by reducing the inherent energy consumption and harvesting the ambient energy. Further scope in continuing the research outcomes of this thesis is also presented.

## 6.1 Summary of the Work

- In **Chapter-2**, Switched-Capacitor assisted Power Gating (SwCap PG) is proposed for reducing the leakage currents of large digital circuits [1]. For the first time, the PG switch is biased in the super turn-off and the super turn-on mode during the off-state and the on-state, respectively. A simple switched-capacitor network reconfigures and biases the PG switch in four possible states with low area and power overhead. During the super turn-off, voltage stress is avoided in the PG switch when the circuit load uses supply voltage equal to the nominal $V_{DD}$ in a given technology, and maximum possible leakage current reduction is achieved by the optimal biasing of the gate voltage. The proposed SwCap PG is experimentally validated in the 180nm CMOS technology. Measurement results of CMOS SwCap PG show that leakage current and $R_{ON}$ reduce by 186-226× and 18%, respectively, as compared to the conventional PG. An alternate solution for the SwCap network using MEMS devices as the switching elements is implemented for additional benefits. Measurement results of MEMS SwCap PG show that leakage current and $R_{ON}$ reduce by 172× and 26%, respectively, compared to the conventional PG. Finally, the applicability of the SwCap PG in the nano-scale CMOS technologies is addressed.
- In **Chapter-3**, the usefulness and the conditions under which NEMS power gating is beneficial over conventional CMOS power gating are evaluated [2]. Nano-electromechanical switches (NEMS) in principal offer close to infinite resistance in the off-state. So by using NEMS switches for power gating, the on-state performance can be improved by adding multiple parallel switches without increasing the leakage current during the off-state. This eventually breaks the trade-off mentioned earlier. NEMS switches have been recently proposed for the power gating application. However, a detailed analysis of the conditions at which the NEMS devices will have an impact is missing. The technique of power gating is analyzed with a NEMS switch using detailed circuit-level simulations to obtain the conditions under which one can obtain net energy savings as compared with FinFET-based power gating. Finally, applicability in the energy reduction on a system-on-chip for a mobile platform made using a 14 nm gate length FinFET device is evaluated.

- In **Chapter-4**, a novel technique of realizing discrete-time (D-T) signal amplification using nano-electromechanical switches (NEMS) is proposed [3]. The amplifier uses mechanical switches instead of traditional solid-state devices and acts as an inherent sample and hold amplifier. The proposed NEMS D-T amplifier operates on a wide dynamic range of signals without consuming any dc power. Moreover, the proposed amplifier does not suffer from the leakage current and the non-linearity associated with the sampling ohmic switch. As a proof of concept, the proposed NEMS D-T amplifier is demonstrated in circuit simulations using the calibrated Verilog-A models of the NEMS device. The simulated amplifier achieves a gain of ~5, handles the maximum differential input signal of 0.65 V, and consumes only 0.6 µW of dynamic power for a sampling frequency of 100 kHz. The non-idealities present in the proposed amplifier are highlighted, and possible ways to overcome them are discussed. Finally, the design considerations required for the NEMS D-T amplifier are described.

- In **Chapter-5**, a new efficient method of extracting power from the piezoelectric transducer using the inductive rectifier is proposed [4]. In the proposed method, the internal capacitance of the transducer is initially pre-charged, and the generated charges in response to the mechanical vibration are accumulated on the capacitor. When the accumulated voltage reaches a maximum allowed value ($V_{MAX}$), the total energy stored in the capacitor is transferred to the output, and the process is repeated. The proposed method ensures that the voltage swing across the transducer is always maximum for extracting the highest possible output power for given CMOS technology. The operational steps are self-timed, and the negative voltage swing across the transducer is avoided. The proposed inductive rectifier is realized using the buck-boost power stage of the DC energy harvesting system for enabling efficient multi-source harvesting. Test chip is fabricated in the 180 nm CMOS technology, having a $V_{MAX}$ of 3.3 V.

The proposed rectifier extracts power in a single stage, even at a lower rectified output voltage ≤ 1 V. For a piezo open-circuit voltage of 1 V, the proposed rectifier extracts 3.68× more power than the maximum output power of a full-bridge rectifier with ideal diodes. This is despite the voltage loss occurring during accumulation in the power stage implementation.

The required area for implementing the proposed energy saving techniques in this thesis are duly summarized in Table 6.1.

Table 6.1: Area requirement for implementing the proposed energy-saving techniques

| Chapter | Proposed method | Area requirement |
|---|---|---|
| 2 | SwCap Power Gating | In 180nm CMOS technology, the SwCap network and the control circuit occupies an area of 0.0065 mm$^2$ (14% of the PG switch area). However, for the same $R_{ON}$, the total area including the PG switch is reduced by 6.5%, due to the super turn-on implementation. |
| 3 | NEMS Power Gating | Minimum estimated area required for implementing the NEMS PG switch for power gating the test logic block considered in this work in 14 nm FinFET technology is 1725 μm$^2$ (81.9% of the logic block area). |
| 4 | NEMS D-T Amplifier | Minimum estimated area required for implementing the NEMS switches in the proposed NEMS D-T amplifier (with differential outputs) is 1250 μm$^2$. |
| 5 | Proposed inductive rectifier for PEH | Total silicon area occupied by the proposed inductive rectifier circuit in the 180nm CMOS technology is 1 mm$^2$. |

*Area overhead depends on the type of MEMS integration (PCB level/die level/3D integration) being used.

## 6.2 Future Scope of Research

### 6.2.1 Switched-capacitor assisted power gating

The proposed SwCap PG always allows for optimal biasing ($V_{GSP(opt)}$) in the super turn-off state for maximum possible leakage reduction, as discussed in Section 2.2 of Chapter-2. If the $V_{GSP}$ of the PMOS PG switch exceeds $V_{GSP(opt)}$, the leakage current starts to increase due to the GIDL effect, as explained in Section 2.2 of Chapter-2. In [5], the optimal bias point for biasing the super turn-off PG switch was obtained by comparing the total leakage current and the GIDL current of two different replica PMOS transistors. The method from [5] requires trans-impedance amplifiers and sense amplifiers. A simpler method of obtaining the optimal bias point ($V_{GSP(opt)}$) is proposed in Fig. 6.1. The scaled replica path mirrors the scaled version of the switch leakage current to bias the ring-oscillator, as shown in Fig. 6.1. The optimal bias point ($V_{GSP(opt)}$) occurs at a voltage that gives the lowest ring oscillator frequency ($f_{OSC}$). The optimal bias point varies with temperature, as mentioned in Section 2.2 of Chapter-2. Performing a run time calibration using the method suggested in Fig. 6.1 allows for the estimation of the optimal bias point in the super turn-off state.

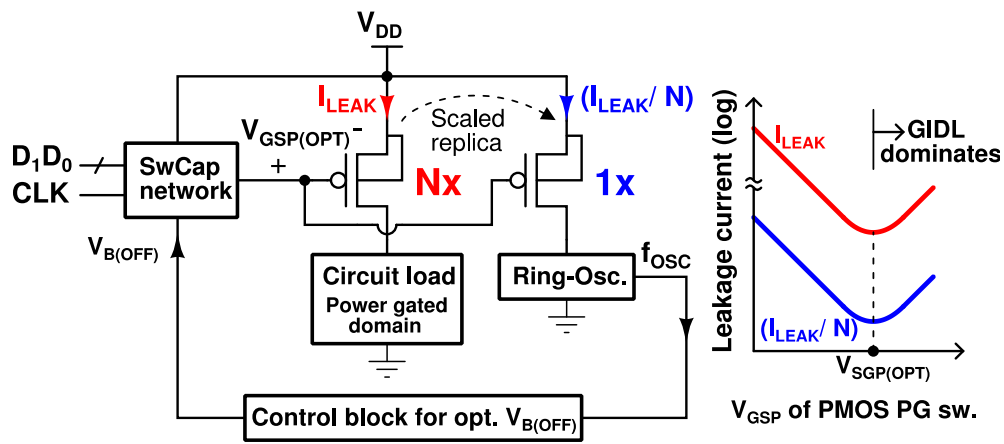

Fig. 6.1. Proposed method of obtaining optimal bias point ($V_{GSP(opt)}$) in the super turn-off state.

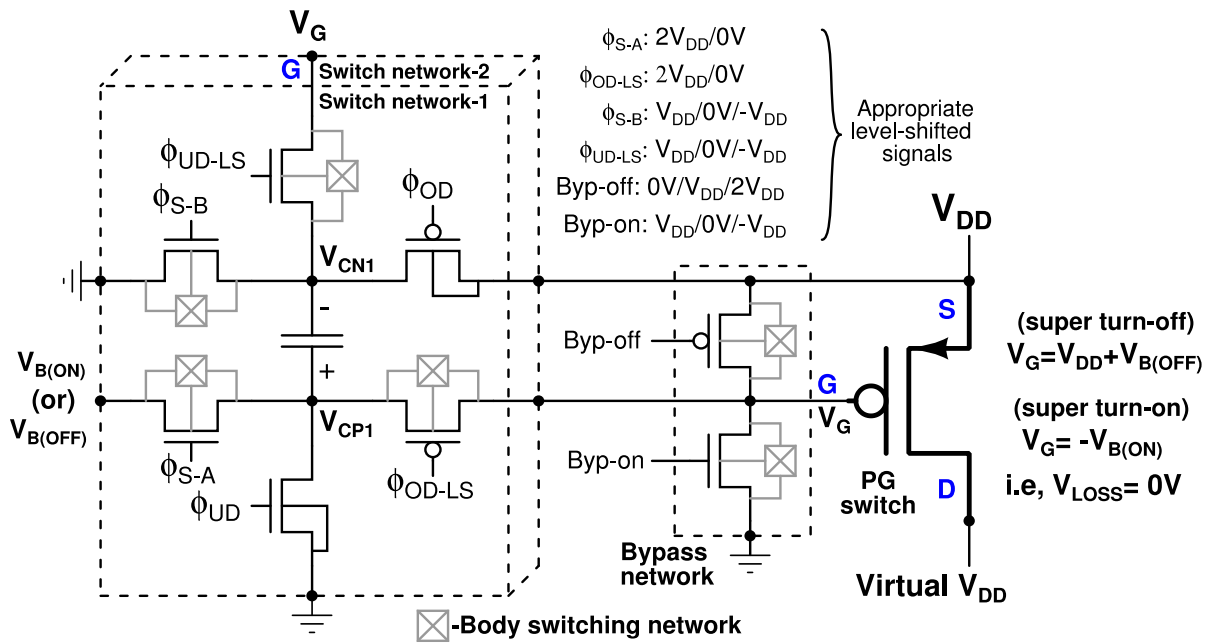

Fig. 6.2. Modified implementation of the switched-capacitor network for avoiding voltage loss in the CMOS SwCap PG.

The voltage swings in the switched-capacitor network exceed beyond the supply rails (above $V_{DD}$ and below 0V). A simpler method of turning off the transistors in the CMOS switch network was proposed in this work. However, it incurs a voltage loss, as discussed in Section 2.4.2 of Chapter-2. An alternate method of implementing the switched-capacitor network to avoid the voltage loss is suggested in Fig. 6.2. In this implementation, the control signals to the switch network are appropriately level shifted and provided, as shown in Fig. 6.2. The body terminals are dynamically switched to the appropriate nodes using the arrangement used in [4].

Hence, by this implementation, the gate voltage in the CMOS SwCap PG is equal to its ideal value of "$V_{DD} + V_{B(OFF)}$" and "$-V_{B(ON)}$" during the super turn-off and super turn-on state, respectively. The additional circuit complexity of generating the voltages "$2V_{DD}$" and "$-V_{DD}$" using a traditional charge pump [6] is shared among all the SwCap network blocks on the chip.

### 6.2.2 NEMS power gating

The usefulness of NEMS power gating with respect to FinFET based gating is evaluated using a digital test circuit in Chapter-3 of this thesis. The method of including the external NEMS PG switches during the ASIC design phase of the CMOS digital circuits is outlined in Section 3.7.2 of Chapter-3. This paves the way for easier incorporation of the electrical characteristics of the NEMS PG switches without any modification to the digital ASIC design flow during the power and timing analysis of the chip.

The shoot-through (or inrush) current is an important consideration during the turn-on event of the PG switch. Multiple techniques have been developed for the management of inrush currents in power gating [R-6], [R-7]. For example, multiple parallel units of the PG switch are turned-on in a staggered manner to reduce the inrush currents. The main goal of Chapter-3 was to evaluate the energy saving performance of NEMS and FinFET gating for a SoC based on the leakage current and $R_{ON}$, the effect of inrush currents and the process variations were not considered. It will be interesting to evaluate the NEMS and FinFET PG switches for inrush currents since NEMS has higher delays than FinFET switches due to its mechanical nature. The effect of in-rush currents and process variations for NEMS and FinFET power gating could be evaluated as a possible future research scope.

### 6.2.3 NEMS discrete-time amplifier

A new technique of obtaining discrete-time (D-T) signal amplification using the NEMS switches is proposed in Chapter-4 of this thesis. The proposed method is validated using the calibrated Verilog-A models of the NEMS switch designed in the MEMS CAD tool (CoventorWare). The designed NEMS switches have to be fabricated in the cleanroom process and experimentally demonstrated. The fabricated NEMS switches required for amplification can be interfaced with the CMOS chip at the PCB level using wire-bonding. However, for reliability, the NEMS switches have to be hermetically packaged and soldered to the PCB. The NEMS D-T amplifier has to drive the input capacitance of the next stage. Hence, an on-chip source-follower stage is required to interface the NEMS D-T amplifier with the subsequent stages [7]. This source-follower stage acts as a voltage buffer and thus avoids any loading.

As discussed in Section 4.5 of Chapter-4, the non-linearity of the proposed NEMS D-T amplifier is influenced by the C-V characteristics of the NEMS capacitive switch, which in turn is dependent on the physical parameters of the switch considered. The effect of physical parameters of the NEMS capacitive switch on the amplifier non-linearity is not much explored as a part of this thesis. The evaluation of NEMS capacitive switch parameters for achieving desired amplifier non-linearity can be an extended research direction for this Chapter. Moreover, the evaluation of the proposed NEMS D-T amplifier for the use in applications like ADC, sensor front-end signal chains etc., can provide an indication of how the NEMS switches affects the overall system performance. This can be simultaneously used to co-design the parameters of the NEMS switch required.

### 6.2.4 Inductive rectifier for piezo-energy harvesting

An efficient inductive rectifier based piezo-energy harvesting is proposed and implemented in Chapter-5. The proposed inductive rectifier utilizes the existing buck-boost power stage of the DC energy harvesting system [8]. Moreover, the work from [8] allows the excess harvested energy to be stored and re-utilized later to power the load. As seen from Chapter-5, the inductor is utilized in the piezo energy harvesting (PEH) only during the exchange of energy between the piezo transducer and the output. The inductor utilization time for the PEH is typically 1% of the vibration time period ($T_{PZ}$). Hence, for the remaining time, the inductor can be utilized for solar and thermal energy harvesting.

Fig. 6.3 shows the simplified power stage for harvesting solar, thermal, and piezo energy simultaneously together. As observed from Fig. 6.3, the inductor is utilized for solar and thermal energy harvesting most of the time. The inductor is utilized by the piezo energy harvesting system approximately for 1% of the vibration time period. The energy from the solar and thermal sources is regulated and delivered to the output load. The excess energy is stored in the storage element (battery or capacitor). In contrast, the energy extracted from the piezo transducer is un-regulated and dumped into the storage element. Hence, the proposed power stage in Fig. 6.3 enables an efficient multi-source energy harvesting interface for sporadic ambient conditions.

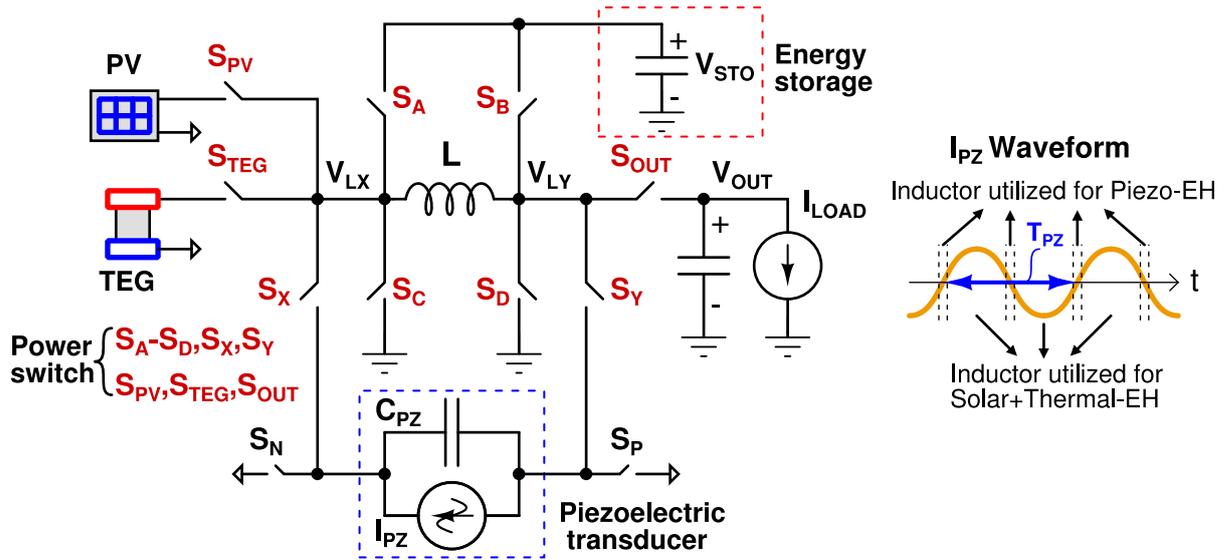

Fig. 6.3. Simplified schematic showing the arrangement for multi-source harvesting of solar, thermal and vibration energy sources together using the same power stage.

The conduction loss in the proposed inductive rectifier could be reduced by using a larger inductor along with sophisticated offset-cancelled ZCD comparators. A fraction of the inductor current after the energy transfer step could be used for pre-charging $C_{PZ}$, before dumping the remaining current to the output. In this way, a separate inductor energizing step could be avoided.

## 6.3 References

[1] S. Sankar, M. Goel, P. -H. Chen, V. R. Rao, and M. S. Baghini, "Switched-Capacitor-Assisted Power Gating for Ultra-Low Standby Power in CMOS Digital ICs," *IEEE Trans. Circuits Syst. I, Reg. Papers*, vol. 67, no. 12, pp. 4281-4294, Dec. 2020.

[2] S. Sankar, U. S. Kumar, M. Goel, M. S. Baghini and V. R. Rao, "Considerations for Static Energy Reduction in Digital CMOS ICs Using NEMS Power Gating," in *IEEE Transactions on Electron Devices*, vol. 64, no. 3, pp. 1399-1403, March 2017.

[3] S. Sankar, M. Goel, M. S. Baghini and V. R. Rao, "A Novel Method of Discrete-Time Signal Amplification Using NEMS Devices," in *IEEE Transactions on Electron Devices*, vol. 65, no. 11, pp. 5111-5117, Nov. 2018.

[4] S. Sankar, P. -H. Chen and M. S. Baghini, "An Efficient Inductive Rectifier Based Piezo-Energy Harvesting Using Recursive Pre-Charge and Accumulation Operation," *in IEEE Journal of Solid-State Circuits (Early Access)*, doi: 10.1109/JSSC.2022.3153590.

[5] A. Valentian and E. Beigne, "Automatic Gate Biasing of an SCCMOS Power Switch Achieving Maximum Leakage Reduction and Lowering Leakage Current Variability," *IEEE J. Solid-State Circuits*, vol. 43, no. 7, pp. 1688-1698, July 2008.

[6] P.-H. Chen, C.-S. Wu, and K.-C. Lin, "A 50 nW-to-10 mW output power tri-mode digital buck converter with self-tracking zero current detection for photovoltaic energy harvesting," *IEEE J. Solid-State Circuits*, vol. 51, no. 2, pp. 523–532, Feb. 2016.

[7] I. Ahmed, J. Mulder and D. A. Johns, "A Low-Power Capacitive Charge Pump Based Pipelined ADC," *IEEE J. Solid-State Circuits*, vol. 45, no. 5, pp. 1016-1027, May 2010.

[8] P.-H. Chen, H.-C. Cheng, and C.-L. Lo, "A Single-Inductor Triple-Source Quad-Mode Energy-Harvesting Interface With Automatic Source Selection and Reversely Polarized Energy Recycling," *IEEE J. Solid-State Circuits*, vol. 54, no. 10, pp. 2671–2679, Oct. 2019.

[9] Michael Keating, David Flynn, Rob Aitken, Alan Gibbons, and Kaijian Shi, "Low Power Methodology Manual: For System-on-Chip Design", Springer, 2007.

[10] Kaijian Shi and Jingsong Li, "A wakeup rush current and charge-up time analysis method for programmable power-gating designs," *in IEEE International SOC Conference*, 2007, pp. 163-165.

# List of Publications

[1] **S. Sankar**, U. S. Kumar, M. Goel, M. S. Baghini and V. R. Rao, "Considerations for Static Energy Reduction in Digital CMOS ICs Using NEMS Power Gating," in *IEEE Transactions on Electron Devices*, vol. 64, no. 3, pp. 1399-1403, March 2017.

[2] **S. Sankar**, M. Goel, M. S. Baghini and V. R. Rao, "A Novel Method of Discrete-Time Signal Amplification Using NEMS Devices," in *IEEE Transactions on Electron Devices*, vol. 65, no. 11, pp. 5111-5117, Nov. 2018.

[3] **S. Sankar**, M. Goel, P. -H. Chen, V. R. Rao, and M. S. Baghini, "Switched-Capacitor-Assisted Power Gating for Ultra-Low Standby Power in CMOS Digital ICs," *IEEE Trans. Circuits Syst. I, Reg. Papers*, vol. 67, no. 12, pp. 4281-4294, Dec. 2020.

[4] **S. Sankar**, P. -H. Chen and M. S. Baghini, "An Efficient Inductive Rectifier Based Piezo-Energy Harvesting Using Recursive Pre-Charge and Accumulation Operation," *in IEEE Journal of Solid-State Circuits (Early Access)*, doi: 10.1109/JSSC.2022.3153590.

# Acknowledgments

I want to thank my supervisor Prof. Maryam Shojaei Baghini, co-supervisor Prof. Ramgopal Rao, and external co-supervisor Prof. Po-Hung Chen (NCTU) for their guidance during the course of my Ph.D. Mayank Goel from Intel also offered his valuable guidance for my research work. The research outcomes from this thesis would not have been possible without their guidance. My research progress committee members - Prof. Dinesh Sharma, Prof. Virendra Singh, and Prof. Madhav Desai, provided some critical feedback and suggestions during the course of my research. This helped me to go in the right direction, and I am very much thankful for that.

I would also like to sincerely thank all the members of the VLSI Lab, Embedded systems lab, and ISL Lab at IIT Bombay, with whom I interacted and greatly benefited. I want to thank the members of the 307 Laboratory at NCTU who hosted and helped me during my academic visit to NCTU Taiwan. I am also very grateful and thankful to interact with wonderful people at IIT Hyderabad and got immense benefit from them.

I would like to acknowledge the financial support from the Ministry of Electronics and Information Technology (MeitY), Government of India, Department of Science and Technology (DST), Government of India. I would also like to thank the National Chip Implementation Centre (CIC), Taiwan, for funding the chip fabrication.

Finally, I would also like to thank all the members who directly or indirectly helped me navigate through the years during my Master's and Ph.D. programme. Without them, it would have been very difficult to achieve this milestone.

**Sivaneswaran Sankar**